\documentclass[aps,prb,reprint,amsmath,amssymb,longbibliography,
  superscriptaddress,floatfix]{revtex4-2}
\usepackage{graphicx}
\usepackage{bm}
\usepackage{amsfonts}
\usepackage{xcolor}
\usepackage{array}[=2016-10-06]
\usepackage{tabularx}
\usepackage{multirow}
\usepackage{makecell}
\usepackage{braket}
\usepackage[percent]{overpic}
\usepackage{adjustbox}
\usepackage{comment}
\usepackage[colorlinks=true,pdfstartview=Fit,
  linkcolor=blue,citecolor=blue,urlcolor=blue]{hyperref}
\usepackage{hypcap}
\hypersetup{pdftitle={Certifying Majorana persistence through Hamiltonian learning from boundary measurements},
  pdfauthor={Tohid Farajollahpour and Alireza Qaiumzadeh}}
\DeclareMathOperator{\Tr}{Tr}
\DeclareMathOperator{\diag}{diag}
\DeclareMathOperator{\sgn}{sgn}

\newcommand{\topinv}{\mathcal{M}}

\newcommand{\norm}[1]{\left\lVert#1\right\rVert}

\DeclareRobustCommand{\secref}[1]{Sec.~\ref{#1}}
\begin{document}
\title{Certifying Majorana persistence through Hamiltonian learning from boundary measurements}
\author{Tohid Farajollahpour}
\email{tohid.farajollahpour@ntnu.no}
\author{Alireza Qaiumzadeh}
\email{alireza.qaiumzadeh@ntnu.no}
\affiliation{Center for Quantum Spintronics, Department of Physics,
Norwegian University of Science and Technology, NO-7491 Trondheim, Norway}
\begin{abstract}

Finite measurements at a chain end can certify local quantum dynamics without determining bulk topology. We combine Hamiltonian learning with commutator bounds to certify the persistence of localized Majorana operators in Kitaev chains. Time-resolved single-shot boundary measurements and local bond calibrations define joint confidence sets for local Hamiltonian coefficients. Under explicit device assumptions, these sets yield state-independent lower bounds on symmetrized autocorrelations with a prescribed failure probability, valid for all compatible continuations. With equal hopping and pairing on the first bond and no additional boundary terms, an exact identity enables inference from the end qubit alone, even with arbitrary time-independent interactions elsewhere. Synthetic benchmarks near a boundary sweet spot certify persistence for up to about 60 times the measurement window. Three- and six-qubit IBM records strengthen conditional guarantees for the programmed coherent Hamiltonian. Applications to quantum-dot models relate tuning precision to persistence times and detuning-induced splitting.
\end{abstract}
\maketitle
\section{Introduction}
\label{sec:introduction}


Certifying a quantum device from restricted access is a recurring task.
An experimenter who controls a few qubits of a simulator, or one end of
a wire, must decide from finite data what the device does and does not
guarantee. We study this task for the Majorana modes of a Kitaev chain
probed at one end. Spatially separated Majorana zero modes in
one-dimensional topological superconductors are candidates for
topological quantum information processing
\cite{Kitaev2001,ReadGreen2000,FuKane2008,Lutchyn2010,Oreg2010,Alicea2012}.
Their accessible boundary signatures, including zero-bias features in
nanowires and magnetic-atom chains \cite{Mourik2012,NadjPerge2014,Jeon2017}, can
arise from topological or trivial mechanisms
\cite{LiuSauStanescuDasSarma2017,PanDasSarma2020,PradaReview2020}, and
establishing bulk topology requires additional evidence~\cite{Yazdani2023}.
Following the operational emphasis of gap-testing and parity-measurement
protocols \cite{TopoGapProtocol,TopoQubitReadout}, we ask what a
finite-time record from one end establishes under explicit assumptions
and with an error probability fixed in advance, and whether a bulk label
can be identified uniformly at fixed time and finite cost.
 
Two platforms make the question concrete. Programmable qubit arrays
realize the Kitaev chain through the Jordan-Wigner map, which on the equal
line, where hopping and pairing amplitudes coincide,
gives the transverse-field Ising model. Edge-qubit autocorrelations of
Majorana edge modes have been measured in a kicked Ising chain on a
superconducting processor \cite{Mi2022}. Gate-programmable quantum-dot
arrays realize the chain with electrons
\cite{SauDasSarma2012,LeijnseFlensbergPRB2012,Liu2022,Tsintzis2022,Bordin2023}.
Two-site \cite{Dvir2023,tenHaaf2024,Zatelli2024} and three-site
\cite{Bordin2025,tenHaaf2025,Bordin2026} chains have been realized, with
spectroscopy of edge and bulk sites, localization tests with an
additional dot, and single-shot parity readout \cite{vanLoo2026}, which
support concrete qubit proposals \cite{Tsintzis2024}. Yet a localized
few-site mode can become a trivial bound state when the chain is extended
\cite{LuethiFate2025,LuethiFalse2025}, so the measured finite device and
its assumed thermodynamic continuation are different inference targets.
We formulate the protocols for qubit simulators, where all boundary
response entries are directly accessible.
 
We use Hamiltonian learning to convert finite boundary records into a
persistence guarantee for a specified boundary-localized Majorana
operator. Under a declared device model, the records and available
calibrations define a joint confidence set for the local Hamiltonian
coefficients and their couplings to the rest of the chain
\cite{Hoeffding1963,JaulinWalter1993,Rump2010}. Bounding the commutator
norm over this set gives a time during which the operator's symmetrized
autocorrelation remains above a chosen threshold, for every initial
state and every compatible continuation, at a stated failure
probability. This uses the approximate-conservation principle underlying
strong and almost strong zero modes
\cite{Fendley2016,Kemp2017,Else2017,BozkurtStrongZero2025}. We also use
finite-range propagation bounds
\cite{LiebRobinson1972,BravyiHastingsVerstraete2006,ChenLucasYin2023}
to quantify limitations on distinguishing bulk topological labels from
boundary records. The certificate concerns the persistence of one
operator; quantum-memory performance requires a separate assessment
\cite{GoldsteinChamon2011,Mazza2013,Ippoliti2016}.
 
Our work builds on restricted-access Hamiltonian learning
\cite{BurgarthMaruyama2009,BurgarthMaruyamaNori2009,BurgarthYuasa2012,
DiFrancoPaternostroKim2009,ZhangSarovar2014,ZhangSarovar2015,
SoneCappellaro2017,WangDong2020,GutaYamamoto2016}, quadratic-chain
tomography \cite{BurgarthMaruyamaNori2011}, and rigorous learning of local
Hamiltonians from local or dynamical data
\cite{Wiebe2014,Bairey2019,AnshuArunachalam2021,HuangLearning}, with links
to inverse Jacobi theory \cite{GesztesySimon1997,DamanikPushnitskiSimon}
and robust free-Hamiltonian reconstruction on a superconducting processor
\cite{Hangleiter2024}. Complementary spectroscopy-based and machine-learned
methods infer or tune Majorana-device parameters
\cite{FulgaAdaptive2013,KochGreplova2023,TaylorSauDasSarma2024,Benestad2024,
KrawczykPawlowski2026,Dourado2026,vanDriel2026,ThammRosenow2023,ThammRosenow2024}.
Related work learns local conservation laws from randomized measurements
\cite{ZhanConservation2024}, reconstructs spectral information through
probe-only Krylov learning \cite{MatsuuraQuantumKrylov2026}, and relates
Majorana locality to protection and braiding \cite{Samuelson2026}.
Here, we propagate a joint coefficient confidence set, including outgoing
couplings, into a commutator bound and a persistence interval for a
specified physical operator. Under the declared device model, the
guarantee holds for every initial state and compatible Hamiltonian.
See Supplemental Material for more details on the assumptions,
measurement access, and guarantees
(Sec.~\ref{SM-sec:supp-related-learning}).
This differs from
testing proximity to a specified Hamiltonian
\cite{GaoHamiltonianCertification2026} and from the Krylov break times
of Ref.~\citenum{MatsuuraKawamuraSalfiIso2026}, which bound the time over
which a truncated effective chain reproduces a return amplitude.
Self-testing approaches certify Majorana-related measurement
observables from their outcome correlations
\cite{Irfan2020,MichalskiMajoranaCertification2026}, addressing a
different property from their dynamical persistence.
Majorana-mode tomography accesses spatial
Majorana strings, directly or through gates mapping them to boundary
readout \cite{HarleShtankoMovassagh2023}. Here the target response is read
out at one boundary, with separately
counted calibration and reference controls where required.

In this work, we derive an exact boundary identity that enables
certification of the end Majorana's persistence using measurements
on the end qubit alone. The identity applies when the first bond has
equal hopping and pairing and no additional boundary terms are
present; the remaining qubits may carry arbitrary time-independent
interactions. Near a boundary sweet spot,
synthetic benchmarks certify persistence for up to about 60 times
the observation window. We also construct chains with a common
boundary region and opposite bulk labels to establish limits on
uniform topological identification at fixed time and a finite
measurement budget. Finally, the bounds relate static
tuning precision to persistence and energy splitting in quantum-dot
models, while three- and six-qubit IBM records tighten conditional
certificates for the programmed coherent Hamiltonian.
 
\section{Boundary experiment and exact response}
\label{sec:setup}
We consider $L$ spinless fermionic sites with open boundaries and set
$\hbar=1$ and the lattice spacing to one. 

\subsection{Model, boundary response, and identifiability}
The Hamiltonian is
\begin{align}
H=&-\sum_{j=1}^{L}\mu_j\left(c_j^\dagger c_j-\frac12\right)
\nonumber\\
&+\sum_{j=1}^{L-1}
\left[-J_jc_j^\dagger c_{j+1}
+\Delta_jc_jc_{j+1}+\mathrm{h.c.}\right].
\label{eq:H}
\end{align}
Here $c_j$ annihilates a fermion on site $j$, $\mu_j$ is its onsite chemical
potential, and $J_j$ and $\Delta_j$ are the hopping and real $p$-wave pairing
amplitudes on bond $(j,j+1)$.

We collect the Hermitian Majoranas $a_j=c_j+c_j^\dagger$ and
$b_j=-i(c_j-c_j^\dagger)$ into
$\bm\gamma=(a_1,b_1,\ldots,a_L,b_L)^T$, with
$\{\gamma_r,\gamma_s\}=2\delta_{rs}$ for $r,s=1,\ldots,2L$.
Equation~\eqref{eq:H} then reads
\begin{equation}
H=\frac{i}{4}\bm\gamma^TK\bm\gamma,
\end{equation}
where $K$ is a real antisymmetric $2L\times2L$ generator. Its only
nonzero site blocks are $K_{jj}=A_j$, $K_{j,j+1}=B_j$, and
$K_{j+1,j}=-B_j^T$, with
\begin{equation}
A_j=\begin{pmatrix}0&-\mu_j\\ \mu_j&0\end{pmatrix},
\qquad
B_j=\begin{pmatrix}0&\Delta_j-J_j\\J_j+\Delta_j&0\end{pmatrix}.
\label{eq:blocks}
\end{equation}
At evolution time $t$, the Heisenberg equation
$\dot{\bm\gamma}(t)=i[H,\bm\gamma(t)]=K\bm\gamma(t)$ has the solution
$\bm\gamma(t)=e^{Kt}\bm\gamma(0)$.  Since $K$ is time independent and
antisymmetric, $e^{Kt}$ is real orthogonal.  

\begin{figure*}[t]
\centering
\includegraphics[width=\textwidth,height=0.57\textheight,keepaspectratio]{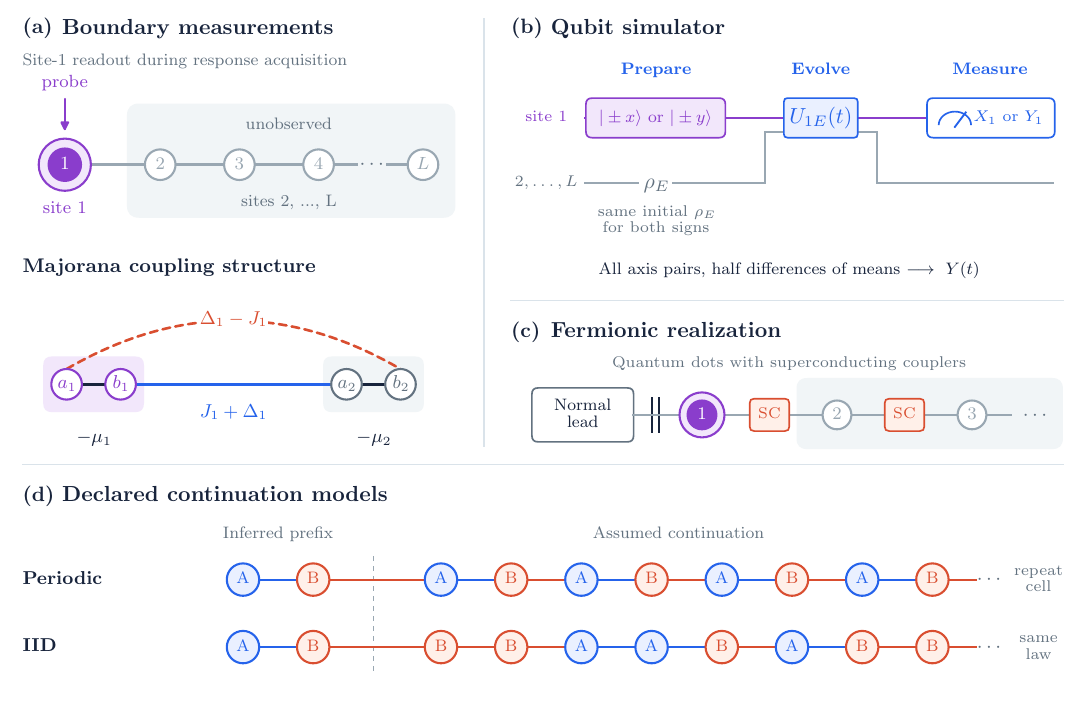}
\caption{Boundary-response acquisition and declared continuations.
(a) Site-1 probing of a coupled chain during response acquisition. The inset shows
the local Majorana couplings.  (b) Jordan-Wigner qubit protocol with
independent preparation and measurement axes and joint evolution
$U_{1E}(t)$ of the boundary qubit and hidden register $E$. (c) Effective
quantum-dot realization probed by normal tunneling. (d) Periodic and IID
continuations. Symbols A and B denote two possible coefficient triples
$(\mu_j,J_j,\Delta_j)$, with each colored outgoing bond included in its
triple. The divider separates the inferred prefix from the assumed
continuation.}
\label{fig:setup}
\end{figure*}

During boundary-response acquisition, the probe or local preparation
and readout address only site $1$ (Fig.~\ref{fig:setup}(a)). Fermionic
parity superselection rules out preparing an
isolated electronic device in an eigenstate of a parity-odd Majorana
operator.  We therefore describe the ideal isolated-chain fermionic
response by the parity-even boundary anticommutator matrix.  Using the
boundary labels $\gamma_{a_1}=a_1$ and $\gamma_{b_1}=b_1$, we define
\begin{equation}
Y_{\alpha\beta}(t)=\frac12
\left\langle\{\gamma_\alpha(t),\gamma_\beta(0)\}\right\rangle,
\qquad \alpha,\beta\in\{a_1,b_1\},
\label{eq:response_definition}
\end{equation}
where $\langle O\rangle=\operatorname{Tr}(\rho O)$ for an arbitrary initial
many-body density operator $\rho$. The boundary response matrix $Y(t)$
collects the overlaps of the time-evolved boundary Majoranas with the
initial ones.
Using the Majorana evolution and the canonical anticommutation
relations, we obtain
\begin{equation}
\{\gamma_\alpha(t),\gamma_\beta(0)\}
=
2(e^{Kt})_{\alpha\beta}\,\mathbb I,
\end{equation}
where $\mathbb I$ is the many-body identity. Consequently,
the boundary anticommutator response is
\begin{equation}
Y(t)=Pe^{Kt}P^T,
\label{eq:exact_response}
\end{equation}
with $P=(I_2\ \ 0)$ selecting $(a_1,b_1)$.
For fixed $K$, this response is independent of the initial
many-body state, including its temperature and correlations.

Within a fixed physical model and sign convention, identifiability asks
whether different coefficient values can produce the same response. For
ideal continuous-time data the connection to $K$ is expressed by the
boundary moments, the time derivatives of $Y(t)$ at $t=0$,
\[
\left.\frac{d^nY(t)}{dt^n}\right|_{t=0}
=
PK^nP^T,
\qquad n=0,1,\ldots,
\]
which link the short-time signal to the Hamiltonian coefficients
\cite{GesztesySimon1997,MoszynskiSwiderski2025,HoKalman1966}. The response
is invariant under orthogonal changes of the hidden Majorana basis,
$K\mapsto OKO^T$ with $O=I_2\oplus O_E$, which can take $K$ outside the
Kitaev form of Eq.~\eqref{eq:blocks}. We therefore infer coefficients
within the physical model and a fixed sign gauge, and the confidence-set
procedure below tests properties shared by every compatible Hamiltonian,
including equivalent branches such as the two-site pair
$(\mu_1,\mu_2,J,\Delta)$ and $(\mu_1,-\mu_2,\Delta,J)$, which are related
by $c_2\mapsto c_2^\dagger$ \cite{JaulinWalter1993,ChristandlRenner2012}.

Bulk labels refer to assumed continuations. For real coefficients the
model belongs to Altland-Zirnbauer class BDI, and we report the parity of
its integer invariant as the class-D $\mathbb Z_2$ label
\cite{AltlandZirnbauer1997,Schnyder2008,KitaevPeriodic2009}. An infinite
uniform chain with $\mu_j=\mu$, $J_j=J$, and $\Delta_j=\Delta\ne0$ has,
when the bulk is gapped, the invariant
\cite{Kitaev2001,KitaevPeriodic2009}
\begin{equation}
\topinv=\sgn(\mu^2-4J^2),
\label{eq:clean_parity}
\end{equation}
where $\topinv=-1$ denotes the topological phase $|\mu|<2|J|$ and
$\topinv=+1$ the trivial phase. At $|\mu|=2|J|$ the bulk gap closes.

\subsection{Boundary protocols}
\label{sec:assumptions}

The information available for coefficient inference depends on which
entries of the boundary response matrix $Y(t)$ are measured. We consider
a Jordan-Wigner spin or qubit implementation
\cite{HarleShtankoMovassagh2023}, with the boundary protocol illustrated
in Fig.~\ref{fig:setup}(b).
The Pauli operators on qubit $j$ are denoted by $X_j,Y_j,Z_j$.
Our Jordan-Wigner convention is
$a_j=(\prod_{k<j}Z_k)X_j$ and $b_j=(\prod_{k<j}Z_k)Y_j$.
The empty product at $j=1$ is the identity, giving $a_1=X_1$ and
$b_1=Y_1$. To obtain $Y_{a_1a_1}(t)$, prepare the
boundary qubit in the opposite eigenstates
$|\pm x\rangle=(|0\rangle\pm|1\rangle)/\sqrt{2}$. The hidden system
$E$ comprises sites $2,\ldots,L$. In both preparations, we use the same
hidden density operator $\rho_E$, uncorrelated with the prepared boundary
qubit. The response identity below holds for any common $\rho_E$. After identical
joint evolution
$U_{1E}(t)=e^{-iHt}$, measure $X_1$. The half difference of the mean
outcomes gives \cite{BurgarthMaruyamaNori2011}
\[
Y_{a_1a_1}(t)
=
\frac12\left[
\langle X_1(t)\rangle_{+x}
-
\langle X_1(t)\rangle_{-x}
\right],
\]
where the subscripts indicate the initial boundary preparation. Repeating
the procedure with independently chosen $X_1$ or $Y_1$ preparation and
measurement axes, using $|\pm y\rangle=(|0\rangle\pm i|1\rangle)/\sqrt{2}$
for the $Y_1$ preparations, gives all four entries, with the preparation
axis selecting the column of $Y(t)$ and the measured observable its row.
For quadratic dynamics,
$\gamma_\alpha(t)=\sum_\mu(e^{Kt})_{\alpha\mu}\gamma_\mu$, and every
hidden Majorana carries a factor $Z_1$ with zero mean in both
preparations, so each mean equals $\pm(e^{Kt})_{\alpha\beta}$ and the
paired difference also cancels a common additive offset. Throughout,
$Y_{\alpha\beta}(t)$ denotes this signed half difference.
The site-1 identity of \secref{sec:a1-identity} is stated directly for the
half difference, so that it holds when the continuation is not quadratic.

In the fermionic realization of Fig.~\ref{fig:setup}(c), the
isolated-chain electron anticommutator defines the scalar channel
$\langle\{c_1(t),c_1^\dagger(0)\}\rangle=v^\dagger Y(t)v$, with
$v=(1,-i)^T/\sqrt2$. In the real quadratic model
$Y_{b_1a_1}=-Y_{a_1b_1}$, so this channel equals
$(Y_{a_1a_1}+Y_{b_1b_1})/2-iY_{a_1b_1}$. It contains the
off-diagonal entry used by the site-1 identity but not the two
diagonal entries separately. Extracting this channel from measured
tunneling conductance requires a justified probe and transport
model \cite{Flensberg2010}, as discussed in
\secref{sec:end-spectroscopy}.

\subsection{Finite-record model and statistical uncertainty}
\label{sec:finite_data}
Finite boundary measurements leave uncertainty in the Hamiltonian
coefficients. We account for it through a simultaneous confidence set,
and a certified property must hold for every coefficient vector in that
set \cite{ChristandlRenner2012,JaulinWalter1993,KlieschRoth2021,Eisert2020}.

The vector $\bm\theta\in\Theta_{\rm adm}\subset\mathbb R^p$ contains
the $p$ coefficients treated as unknown in the Kitaev-chain
Hamiltonian introduced above. Depending on the inference model,
these include onsite potentials $\mu_j$, hopping amplitudes $J_j$,
and pairing amplitudes $\Delta_j$. The compact parameter domain
$\Theta_{\rm adm}$ specifies their allowed joint values.
The generator $K(\bm\theta)$ depends affinely on these coefficients,
whereas the sampled response
$\bm f(\bm\theta)\in\mathbb R^D$ is generally nonlinear, with entries
\[
f_\iota(\bm\theta)
=[Pe^{K(\bm\theta)t_\ell}P^T]_{\alpha\beta}.
\]
The $D$ prespecified coordinates are arranged in a fixed order, with
$\iota=(\ell,\alpha,\beta)$ identifying the sampling time,
measurement axis, and preparation axis, respectively.

For each coordinate in the qubit protocol, the two opposite preparations
yield $N_{\pm}$ binary outcomes per sign,
$\mathsf{o}_{\iota,s}^{(\pm)}\in\{-1,+1\}$, where $s=1,\ldots,N_{\pm}$
indexes repetitions. The protocol assumes independent outcomes across
repetitions, signs, and separately acquired coordinates, with a fixed
outcome distribution for each sign and coordinate. The half difference
of the two sample means estimates the corresponding response entry,
\begin{equation}
\widehat y_\iota=\frac1{2N_{\pm}}\sum_{s=1}^{N_{\pm}}
\left(\mathsf{o}_{\iota,s}^{(+)}-\mathsf{o}_{\iota,s}^{(-)}\right)
=f_\iota(\bm\theta_\star)+\epsilon_\iota+b_\iota,
\label{eq:shots}
\end{equation}
where $\bm\theta_\star$ is the true coefficient vector in the assumed
quadratic model. The shot noise $\epsilon_\iota$ is the zero-mean
fluctuation of $\widehat y_\iota$ about its expectation, while the
deterministic bias $b_\iota$ is the difference between that expectation
and the model prediction. Each response estimate uses $2N_{\pm}$ shots.
Throughout, shots are counted per preparation sign and coordinate, and the
total number $N_{\rm tot}$ of binary outcomes of a record includes
all signs, coordinates, and calibration stages. 

\section{Persistence from learned Hamiltonian sets}
\label{sec:finite_majorana}
An inferred coefficient set can certify slow evolution of a localized
operator in the finite device. We call the first $m$ sites the prefix and
the rest of the chain its
continuation. The guarantee requires a simultaneous confidence set for the
prefix coefficients and all couplings to the continuation, and inferring
that set requires a forward model that accounts for the continuation. We
model the complete device (\secref{sec:three-site}), add a validated
truncation bound (\secref{sec:truncated-inference}), or, for the end
Majorana, use an identity that needs no continuation model
(\secref{sec:a1-identity}).

\subsection{Joint confidence sets}
\label{sec:confidence-sets}
The record used to form the confidence set has $N_{\rm pil}$ shots per
sign and coordinate,
response estimate $\widehat{\bm y}^{\rm pil}$, and failure allocation
$\delta_{\rm pil}\in(0,1)$. A prescribed bias bound
$|b_\iota^{\rm pil}|\le\epsilon_{\rm pil}$ for every coordinate,
with $\epsilon_{\rm pil}\ge0$, gives the common response tolerance
\[
\xi_{\rm pil}
=\sqrt{\frac{\log(2D/\delta_{\rm pil})}{N_{\rm pil}}}
+\epsilon_{\rm pil}.
\]
The square-root term accounts for finite-shot uncertainty simultaneously
over the $D$ coordinates, and the second term allows for systematic bias.
The compatible coefficient set is
\[
\mathcal C_{\rm pil}
=\left\{\bm\theta\in\Theta_{\rm adm}\ \middle|\
\|\bm f(\bm\theta)-\widehat{\bm y}^{\rm pil}\|_\infty
\le\xi_{\rm pil}\right\}.
\]
Here $\|\cdot\|_\infty$ denotes the largest absolute coordinate.
For a fixed admissible domain and forward model, the stated sampling
assumptions, together with
$\bm\theta_\star\in\Theta_{\rm adm}$ and valid bias bounds,
give the coverage guarantee
\[
\Pr(\bm\theta_\star\in\mathcal C_{\rm pil})
\ge1-\delta_{\rm pil},
\]
uniformly over the admissible domain, by Hoeffding's inequality
\cite{Hoeffding1963} and a union bound. In the calibrated benchmarks,
independent local measurements supply simultaneous bond intervals with
coverage at least $1-\delta_{\rm cal}$. These intervals constrain the
admissible coefficient set and are propagated through both the response
model and the final certificate. Their failure allocation is added to
that of the boundary batch.

A validated branch-and-bound search encloses $\mathcal C_{\rm pil}$ in a
union of retained parameter boxes \cite{Moore1966,JaulinWalter1993,Rump2010}.
A box is discarded only when validated response bounds exclude every
coefficient vector it contains, and unresolved boxes are retained when the
search reaches its computational limit, so the reported union can be
larger than the compatible set. The retained boxes bound the residual of
Eq.~\eqref{eq:certified_residual} or enter the conditional label tests of
\secref{sec:certificate}. See Supplemental Material for more
details on the validated response bounds and search
(Sec.~\ref{SM-sec:supp-pilot}).

\subsection{Commutator bound and certified residual}
For a fixed prefix depth $1\le m<L$, let $K_m$ be the
$2m\times2m$ principal block of $K$ describing the first $m$ sites.
A real unit vector $\bm g\in\mathbb R^{2m}$ assigns coefficients to
their Majorana operators, with $\|\bm g\|_2=1$ and $\|\cdot\|_2$
denoting the Euclidean vector norm. Grouping these coefficients by site,
we write $\bm g_j=(g_{a_j},g_{b_j})^T$ for $j=1,\ldots,m$ and define
the finite-support Majorana-type operator
\begin{equation}
\Gamma_m(\bm g)
=
\sum_{j=1}^{m}
\left(g_{a_j}a_j+g_{b_j}b_j\right).
\end{equation}
The Majorana anticommutation relations and the real unit normalization
of $\bm g$ imply
$\Gamma_m^\dagger=\Gamma_m$ and $\Gamma_m^2=I$. For a fixed candidate
$\bm g$, its Heisenberg evolution under the full
time-independent Hamiltonian $H$ is
$\Gamma_m(t)=e^{iHt}\Gamma_m(\bm g)e^{-iHt}$. The equation of motion
and unitary invariance of the operator norm give
\[
\left\|\frac{d\Gamma_m(t)}{dt}\right\|
=\|[H,\Gamma_m(\bm g)]\|,
\]
where $\|\cdot\|$ denotes the many-body operator norm.
Let $B_m$ denote the outgoing bond block coupling sites $m$ and $m+1$,
and let $\mathcal C_m$ be a simultaneous confidence set for the
compatible pairs $(K_m,B_m)$ inferred from the response and any
independent calibration data. For the
quadratic Hamiltonian, linearity of $\Gamma_m$ in the Majoranas gives
$\|[H,\Gamma_m(\bm g)]\|=\|K\overline{\bm g}\|_2$, where
$\overline{\bm g}$ is the zero extension of $\bm g$ to the full chain, and
block tridiagonality splits $K\overline{\bm g}$ into $K_m\bm g$ on the
prefix and $-B_m^T\bm g_m$ on site $m+1$. Parity-even terms supported on
sites $j>m$, interacting or not, commute with $\Gamma_m$ and leave the
residual unchanged. We therefore define the certified residual
$\varepsilon_m(\bm g,\mathcal C_m)$ as the largest
commutator norm allowed by $\mathcal C_m$,
\begin{equation}
\varepsilon_m(\bm g,\mathcal C_m)
=
\sup_{(K_m,B_m)\in\mathcal C_m}
\left[
\|K_m\bm g\|_2^2
+
\|B_m^T\bm g_m\|_2^2
\right]^{1/2}.
\label{eq:certified_residual}
\end{equation}
For every coefficient choice in $\mathcal C_m$,
this gives $\|d\Gamma_m(t)/dt\|\le\varepsilon_m(\bm g,\mathcal C_m)$.
Spatial localization is quantified separately through the weight on the first
$\ell$ sites,
\begin{equation}
W_\ell(\bm g)
=
\sum_{j=1}^{\ell}\|\bm g_j\|_2^2,
\qquad 1\le\ell\le m.
\label{eq:localization_weight}
\end{equation}
The normalization makes $W_\ell$ a fraction of the total coefficient
weight. A value close to one for $\ell\ll m$ indicates concentration
near the accessible boundary.
For $m\ge2$ we fix in advance a depth $1\le\ell<m$ and a minimum weight
$w_{\min}\in(0,1]$. One possible choice minimizes the commutator norm
at the nominal coefficients, subject to $\|\bm g\|_2=1$ and
$W_\ell(\bm g)\ge w_{\min}$. Once selected, the candidate is held
fixed while the residual is bounded over $\mathcal C_m$.
Uniform evaluation over the entire confidence set $\mathcal C_m$ permits
data-dependent selection of $\bm g$ under the same failure allocation.

The residual bound gives a persistence guarantee. Suppose that $\mathcal C_m$
contains the true prefix coefficients and outgoing bond with probability
at least $1-\delta$ over repetitions of the data acquisition, where
$0<\delta<1$. For the selected candidate, write
$\Gamma=\Gamma_m(\bm g)$ and
$\varepsilon=\varepsilon_m(\bm g,\mathcal C_m)$.
Integrating the equation of motion and using
$[\Gamma(t)-\Gamma]^2=2I-\{\Gamma(t),\Gamma\}$, we find that on this
coefficient-coverage event the operator obeys \cite{Kemp2017,Else2017}
\begin{align}
\|\Gamma(t)-\Gamma\|
&\le\varepsilon|t|,
\label{eq:operator_persistence}\\
\frac12\{\Gamma(t),\Gamma\}
&\succeq
\left(1-\frac12\varepsilon^2t^2\right)I,
\label{eq:operator_overlap}
\end{align}
simultaneously for every continuation beyond the outgoing bond and all
real $t$. 

For any initial many-body density operator $\rho$, define the
normalized symmetrized autocorrelation
\[
C_\Gamma(t;\rho)
=\frac12\operatorname{Tr}\!\left[\rho\{\Gamma(t),\Gamma\}\right].
\]
Since $\Gamma^2=I$, its initial value is $C_\Gamma(0;\rho)=1$.
For quadratic dynamics $\Gamma(t)$ remains linear in the Majoranas, so the
anticommutator is a multiple of the identity,
$\frac12\{\Gamma(t),\Gamma\}=\overline{\bm g}^Te^{Kt}\overline{\bm g}\,\mathbb I$.
Hence $C_\Gamma(t;\rho)=\overline{\bm g}^Te^{Kt}\overline{\bm g}$ for
every $\rho$, and the state independence of
Eqs.~\eqref{eq:operator_persistence} and \eqref{eq:operator_overlap} is
automatic here.
For any target $\eta_{\rm ac}\in[0,1)$,
Eq.~\eqref{eq:operator_overlap} therefore guarantees
$C_\Gamma(t;\rho)\ge\eta_{\rm ac}$ throughout
$|t|\le\tau_{\eta_{\rm ac}}$, where for $\varepsilon>0$,
\begin{equation}
\tau_{\eta_{\rm ac}}
=
\frac{\sqrt{2(1-\eta_{\rm ac})}}
{\varepsilon_m(\bm g,\mathcal C_m)}.
\label{eq:certified_persistence_time}
\end{equation}
When the commutator bound vanishes, we set
$\tau_{\eta_{\rm ac}}=\infty$.

An existence certificate lets the candidate depend on the coefficients. On
the coverage event, every compatible device then contains an operator with
$W_\ell\ge w_{\min}$ and residual at most
$\varepsilon_m^{\rm ex}=\sup_{\mathcal C_m}\min_{\bm g}\|K\overline{\bm g}\|_2$,
where the minimum runs over unit vectors with $W_\ell(\bm g)\ge w_{\min}$.

\subsection{Equal-line candidates}
\label{sec:equal-line}
On the equal line, $\Delta_j=J_j>0$, including the outgoing bond at $j=m$,
we consider the normalized candidate
\[
\Gamma_m
=\frac{1}{\mathcal N_m}\sum_{j=1}^{m}\alpha_j a_j,~~~
\mathcal N_m
=\left(\sum_{j=1}^{m}\alpha_j^2\right)^{1/2},~~~
\alpha_1=1,
\]
where $\alpha_j\in\mathbb R$ are the unnormalized amplitudes.
Cancellation of the onsite and neighboring-bond contributions to each
interior $b_j$ gives the zero-energy recurrence
\cite{Kitaev2001,DeGottardi2013}
\begin{equation}
\alpha_{j+1}
=-\frac{\mu_j}{2J_j}\alpha_j,
\qquad j=1,\ldots,m-1.
\label{eq:oneway_recurrence}
\end{equation}
The only remaining term of $[H,\Gamma_m]$ is proportional to
$\mu_m\alpha_mb_m/\mathcal N_m$, so the exact residual is
$\varepsilon_m=|\mu_m\alpha_m|/\mathcal N_m$. For uniform coefficients
$\mu_j=\mu$ and $J_j=\Delta_j=J>0$, define
$r=|\mu/(2J)|$. The recurrence gives $|\alpha_j|=r^{j-1}$, and hence
\begin{equation}
\frac{\varepsilon_m}{J}
=\frac{2r^m}{\sqrt{\sum_{n=0}^{m-1}r^{2n}}},
\qquad
W_1=\frac{1}{\sum_{n=0}^{m-1}r^{2n}}.
\label{eq:uniform_residual}
\end{equation}
For $r<1$, increasing the prefix depth makes the residual exponentially
small while $W_1$ approaches $1-r^2$. The candidate remains
localized near the accessible boundary. At $r=1$, the residual decreases
as $2J/\sqrt m$, but $W_1=1/m$ because the candidate spreads uniformly.
For $r>1$, its boundary weight vanishes and the residual approaches a
nonzero value. The three regimes show why a small residual must be
accompanied by a localization condition.

Here and below $\tau_{0.9}$ denotes the time of
Eq.~\eqref{eq:certified_persistence_time} for the autocorrelation
threshold $\eta_{\rm ac}=0.9$. 
For a common nominal onsite value $\bar\mu$ satisfying
$|\bar\mu|<2J$, onsite uncertainty
$|\mu_j-\bar\mu|\le2J\eta_\mu$ spoils the cancellations in
Eq.~\eqref{eq:oneway_recurrence}, even when the equal bonds are
known exactly. For $\eta_\mu>0$, the certificate of a candidate
frozen at the nominal coefficients saturates at
$\tau_{0.9}J\to\sqrt{0.2}/(2\eta_\mu)$ as $m\to\infty$,
about $224$, $22$, and $4.5$ for
$\eta_\mu=0.001$, $0.01$, and $0.05$.
The existence
certificate has no such floor. For
signed ratios $r_j=\mu_j/(2J)$ with $|r_j-\bar r|\le\eta_\mu$ and
$r_+=|\bar r|+\eta_\mu<1$, a localization depth $1\le\ell<m$, and a threshold $0<w_{\min}\le1-r_+^{2\ell}$, the profile-dependent recurrence candidates satisfy
\begin{equation}
\frac{\varepsilon_m^{\rm ex}}{J}\le2r_+^m,
\qquad
W_\ell\ge1-r_+^{2\ell},
\label{eq:equal_line_existence_bound}
\end{equation}
because $\mathcal N_m\ge1$ and $|\alpha_{j+1}/\alpha_j|\le r_+$. The
existence certificate thus improves exponentially with depth despite
onsite uncertainty.

\subsection{Three-site benchmark}
\label{sec:three-site}
\label{sec:persistence-benchmarks}
Our base benchmark is a synthetic three-site equal-line qubit device with
ideal readout (Supplemental Material, Sec.~\ref{SM-sec:supp-three-site}).
The construction of coefficient confidence sets, including the
treatment of measurement and calibration uncertainty, is described
in Secs.~\ref{sec:finite_data} and~\ref{sec:confidence-sets}.
Its onsite coefficients vary
independently over $\mu_j/J_{\rm ref}\in[0.5,3.5]$, where $J_{\rm ref}$ is
a fixed reference energy. All benchmark energies and times are given in
units of $J_{\rm ref}$ and $1/J_{\rm ref}$. The bond strengths
$\kappa_j=2J_j$ are calibrated before the boundary
experiment (Supplemental Material, Fig.~\ref{SM-fig:supp-three-site}). Each bond is isolated,
its two qubits are prepared in
$|00\rangle$, and $Z_j$ is measured after a known time, which for bond
$(2,3)$ requires preparation on sites 2 and 3 and readout at site 2.
These outcomes are counted in the budget. Sites 1 and
2 form the prefix, $\kappa_2$ is the outgoing bond, and $\mu_3$ is fitted
as a nuisance coefficient, enclosed together with the prefix
coefficients but not certified, so the forward model is the complete
three-site device. Each record samples all four response entries at
$tJ_{\rm ref}=0.25,0.5,\ldots,3$, within the observation window
$T=3/J_{\rm ref}$, with $N_\pm=2\times10^4$ shots per sign and coordinate,
$1.92\times10^6$
boundary outcomes and $4\times10^5$ calibration outcomes in all, with a
coverage failure probability of at most $0.02$ per record. The three
example devices and their learned onsite hulls are listed in Supplemental Material,
Table~\ref{SM-tab:compact-three-site}.

\begin{figure*}[t]
\centering
\includegraphics[width=\textwidth,height=0.57\textheight,keepaspectratio]{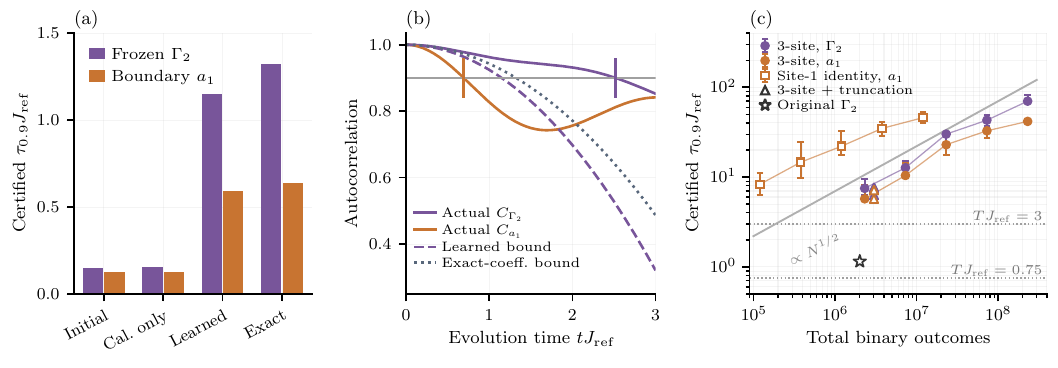}
\caption{Certified persistence. 
(a) Certified $\tau_{0.9}$ for the frozen two-site operator $\Gamma_2$ and
the boundary operator $a_1$ in the first three-site record of
\secref{sec:three-site}, from the commutator bound at four levels of
coefficient information.
(b) Autocorrelations of the simulated device (solid), with the learned
(dashed) and exact-coefficient (dotted) commutator bounds for $\Gamma_2$.
Vertical ticks mark the
first crossings of $0.9$, and the window ends at $TJ_{\rm ref}=3$.
(c) Certified times near the boundary sweet spot versus total binary
outcomes. Complete three-site inference ($TJ_{\rm ref}=3$) applies to the
rule-selected two-site operator and to $a_1$. Results for the site-1 identity
certificate for $a_1$ ($TJ_{\rm ref}=0.75$) are pooled over four
continuations. Three-site inference with a truncation bound
($TJ_{\rm ref}=0.5$) is applied to 40-site devices with three different tails
(triangles). Markers show medians and bars the range over records. The
coverage failure probability is at most $0.02$ per three-site record and
$0.01$ per identity record. The star marks the original benchmark, dotted
lines the windows, and the grey line $N^{1/2}$ scaling.}
\label{fig:persistence_window}
\end{figure*}

For the first example, $\bm\mu/J_{\rm ref}=(0.7,1.0,1.3)$ and
$\bm\kappa/J_{\rm ref}=(1.92,2.08)$, the confidence set bounds the
persistence of the fixed two-site operator
$\Gamma_2=(a_1-\alpha a_2)/\sqrt{1+\alpha^2}$, with $\alpha\simeq0.358$
and boundary weight $W_1\simeq0.886$ selected from the learned hull, and
of the bare boundary operator $a_1$. On the equal line the
outgoing term of Eq.~\eqref{eq:certified_residual} vanishes for every
candidate without a $b$ component on its terminal site, so the guarantee
for $\Gamma_2$ depends only on $(\mu_1,\mu_2,\kappa_1)$.
Figure~\ref{fig:persistence_window}(a) compares four levels of coefficient
information. The initial domain, calibrated bonds alone, the learned set,
and the exact coefficients give $\tau_{0.9}J_{\rm ref}\simeq0.149$, $0.154$,
$1.151$, and $1.325$. The learned certificate reaches $87\%$ of the
exact-coefficient commutator bound, whereas the autocorrelation of the
simulated device first falls below $0.9$ at $tJ_{\rm ref}\simeq2.52$
[Fig.~\ref{fig:persistence_window}(b)]. For $a_1$ the learned and
exact-coefficient bounds are $0.590$ and $0.639$, against a first crossing
at $0.70$. Because the candidate was selected from the full record, the
earlier-stage values are retrospective evaluations of the same operator.
The remaining gap reflects both uncertainty in the inferred coefficients
and the conservativeness of the commutator envelope. With
independently uncertain hopping and pairing, the generic case, the
outgoing bond contributes $|\Delta_2-J_2|\alpha/\sqrt{1+\alpha^2}\in
[0.050,0.056]J_{\rm ref}$ to the residual, and the learned set certifies
$1.336$, compared with $1.652$ at the exact coefficients and an actual
first crossing at $4.39$
(Supplemental Material, Sec.~\ref{SM-sec:supp-independent-bonds}). With calibrated symmetric
readout flips at probability $15\%$, twenty independent equal-line
records of the first example give
$\tau_{0.9}(\Gamma_2)J_{\rm ref}\in[1.043,1.081]$, with coverage at least
$98\%$ per record (Supplemental Material, Sec.~\ref{SM-sec:supp-readout-noise}).

\subsection{Persistence beyond the observation window}
\label{sec:beyond-window}
In the benchmark above, both the certified times and the simulated
decay lie inside the window. At the boundary sweet spot
$\mu_1=0$, $J_1=\Delta_1$, the
end Majorana is conserved, since
$[H,a_1]\propto\mu_1b_1+(J_1-\Delta_1)b_2$ for every continuation.
For a fixed Hamiltonian and an informative short window, the data can
constrain $|\mu_1|$ to a scale of order $1/(T\sqrt{N_{\rm tot}})$,
extending the certified time as $T\sqrt{N_{\rm tot}}$ at fixed record design.

For the complete three-site model, whose chain ends at site 3, we rerun
the same pipeline at
$\bm\mu/J_{\rm ref}=(0.005,1.0,1.3)$ with the original bonds, a declared
domain $\mu_1/J_{\rm ref}\in[-1,2]$, and calibrated bonds included in the
budget (Supplemental Material, Sec.~\ref{SM-sec:supp-sweet-spot}). At
$2.32\times10^6$ total binary outcomes, the
median certified times are $5.7/J_{\rm ref}$ for $a_1$ and $7.5/J_{\rm ref}$
for the rule-selected two-site operator. At $2.32\times10^8$ outcomes
the medians reach
$42/J_{\rm ref}$ and $70/J_{\rm ref}$, that is, 14 and 23 times the window
[Fig.~\ref{fig:persistence_window}(c)]. The five-record medians at each
budget have fitted shot exponents $0.44$ and $0.49$, respectively.
These are finite-budget fits without exponent-uncertainty estimates. All
25 retained sets cover the true coefficients.
For $a_1$ the growth must eventually saturate at the exact-coefficient
value $\sqrt{0.2}/|\mu_1|\simeq89/J_{\rm ref}$.

\subsection{Site-1 identity for the end Majorana}
\label{sec:a1-identity}
For $a_1$ the Heisenberg
equation closes on boundary data. On an equal-line first bond,
$\dot a_1=-\mu_1b_1$ for every continuation acting on sites $j\ge2$,
quadratic or not, and the persistence bound of Eq.~\eqref{eq:operator_overlap} applied to
$b_1$ itself, whose commutator norm is
$\Omega_1=(\mu_1^2+\kappa_1^2)^{1/2}$, gives
$Y_{b_1b_1}(s)\ge1-\frac12\Omega_1^2s^2$. Hence
\begin{align}
Y_{a_1b_1}(t)&=-\mu_1\int_0^tY_{b_1b_1}(s)\,ds,
\nonumber\\
t-\tfrac16\Omega_1^2t^3&\le\int_0^tY_{b_1b_1}(s)\,ds\le t
\qquad(\Omega_1t\le2).
\label{eq:a1_identity}
\end{align}
Beyond
$\Omega_1t=2$ the lower bound continues as $8/(3\Omega_1)-t$ (Supplemental Material,
Sec.~\ref{SM-sec:supp-a1-identity}). Measuring $Y_{a_1b_1}$ at a few short
times therefore bounds $|\mu_1|$, and thus $\tau_{0.9}(a_1)$, using only
site-1 data, the declared domain $|\mu_1|\le2J_{\rm ref}$, and the
declared bound $\kappa_1\le2.4J_{\rm ref}$. No model of the continuation
and no access beyond site 1 enter. We simulate records at
$tJ_{\rm ref}=0.25,0.5,0.75$ from the bare device near the boundary sweet spot and from
40-site devices with the same first three sites and topological, trivial,
or disordered tails. With $1.2\times10^5$ outcomes the median certified
time over all continuations is $8.3/J_{\rm ref}$, eleven times the window.
With $1.2\times10^7$ outcomes it is $46/J_{\rm ref}$, about 61 times the
window, and the medians for the four continuations lie between $43/J_{\rm ref}$
and $46/J_{\rm ref}$. All 100 records cover $\mu_1$ (Supplemental Material,
Sec.~\ref{SM-sec:supp-tail-independent}).

In qubit form the equal-line chain is the transverse-field Ising model
$H=\sum_j\mu_jZ_j/2-\sum_j\kappa_jX_jX_{j+1}/2$. The identity involves
only its site-1 terms, so it holds for the exact
bond-1 cut $H=\mu_1Z_1/2-\kappa_1X_1X_2/2+H_{\ge2}$, with no other term
acting on qubit 1 and any time-independent Hermitian $H_{\ge2}$ on qubits
$j\ge2$. Bounded additional terms are handled by Eq.~\eqref{SM-eq:supp-boundary-perturbation-time}
in Supplemental Material. The coefficients must
also remain fixed across shots and over the certified time. Since
$Y_{a_1b_1}(t)$ is odd in $\mu_1$, symmetric shot-to-shot variation can
cancel this signal. For bounded offsets held
constant during each shot, separate diagonal-response measurements can
bound $\mathbb E\mu_1^2$ and hence the shot-averaged autocorrelation
(Supplemental Material, Eq.~\eqref{SM-eq:supp-mixture-moment-bound}). Exact many-body
benchmarks with interactions,
parity-breaking fields confined to the continuation, and disordered
couplings give median certificates between $40/J_{\rm ref}$ and
$50/J_{\rm ref}$ from $1.2\times10^7$ outcomes, and all 36 records cover
$\mu_1$ (Supplemental Material, Sec.~\ref{SM-sec:supp-interacting}). Outside the quadratic model the
signed response depends on the common hidden preparation, but the
persistence bound holds for every initial state. This bound
can be nearly attained by the admissible disconnected first bond,
$\kappa_1=0$, for which $a_1$ precesses at frequency $\mu_1$.

\textit{Independent-shot scaling.} The site-1 certificate has fitted shot
exponents $0.37$ in the benchmarks above and $0.40$ at fixed window. A
boundary-field comparison bounds
the possible asymptotic exponent for independent shots. For any
protocol using site-1 preparation and readout, at most time $T$ per
evolution, and $N_{\rm tot}$ fresh-reset shots, a certified time $\hat\tau$
valid with probability $1-\delta$ for both devices with
$\kappa_1=0$ and $\mu_1\in\{0,\nu\}$ obeys, on the conserved device,
\begin{equation}
\Pr\bigl[\hat\tau\le\arccos(0.9)\,T\sqrt{N_{\rm tot}}\bigr]\ge\tfrac12-\delta .
\label{eq:incoherent_limit}
\end{equation}
Both devices have the same continuation. In the device with
$\mu_1=\nu$, $a_1$ precesses and a valid
certificate cannot exceed $\arccos(0.9)/\nu$. Their evolutions differ by
at most $\nu T/2$ in operator norm, so each
shot has output fidelity at least $1-\nu^2T^2/8$, and the total-variation
distance of the full, possibly adaptive, record is at most
$\nu T\sqrt{N_{\rm tot}}/2$ \cite{Watrous2018,Tsybakov2009}. Choosing
$\nu=1/(T\sqrt{N_{\rm tot}})$ makes it at most $1/2$, which gives
Eq.~\eqref{eq:incoherent_limit}. It applies to any model
class containing this pair, including one with a known continuation.
It is therefore a boundary-field resolution limit, rather than a cost
specific to continuation-independent inference. The fitted exponents below
$1/2$ reflect the finite budgets and saturation at nonzero $\mu_1$.

\subsection{Robustness and calibration controls}
\label{sec:robustness}
The long-time guarantee depends on the error bounds supplied to the
inference. A synthetic study propagates first-bond mismatch and bounded
response bias through both the compatible set and the final residual.
At $1.2\times10^7$ outcomes, bounds
$|\Delta_1-J_1|\le0.005J_{\rm ref}$ and $|b_\iota|\le0.001$ reduce
the median fixed-Hamiltonian guarantee from $45.5/J_{\rm ref}$ to
$21.3/J_{\rm ref}$, still beyond the $0.75/J_{\rm ref}$ measurement
window (Supplemental Material, Fig.~\ref{SM-fig:supp-robustness}). A separate
diagonal-response analysis certifies the autocorrelation averaged over
quasistatic onsite fluctuations, including bounded first-bond mismatch.
The two guarantees refer to different statistical models of the device
(Supplemental Material, Sec.~\ref{SM-sec:supp-robustness}).

Direct onsite Ramsey calibration~\cite{Ramsey1950} provides a useful
comparison when the neighboring coupling can be isolated. With ideal
isolation and transfer, it gives tighter bounds in the tested example
at the same shot budget, evolution window, and confidence level.
An uncertainty allowance for transferring the isolated onsite
coefficient to the coupled device reduces that gain
(Supplemental Material, Sec.~\ref{SM-sec:supp-ramsey-comparison}). On an equal-line first bond,
the site-1 identity uses the coupled device throughout and requires no control or model of
the continuation. The comparison holds the supplied error envelopes
fixed. Additional experimental calibration of those envelopes would
contribute to the measurement budget.

\subsection{Finite-prefix inference with a truncation bound}
\label{sec:truncated-inference}
The full inference can also cover an unknown continuation within
declared equal-line coefficient bounds. For an
equal-line model of the first $n\ge1$ sites and any continuation obeying
declared bounds $0\le\mu_j\le\mu_{\max}$ and $0<\kappa_j\le\kappa_{\max}$,
the quantity $G=\mu_{\max}+\kappa_{\max}$ bounds the norms of the full
and truncated generators. The nondecreasing propagation envelope is
\[
\overline{\mathfrak B}_d(x)
=\sup_{0\le u\le x}
\min\left\{1,\,2\sum_{k=d}^{\infty}
|\operatorname{BesJ}_k(u)|\right\},
\]
for integers $d\ge0$ and $x\ge0$, where $\operatorname{BesJ}_k$ is the
Bessel function of the first kind. For $G\ge\|K\|_2$, the Chebyshev
expansion \cite{TalEzerKosloff1984}
\[
e^{Kt}=\operatorname{BesJ}_0(Gt)I+
2\sum_{n\ge1}(-i)^n\operatorname{BesJ}_n(Gt)\operatorname{Cheb}_n(iK/G)
\]
has an $n$th term of graph range $n$ and norm at most
$2|\operatorname{BesJ}_n(Gt)|$, where $\operatorname{Cheb}_n$ is the
Chebyshev polynomial of the first kind. The
envelope therefore bounds propagation across at least $d$ links of the
Majorana coupling graph up to dimensionless time $x$, with
$\overline{\mathfrak B}_0(x)=1$.
The distances from the measured boundary $\{a_1,b_1\}$ to the
cut-bond endpoints $a_{n+1}$ and $b_n$ are $2n-1$ and $2n-2$,
respectively. Duhamel's formula across bond $(n,n+1)$ therefore gives
\begin{align}
\sup_{0\le t\le T}&\|Y(t)-Y_n(t)\|_2\le E_n(T),
\nonumber\\
E_n(T)&=\kappa_{\max}\!\int_0^T\overline{\mathfrak B}_{2n-1}[G(T-s)]\,
\overline{\mathfrak B}_{2n-2}(Gs)\,ds.
\label{eq:truncation_bound}
\end{align}
The quantity $E_n(T)$ bounds the response error from the omitted
continuation throughout the observation window
(Supplemental Material, Sec.~\ref{SM-sec:supp-truncation}). Adding $E_n(T)$ to $\epsilon_{\rm pil}$
restores coverage for every such continuation. For
$\mu_{\max}=3.5J_{\rm ref}$ and $\kappa_{\max}=2.4J_{\rm ref}$,
$E_n(T)\le2\times10^{-3}$, a tenth of the shot radius of the benchmark,
requires $n=3$ for $TJ_{\rm ref}\le0.5$, $n=4$ for $TJ_{\rm ref}=1$, and
$n=8$ for $TJ_{\rm ref}=3$. The return-trip light cone of the original
window would thus force five additional modeled sites. With
$TJ_{\rm ref}=0.5$ and $E_3(T)\le4.71\times10^{-5}$, the unchanged three-site
search applied to records from the three 40-site devices certifies
$\tau_{0.9}(a_1)J_{\rm ref}\ge7.15$, $7.15$, and $5.7$ and
$\tau_{0.9}(\Gamma_2)J_{\rm ref}\ge7.6$, $7.6$, and $6.3$ for the
topological, trivial, and disordered tails, respectively, from
$2.32\times10^6$ outcomes, 11 to 15 times the window. All three sets cover
the coefficients of the first three sites.

The three inference designs use different observation windows and sampling
coordinates. The global searches include separate bond calibration,
whereas the site-1 identity uses only boundary-response measurements.
Their certified times therefore reflect both the model assumptions and
the measurement design.

\label{sec:depth-cost}
Deeper candidates have exponentially smaller exact residuals
[Eq.~\eqref{eq:uniform_residual}], but their coefficients are harder to
learn from the boundary. For the studied profiles, a local
Fisher-information estimate predicts that with boundary-only access the
shots needed for a fixed residual allowance grow 4- to 10-fold per added
site beyond $m\simeq4$ to $6$, whereas local bond calibration
substantially reduces that growth. This is a cost proxy, not a validated
bound (Supplemental Material, Sec.~\ref{SM-sec:supp-depth-cost}). A light-cone-aware affine
contractor, with linear-programming bounds
certified in rational arithmetic \cite{NeumaierShcherbina2004}, extends
the validated inference to complete five- and six-site
devices, where it certifies depth-four and depth-five candidates within
$8\%$ of their exact-coefficient bounds
(Supplemental Material, Sec.~\ref{SM-sec:supp-contractor}).

\section{Finite-resolution obstruction for the bulk label}
\label{sec:nogo}

We now ask whether a finite boundary record can determine the
bulk topological label. We compare two connected semi-infinite
chains that share the first $m$ sites and the outgoing bond but
have continuations, or tails, with opposite bulk topological labels
(Fig.~\ref{fig:matched_prefix_summary}(a)).
The discrimination bounds below cover the specified boundary-record
and independent-reset channel protocols.

\subsection{Matched continuations and boundary response}

The equal-line recurrence of Eq.~\eqref{eq:oneway_recurrence} provides a
direct criterion for choosing the tails. Its logarithmic ratios and their
asymptotic mean are
\begin{equation}
z_j=\log\left|\frac{\mu_j}{2J_j}\right|,
\qquad
\lambda=\lim_{m\to\infty}\frac1m\sum_{j=1}^{m}z_j
\label{eq:scalar_lyapunov}
\end{equation}
whenever the limit exists. The exponent $\lambda$ is the asymptotic
spatial growth rate of the recurrence amplitudes. A negative value gives a
normalizable boundary mode, while a positive
value gives a growing, nonnormalizable solution \cite{DeGottardi2013}. For
the eventually uniform gapped tails used
below, these alternatives correspond to $\topinv=-1$ and $\topinv=+1$.
Disordered continuations require the additional assumptions discussed
in \secref{sec:certificate}.

The construction uses $J>0$, a positive dimensionless shared-prefix
ratio $r_{\rm pre}$, and uniform-tail ratios
$0<r_{\rm top}<1<r_{\rm triv}$, all independent of the prefix length $m$.
For every integer $m\ge1$ and $\sigma\in\{{\rm top},{\rm triv}\}$,
the corresponding semi-infinite chain has
$J_j=\Delta_j=J$ on every bond and the onsite profile
\begin{equation*}
\frac{\mu_j^{(m,\sigma)}}{2J}=
\begin{cases}
r_{\rm pre},&1\le j\le m,\\
r_\sigma,&j\ge m+1,
\end{cases}
\end{equation*}
where $r_\sigma$ is $r_{\rm top}$ or $r_{\rm triv}$ according to $\sigma$.
Denote the Majorana generator by $K_\sigma^{(m)}$ and the boundary
response by $Y_\sigma^{(m)}(t)=Pe^{K_\sigma^{(m)}t}P^T$.
We omit the superscript when $m$ is fixed. The two chains have opposite bulk labels,
$\topinv(K_{\rm top})=-1$ and $\topinv(K_{\rm triv})=+1$, while their
coefficients agree through site $m$ and on the bond $(m,m+1)$. Their tail
gaps $2J|1-r_\sigma|$ and tail Lyapunov exponents $\log r_\sigma$ are
independent of $m$. The family
$\mathcal K_{\rm mp}=\{K_\sigma^{(m)}\mid m\ge1,\sigma\in\{{\rm top},{\rm triv}\}\}$
is uniformly bounded. For two chain generators $K,K'\in\mathcal K_{\rm mp}$,
we quantify the largest difference between their boundary responses over
the observation window $0\le t\le T$ by
\begin{equation}
d_T(K,K')=
\sup_{0\le t\le T}\norm{P(e^{Kt}-e^{K't})P^T}_2.
\label{eq:finite_window_metric}
\end{equation}

The common prefix suppresses the influence of the different tails on
the short-time boundary response. For the matched chains defined above,
the boundary moments satisfy
\begin{equation}
PK_{\rm top}^{n}P^T=PK_{\rm triv}^{n}P^T,
\qquad n=0,1,\ldots,4m-1,
\label{eq:moment_matching}
\end{equation}
while the first difference for this positive-ratio pair occurs at order
$4m$,
\begin{equation}
PK_{\rm top}^{4m}P^T\ne PK_{\rm triv}^{4m}P^T.
\label{eq:first_unmatched_moment}
\end{equation}

The response difference over a finite observation window is bounded
in terms of the coefficient scales
$G_\star=2J[1+\max(r_{\rm pre},r_{\rm top},r_{\rm triv})]$
and $\Lambda_\star=2J|r_{\rm triv}-r_{\rm top}|$.
For every decay parameter $\zeta>0$ and observation window $T\ge0$,
the responses satisfy \cite{Iserles2000}
\begin{align}
& B_{m,T}(\zeta)
=T\Lambda_\star
e^{-2\zeta(2m-1)+e^\zeta G_\star T},
\nonumber\\
&\sup_{0\le t\le T}
\|Y_{\rm top}(t)-Y_{\rm triv}(t)\|_2
\le B_{m,T}(\zeta).
\label{eq:locality_nogo}
\end{align}
Because $G_\star$ and $\Lambda_\star$ are independent of $m$, the bound
tends to zero as the shared prefix grows at any fixed finite $T$ and
fixed $\zeta>0$. The two responses can therefore be made arbitrarily
close while the opposite bulk labels and nonzero tail gaps remain fixed.
In terms of the response distance, this gives
\begin{equation}
\inf_{\substack{K_0,K_1\in\mathcal K_{\rm mp}\\
\topinv(K_0)\ne\topinv(K_1)}}d_T(K_0,K_1)=0.
\label{eq:zero_uniform_separation}
\end{equation}

The statement concerns the worst case over the declared model class.
For every window $T$, every response tolerance, and every shot budget
there is a pair in $\mathcal K_{\rm mp}$ whose labels differ while both
responses lie within that tolerance
of their midpoint, so that a label required to hold for every response
compatible with the record must be inconclusive. The statement does not
say that a given pair is indistinguishable at every budget. Each fixed
pair remains distinguishable in exact continuous-time data through its
first unequal moment at order $4m$, and \secref{sec:channel-discrimination}
quantifies the shots this takes. Off the equal line, counting walks on the
Majorana coupling graph gives
agreement of the moments only through order $2m$, because both
bond entries of Eq.~\eqref{eq:blocks} are then nonzero and the boundary
reaches site $m+1$ in $m$ steps. The equal-line structure, in which the
Majorana graph is a single path, doubles the matched order.

The exponential envelope proves the asymptotic statement but is loose at
the demonstrated depths. For $G_\sigma=2J[1+\max(r_{\rm pre},r_\sigma)]$,
Duhamel's formula with the propagation envelope $\overline{\mathfrak B}_d$
of \secref{sec:truncated-inference} gives the sharper finite-depth bound
\begin{align}
\sup_{0\le t\le T}&\|Y_{\rm top}(t)-Y_{\rm triv}(t)\|_2
\nonumber\\
&\le\Lambda_\star\!\int_0^T\overline{\mathfrak B}_{2m-1}[G_{\rm top}(T-s)]\,
\overline{\mathfrak B}_{2m-1}(G_{\rm triv}s)\,ds .
\label{eq:bessel_window}
\end{align}
For the illustrated ratios
$(r_{\rm pre},r_{\rm top},r_{\rm triv})=(0.70,0.60,1.40)$,
an upper Riemann sum of rigorous grid majorants gives bounds rounded upward to
$1.80$, $3.55\times10^{-2}$, and $8.80\times10^{-9}$ at $TJ=8$ for
$m=8$, $10$, and $14$. Because
$\operatorname{BesJ}_n(x)$ decreases faster than
exponentially once $n$ exceeds $x$, the integrand is negligible unless the
out-and-back graph distance
$2(2m-1)$ is comparable to $G_\star T$. A record of duration $T$ is thus
effectively blind to coefficients beyond a return-trip depth of about
$G_\star T/4$ sites, roughly ten sites for this pair at $TJ=8$, and
maintaining separation at fixed response resolution requires a window that
grows at least linearly with $m$. The derivations of
Eqs.~\eqref{eq:moment_matching} through \eqref{eq:bessel_window}
are given in Supplemental Material, Sec.~\ref{SM-sec:supp-nogo-proof}. Below, $B_{m,T}(\zeta)$ may everywhere
be replaced by the right side of Eq.~\eqref{eq:bessel_window}.

\begin{figure*}[t]
\centering
\includegraphics[width=\textwidth,height=0.57\textheight,keepaspectratio]{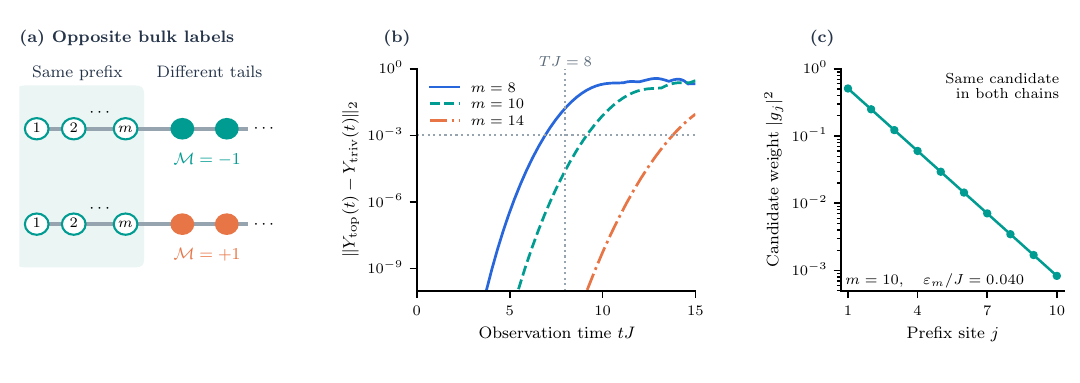}
\caption{Matched-prefix chains with opposite bulk labels. 
(a) A shared prefix and outgoing bond followed by different constant tails.
(b) Boundary-response distances for $L=64$ truncations and $m=8,10,14$,
with $(r_{\rm pre},r_{\rm top},r_{\rm triv})=(0.70,0.60,1.40)$. (c) Site
weights of the normalized
truncated Majorana candidate for $m=10$, with its residual indicated.}
\label{fig:matched_prefix_summary}
\end{figure*}

For $r_{\rm pre}<1$, both continuations admit the same boundary-localized
truncated candidate and the same small residual, despite their opposite
bulk labels (Fig.~\ref{fig:matched_prefix_summary}(c)).

Figure~\ref{fig:matched_prefix_summary}(b) shows that increasing the
shared-prefix depth from $m=8$ to $10$ and $14$ delays the first crossing
of the indicated response-resolution threshold in this example.
The comparison of sampled distances with rigorous response envelopes is
given in Supplemental Material, Sec.~\ref{SM-sec:supp-nogo-numerics}.

\subsection{Finite-shot and channel bounds}
\label{sec:channel-discrimination}
Fix the matched pair at prefix depth $m$ and a set $\mathcal R$ of $D\ge1$
measurement settings chosen before collecting data, with sampling times
$0<t_\ell\le T$. In each shot a preparation sign $S=\pm1$ is drawn with
equal probability and only the signed outcome $X=SO$ of the measured
$O=\pm1$ is retained. With the same hidden state for both signs,
$\mathbb E_hX=y_{h,\iota}=[Y_h(t_\ell)]_{\alpha\beta}$ under hypothesis
$h\in\{{\rm top},{\rm triv}\}$, and $N_{\rm sgn}$ signed outcomes per
setting give $DN_{\rm sgn}$ shots in total. Assume
$|y_{h,\iota}|\le y_{\max}<1$ for both hypotheses and every setting. For
equal priors, let $P_{\rm succ}(\psi)$ be the probability that a possibly
randomized decision rule $\psi$ returns the correct label. The two-point
testing bound \cite{Tsybakov2009,PolyanskiyWu2025} gives, for every
$\zeta>0$,
\begin{equation}
P_{\rm succ}(\psi)
\le \frac12+
\min\left\{
\frac12,
B_{m,T}(\zeta)\sqrt{\frac{N_{\rm sgn}D}
{8(1-y_{\max}^2)}}
\right\},
\label{eq:learner-success-bound}
\end{equation}
so a success probability of at least $1/2+\eta_{\rm adv}$, with
$0<\eta_{\rm adv}\le1/2$, requires
\begin{equation}
N_{\rm sgn}\ge
\frac{8(1-y_{\max}^2)\eta_{\rm adv}^2}
{D T^2\Lambda_\star^2}
e^{4\zeta(2m-1)-2e^{\zeta} G_\star T}.
\label{eq:learner-shot-cost}
\end{equation}
With $\zeta$ optimized, the right side scales as $[2d/(eG_\star T)]^{4d}$
with $d=2m-1$. The required shot count grows faster than exponentially
with the shared-prefix depth, for every decision rule on the specified
signed record, including rules trained on data independent of the test
label. The bound is valid but not tight. For the pair of
Fig.~\ref{fig:matched_prefix_summary} with $D=64$ coordinates over $TJ=8$,
Eq.~\eqref{eq:bessel_window} gives a nontrivial bound only at $m=14$,
$P_{\rm succ}\le0.5000013$ for $N_{\rm sgn}=10^3$, while a pair-specific
calculation with the actual response values bounds the success at
$0.7048$ for $m=8$ and $0.50031$ for $m=10$, compared with pair-specific
small-signal approximations to Bayes-optimal success of $0.614$ and $0.50017$
(Supplemental Material, Sec.~\ref{SM-sec:supp-stress-test}). The corresponding Bayes
advantage above chance is about $56\%$ of the pair-specific Pinsker
advantage, showing that the Pinsker step in
Eq.~\eqref{eq:learner-success-bound} can itself be loose.

The obstruction extends to boundary probes entangled with a retained
reference. For equal-length truncations of the matched pair interrogated
by free evolution with a fresh maximally mixed hidden reset for every use,
the boundary channel $\Phi_K^t$ has Bloch matrix
$Y(t)\oplus\det Y(t)$, so $Y(t)$ determines it, and the unhalved diamond
norm, which bounds
distinguishability with an arbitrary retained reference, satisfies (Supplemental Material,
Sec.~\ref{SM-sec:supp-quantum-channel}) \cite{Sacchi2005}
\begin{equation}
\|\Phi_{\rm top}^t-\Phi_{\rm triv}^t\|_\diamond
\le\min\{2,2\|Y_{\rm top}(t)-Y_{\rm triv}(t)\|_2\}.
\label{eq:boundary-channel-distance}
\end{equation}
For $N_{\rm use}$ interrogations at prescribed times in $[0,T]$, with
arbitrary accessible processing between uses and joint final readout, the
Helstrom bound and a hybrid argument \cite{Watrous2018} limit the
equal-prior success probability to
\begin{equation}
P_{\rm succ}^{(N_{\rm use})}
\le\min\{1,\tfrac12+\tfrac12N_{\rm use}B_{m,T}(\zeta)\},
\label{eq:boundary-channel-uses}
\end{equation}
so an advantage $\eta_{\rm adv}$ requires
$N_{\rm use}\ge2\eta_{\rm adv}/B_{m,T}(\zeta)$ uses.

\subsection{Labels under declared continuations}
\label{sec:certificate}
Uniform bulk identification requires an assumption connecting the learned
region to the unseen continuation. For exact repetition
of a two-site equal-line cell, Eq.~\eqref{eq:oneway_recurrence} gives
$r_{\rm cell}=\mu_1\mu_2/(\kappa_1\kappa_2)$ and
$\topinv=\sgn(r_{\rm cell}^2-1)$. A label is returned only if the cell margin
$\kappa_1\kappa_2-|\mu_1\mu_2|$ stays strictly on one side of
zero throughout the retained set. A positive margin certifies
$\topinv=-1$, and a negative margin certifies $\topinv=+1$. Two of the three benchmark devices give strict labels and the
near-transition device is inconclusive. Off the equal line, a periodic
cell is topological when both eigenvalues of the cell transfer matrix of
one of the two zero-mode recurrences lie inside the unit circle, and
trivial when they lie on opposite sides of it
\cite{DeGottardi2013,DeGottardiPRB2013}. Strict Jury inequalities
\cite{Jury1962,Jury1964}, evaluated on every retained box, test these
conditions.

For a declared IID law on the equal line with bounded $z_j$, the exponent
of Eq.~\eqref{eq:scalar_lyapunov} is $\lambda=\mathbb E[z_1]$ almost
surely, and under the mobility-gap hypotheses of the bulk-edge
correspondence for disordered chiral chains its sign gives the bulk label
\cite{GrafShapiro2018,MotrunichDamleHuse2001,BrouwerDuckheim2011,DeGottardi2013}.
A confidence
interval for $\lambda$ adds the mean reconstruction radius of the $z_j$ to
the Hoeffding radius $h_m=W_z[\log(2/\delta_{\rm stat})/(2m)]^{1/2}$ of
the $m$-site disordered prefix, where $W_z$ is the declared width of the
support of $z_j$ and $\delta_{\rm stat}$ the failure allocation for
disorder sampling \cite{Hoeffding1963}. Its sign determines the
conditional label. Even with perfect reconstruction, a correct conclusive
label with probability at least $1-\delta_{\rm stat}$ is ensured once
$|\lambda|\ge2h_m$, which for $W_z=1$, $|\lambda|=0.5$, and
$\delta_{\rm stat}=0.01$ needs $m\ge43$, beyond the few-site benchmarks.
In both tests, $\widehat{\topinv}\in\{-1,+1,\perp\}$ denotes a label or an
inconclusive output, and by a union bound the probability of a conclusive
wrong label is at most the sum of the failure allocations used,
$\delta_{\rm pil}+\delta_{\rm cal}+\delta_{\rm stat}$. The frequency of
conclusive outputs is
a separate performance measure. These labels describe the assumed
continuation, rather than a topology assigned to the finite measured
device.

\begin{figure*}[t]
\centering
\includegraphics[width=\textwidth,height=0.57\textheight,keepaspectratio]{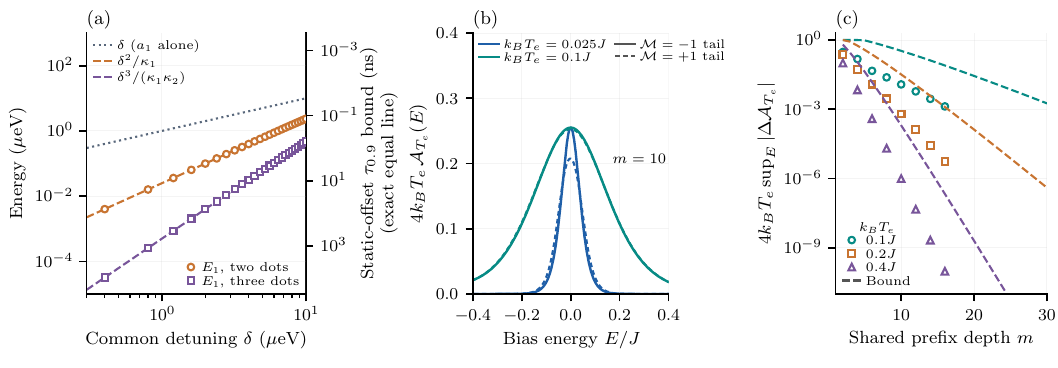}
\caption{Quantum-dot persistence bounds and thermally broadened end spectra. 
(a) Three-site quantum-dot chain with the couplings of
Ref.~\citenum{tenHaaf2025}, $\kappa_1=40\,\mu$eV and $\kappa_2=50\,\mu$eV.
Lines are the residual bounds of
Eq.~\eqref{eq:splitting_hierarchy} on the exact equal line for a static
common detuning $\delta$,
at depths one, two, and three. Markers are the
exact splitting $E_1$ of the two-dot chain on the first bond and of the
three-dot chain. The right axis converts the residual bound into a static-offset
persistence bound $\tau_{0.9}=\hbar\sqrt{0.2}/\varepsilon$ for the
profile-dependent operator.
(b) Thermally broadened end spectra, Eq.~\eqref{eq:thermal_spectrum}, of
the matched-prefix pair of Fig.~\ref{fig:matched_prefix_summary} with
$m=10$ and $L=600$, for topological (solid) and trivial (dashed) tails.
(c) Largest difference of the two end spectra, relative to the largest
possible thermal peak height, versus the shared depth $m$. Markers show
actual values and the dashed line the bound of Eq.~\eqref{eq:spectral_obstruction}.
}
\label{fig:experiments}
\end{figure*}

\section{Relation to experiments}
\label{sec:experiments}
We apply the persistence bounds to few-site quantum-dot models using
experimentally reported couplings. We then examine how thermal
broadening limits the distinction between chains with opposite bulk
labels in end tunneling spectra. Finally, we assess Hamiltonian
learning from boundary-response measurements on a digital qubit
processor.
\subsection{Three-site quantum-dot chains}
\label{sec:three-site-experiments}

In the free-fermion solution of the equal-line chain
\cite{LiebSchultzMattis1961,Pfeuty1970,Kitaev2001}, $K$ is tridiagonal in
the order
$(a_1,b_1,\ldots,a_L,b_L)$, and its only perfect matching pairs each $a_j$
with $b_j$. Hence $\operatorname{Pf}K=\prod_j(-\mu_j)$, and the
single-particle excitation energies $0\le E_1\le\cdots\le E_L$, the
positive eigenvalues of $iK$, obey
\begin{equation}
\prod_{k=1}^{L}E_k=\prod_{j=1}^{L}|\mu_j| .
\label{eq:pfaffian_identity}
\end{equation}
Because $E_1$ is the smallest singular value, every normalized candidate
bounds it, $E_1\le\|K\overline{\bm g}\|_2$. The recurrence candidate of
Eq.~\eqref{eq:oneway_recurrence} therefore gives, for each depth $m\le L$,
\begin{equation}
E_1\le\varepsilon_m\le\frac{\prod_{j=1}^{m}|\mu_j|}{\prod_{j=1}^{m-1}\kappa_j}.
\label{eq:splitting_hierarchy}
\end{equation}
Since the recurrence coefficients depend on the onsite profile, this
supplies, as in Eq.~\eqref{eq:equal_line_existence_bound}, a localized
operator for each compatible profile together with an upper bound on
$E_1$, the even-odd ground-state splitting in this model. The splitting
vanishes whenever a
single dot stays at $\mu_j=0$, and a common detuning $\delta$ splits the
ground state at most as $\delta^2/\kappa_1$ in a two-site chain and as
$\delta^3/(\kappa_1\kappa_2)$ in a three-site chain.

This bound reproduces the protection hierarchy reported in quantum-dot
experiments and explained there with the same model. Two-site chains are
quadratically protected against a common shift of the dot levels, and
three-site chains are more stable \cite{Bordin2025}. At zero
superconducting phase difference, the outer zero-bias peaks of three-site
chains survive the detuning of any one or two dots and split only when all
three are detuned \cite{tenHaaf2025,Bordin2026}. Detuning an outer dot
moves zero-bias weight onto the middle dot \cite{tenHaaf2025}, as the
recurrence amplitude $\alpha_2=-\mu_1/\kappa_1$ predicts. For the reported
couplings $J_1=\Delta_1\simeq20\,\mu$eV and $J_2=\Delta_2\simeq25\,\mu$eV
\cite{tenHaaf2025}, a common detuning of $5\,\mu$eV gives, in the
quadratic model, $E_1=0.062\,\mu$eV, within $1.3\%$ of the cubic bound
[Fig.~\ref{fig:experiments}(a)]. In the phase-$\pi$ configuration,
detuning the left dot splits only the
left peak, as observed and reproduced with Kitaev-model simulations in
Ref.~\citenum{tenHaaf2025}.

The same residuals bound persistence under static level offsets. Suppose
that every dot level obeys $|\mu_j|\le\delta$ and that the bonds
are on the equal line. As in Eq.~\eqref{eq:equal_line_existence_bound}, every
compatible device then contains an operator of depth $m$ with
$W_1\ge1-(\delta/\kappa_{\min})^2$ and residual at most
$\delta^m/\prod_{j<m}\kappa_j$. With $\hbar$ restored,
$\tau_{0.9}\ge\hbar\sqrt{0.2}/\varepsilon$, which is $0.294\,$ns divided by
the residual in $\mu$eV. For the couplings above and $\delta=5\,\mu$eV, the
bound is $58.8\,$ps for the outer-dot Majorana alone, $0.47\,$ns at depth
two, and $4.7\,$ns at depth three. For $\delta=1\,\mu$eV the depth-three
value is $0.588\,\mu$s
(Supplemental Material, Table~\ref{SM-tab:supp-experiment-guarantees}). Each added dot thus
multiplies the guarantee by about $\kappa/\delta$. Further dots beyond the
third leave the depth-three guarantee unchanged if the new bond is on the
equal line, since the outgoing term of Eq.~\eqref{eq:certified_residual}
then vanishes. These bounds address static level offsets in the quadratic model. Coulomb
interactions, finite Zeeman energy, the operating-point dependence of the
effective couplings, charge noise, and quasiparticle poisoning require
separate treatment
\cite{Tsintzis2022,Liu2022,Tsintzis2024,PanDasSarmaLiu2025}.

Away from the equal line, the bond sums are $s_j=J_j+\Delta_j>0$
and the mismatches are $d_j=\Delta_j-J_j$. For a chain ending at site 3,
$L=m=3$, the full chiral recurrence cancels the interior residuals.
A simultaneous coefficient box with $|\mu_j|\le\delta$,
$s_j\ge s_{j,-}>0$, and $|d_j|\le d_{\max}$ then gives each compatible
profile a normalized candidate satisfying
\begin{align}
E_1\le\varepsilon_3\le{}
\frac{\delta^3}{s_{1,-}s_{2,-}}+\delta d_{\max}
\left(\frac{1}{s_{1,-}}+\frac{1}{s_{2,-}}\right).
\label{eq:off-line-dot-bound}
\end{align}
The mismatch contribution is therefore proportional to
$\delta d_{\max}$ for three-site chains, rather than $d_{\max}$ alone.
For $\delta=d_{\max}=1\,\mu$eV and
$(s_{1,-},s_{2,-})=(40,50)\,\mu$eV, this gives
$\tau_{0.9}\ge6.46\,$ns with $W_1>0.9989$.
Section~\ref{SM-sec:supp-off-line-dot-bound} of Supplemental Material derives this bound, the
localization bound, the outgoing leakage for a truncated prefix, and the
uncertainty terms for a fixed candidate.

Two further experimental tools have counterparts here. The additional-dot
test of Ref.~\citenum{Bordin2026}, proposed in Ref.~\citenum{SoutoQD2023},
checks the localization of the end Majorana on the outer dot with a
resolution of about $1\,\mu$eV. The site-1 identity of
\secref{sec:a1-identity} addresses the same object in the time domain. Spectroscopy on
every dot \cite{tenHaaf2025} and additional probe dots \cite{Bordin2026}
go beyond site-1 access.

\subsection{End spectroscopy as a finite-window measurement}
\label{sec:end-spectroscopy}

Chain ends are probed by tunneling spectroscopy, either with a scanning
tunneling microscope on magnetic-atom chains \cite{NadjPerge2014,Jeon2017}
or with a normal lead on the outer dot \cite{Dvir2023,tenHaaf2025}.
In the quadratic model, the isolated-chain site-1 spectral function
is the Fourier transform of the electron anticommutator
$v^\dagger Y(t)v$. We consider this spectral function convolved
with the thermal kernel $-\partial_E f_{T_e}(E)$
\cite{BruusFlensberg2004}. Relating it to measured differential
conductance requires a justified probe and transport model,
including the effects of lead coupling and Andreev reflection
\cite{Flensberg2010}. Fourier transformation turns the thermal
convolution into a weight on the time axis,
\begin{align}
&\mathcal A_{T_e}(E)=\frac{1}{2\pi\hbar}\int_{-\infty}^{\infty}
e^{iEt/\hbar}\,v^\dagger Y(t)v\,\phi_{T_e}(t)\,dt,
\nonumber\\
&\phi_{T_e}(t)=\frac{\pi k_BT_et/\hbar}{\sinh(\pi k_BT_et/\hbar)}.
\label{eq:thermal_spectrum}
\end{align}
The weight decays
exponentially beyond $\hbar/(\pi k_BT_e)$, which is $0.12\,$ns at
$20\,$mK. An end spectrum therefore records this channel over an
effective window of this length, comparable to the benchmark duration
$T=3\hbar/J\simeq0.20\,$ns at $J=10\,\mu$eV.
For $J$ between $10$ and $20\,\mu$eV, the effective window corresponds to
$TJ\simeq2$ to $4$.

The matched-prefix bound then carries over. Let $\mathcal D_m(t)$ denote
the right side of Eq.~\eqref{eq:bessel_window} with $T$ replaced by $t$,
and let $\Delta\mathcal A_{T_e}$ be the difference of the two thermally
broadened spectra. Relative to the largest possible thermal peak height
$1/(4k_BT_e)$, using $|v^\dagger\Delta Y(t)v|\le\min\{2,\mathcal D_m(t)\}$
in Eq.~\eqref{eq:thermal_spectrum} gives
\begin{align}
&4k_BT_e\sup_E|\Delta\mathcal A_{T_e}(E)|
\nonumber\\
&\qquad\le\frac{4k_BT_e}{\pi\hbar}\int_0^\infty\phi_{T_e}(t)
\min\{2,\mathcal D_m(t)\}dt.
\label{eq:spectral_obstruction}
\end{align}
Because $\mathcal D_m(t)$ is negligible before the return time, which grows
linearly with $m$, the right side decreases exponentially with $m$ at
fixed $T_e$. For the pair of Fig.~\ref{fig:matched_prefix_summary}, it
falls below $10^{-2}$ for $m\ge13$ at $k_BT_e=0.2J$ and for $m\ge7$ at
$0.4J$. The actual spectra are closer
[Fig.~\ref{fig:experiments}(b),(c)]. Once ten sites are shared, they differ
by $1.2\times10^{-2}$, $6.3\times10^{-4}$, and $9.9\times10^{-7}$ of the
peak height at $k_BT_e=0.1J$, $0.2J$, and $0.4J$
(Supplemental Material, Table~\ref{SM-tab:supp-spectroscopy}), although the
semi-infinite continuations have opposite bulk labels. The topological
continuation supports a zero mode, whose finite open truncation has an
exponentially small splitting. The lowest energy of the
trivial-tail chain is $0.023J$, and the difference grows to
$4.8\times10^{-2}$ when $k_BT_e=0.025J$ becomes comparable to it. As in
the time domain, the obstruction holds at fixed temperature and
spectral precision, while a fixed pair can be resolved at finer precision
or lower temperature.

For $J$ between $10$ and $20\,\mu$eV and $T_e$ between $20$ and $50\,$mK,
$k_BT_e/J$ lies between
$0.09$ and $0.43$. Within the weak-probe model, this gives one
quantitative reason why a zero-bias peak at a chain end does not fix the
bulk label \cite{Yazdani2023}. The end spectra of chains with opposite
labels can agree to a part in $10^2$ at experimental temperatures.
Spatially resolved
spectroscopy along the chain
\cite{NadjPerge2014,Yazdani2023} or probes on interior dots
\cite{tenHaaf2025} can provide additional information, while a declared
continuation model restricts the inference target
(\secref{sec:certificate}). Spin-resolved end spectroscopy
\cite{Jeon2017} lies outside the spinless model and is not covered by
Eq.~\eqref{eq:spectral_obstruction}.

\begin{figure}[!tbp]
\centering
\includegraphics[width=\columnwidth]{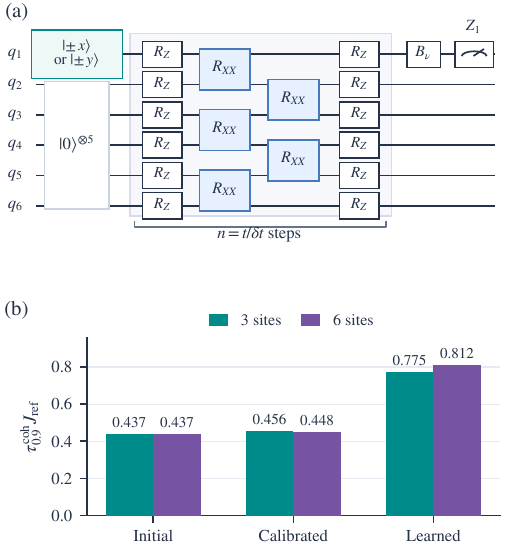}
\caption{
(a) Six-site logical circuit with boundary preparation
$|\pm x\rangle$ or $|\pm y\rangle$ and readout axis
$\nu\in\{x,y\}$. The hidden reset is $|0\rangle^{\otimes5}$,
and $B_x=H$ or $B_y=HS^\dagger$ precedes $Z_1$ measurement.
Each symmetric step contains $R_Z(\mu_j\delta t/2)$,
$R_{XX}(-\kappa_j\delta t)$, and $R_Z(\mu_j\delta t/2)$ layers,
with commuting $XX$ gates grouped for clarity. For
$\delta tJ_{\rm ref}=0.25$, times $tJ_{\rm ref}=1.5,3,4$
use $n=6,12,16$ steps. The three-site circuit has three wires.
(b) Certified time bounds for the same fixed
$\Gamma_2=(a_1-\alpha a_2)/\sqrt{1+\alpha^2}$, with
$\alpha\simeq0.3443$ and $W_1\ge0.8940$, over the initial,
calibration-only, and learned sets. The different reference designs and
budgets preclude a hardware-performance ranking.}
\label{fig:ibm-learning}
\end{figure}

\subsection{Processor measurements}
\label{sec:processor-measurements}
We acquired three- and six-site records on an IBM Quantum
processor, with about $9.6\times10^5$ and $2.9\times10^5$
binary outcomes, respectively (Fig.~\ref{fig:ibm-learning}). In both
studies, every onsite coefficient is unknown within
$\mu_j/J_{\rm ref}\in[0.5,1.5]$, and every equal-line
bond is uncertain within $\kappa_j/J_{\rm ref}\in[1.6,2.4]$ before
calibration. The target record uses preparation and readout at site 1.
Known onsite reference controls throughout the register and isolated-pair
bond calibrations require additional local access. Measured reference
responses provide the observation model for each study. The three-site study
interpolates a tensor grid of measured controls. The six-site study uses
separate coordinate scans, each varying one onsite coefficient while
leaving all other unknown coefficients at their target settings.

Both records yield partial joint confidence sets. The first two onsite
ranges narrow, while the third remains unresolved in the three-site
record and sites 3 through 6 remain unresolved in the six-site record.
For the same frozen two-site operator $\Gamma_2$, learning raises the
coherent-model guarantee from
$\tau_{0.9}^{\rm coh}J_{\rm ref}\ge0.4567$ after bond calibration to
$0.7754$ for three sites, and from $0.4480$ to $0.8126$ for six sites.
The gain comes from reduced coefficient uncertainty, while the device
dynamics remain unchanged.

Each result has conditional failure allocation $0.02$ under the stated
hardware-transfer assumptions and refers to the programmed continuous
coherent Hamiltonian. Sections~\ref{SM-sec:supp-ibm-three-site} and
\ref{SM-sec:supp-ibm-six-site} of Supplemental Material give these
assumptions and the acquisition,
confidence-set construction, and calibration details.

\section{Discussion}
\label{sec:discussion}

Under the declared device model, finite boundary records can certify
how long a localized Majorana-type operator keeps its symmetrized
autocorrelation above a threshold at a stated failure probability. Once a simultaneous confidence set covers the inferred
prefix and its outgoing couplings, the commutator guarantee holds for
every allowed continuation beyond them. For the end Majorana, the site-1 identity removes the need for a
continuation model when the first bond has equal hopping and pairing
and no additional boundary terms are present. Near the boundary sweet
spot, finite records can establish persistence far beyond their own
evolution window.

Bulk inference behaves differently. Equal-line chains with matched
prefixes and opposite bulk labels have
boundary moments that agree through order $4m-1$. At fixed time and finite
budget, some such pair defeats any
uniformly reliable label over the stated class, although every fixed pair
can eventually be resolved. The same localized operator and persistence
certificate can occur in both sectors. A uniform bulk label therefore
needs a declared relation to the unseen continuation, as in the periodic
and IID tests.

The residual also bounds the detuning-induced splitting of few-dot
chains. Thermal smearing produces a finite-resolution obstruction
for isolated-chain end spectra; applying this bound to measured
tunneling conductance requires a justified probe and transport model. Qubit simulators \cite{Mi2022} provide a
setting for implementing the
boundary protocol. Conditional processor learning is
demonstrated by the three- and six-site records in
Fig.~\ref{fig:ibm-learning}. Both strengthen the coherent-model guarantee
for the same localized operator while leaving several coefficients
unresolved.
A persistent boundary operator does not by itself store a qubit
\cite{GoldsteinChamon2011,Mazza2013,Ippoliti2016}, so memory performance
requires a separate assessment.

\begin{acknowledgments}
This work was supported by the Research Council of Norway through Grant Nos. 353919 and 361800 ``QTransMag'', and Grant No. 262633 ``QuSpin''. 
We acknowledge the use of IBM Quantum services
for this work. We have used ChatGPT and Codex (OpenAI) to assist with developing and
debugging numerical and plotting code, preparing figures, and
improving the clarity of the manuscript.

\end{acknowledgments}

\bibliography{references}

\clearpage
\onecolumngrid
\raggedbottom
\setcounter{section}{0}
\setcounter{subsection}{0}
\setcounter{equation}{0}
\setcounter{figure}{0}
\setcounter{table}{0}
\setcounter{theorem}{0}
\renewcommand{\thesection}{S\arabic{section}}
\renewcommand{\thesubsection}{\thesection\,\Alph{subsection}}
\renewcommand{\thesubsubsection}{\thesubsection\,\arabic{subsubsection}}
\makeatletter
\renewcommand{\p@subsection}{}
\renewcommand{\p@subsubsection}{}
\makeatother
\renewcommand{\theequation}{S\arabic{equation}}
\renewcommand{\thefigure}{S\arabic{figure}}
\renewcommand{\thetable}{S\arabic{table}}
\renewcommand{\thetheorem}{S\arabic{theorem}}
\renewcommand{\theHsection}{supp.\arabic{section}}
\renewcommand{\theHsubsection}{\theHsection.\arabic{subsection}}
\renewcommand{\theHsubsubsection}{\theHsubsection.\arabic{subsubsection}}
\renewcommand{\theHequation}{supp.\arabic{equation}}
\renewcommand{\theHfigure}{supp.\arabic{figure}}
\renewcommand{\theHtable}{supp.\arabic{table}}
\begin{center}
{\large\bfseries Supplemental Material}\par\medskip
{\bfseries Certifying Majorana persistence through Hamiltonian learning from boundary measurements}\par\medskip
Tohid Farajollahpour and Alireza Qaiumzadeh
\end{center}
\noindent

\par\medskip
\noindent\textbf{Contents}\par\smallskip
{\small
\hyperref[SM-sec:supp-nogo-section]{\ref*{SM-sec:supp-nogo-section}\quad Finite-resolution topology obstruction}\hfill\pageref*{SM-sec:supp-nogo-section}\par
\hspace*{1.5em}\hyperref[SM-sec:supp-nogo-proof]{\ref*{SM-sec:supp-nogo-proof}\quad Moment and response bounds}\hfill\pageref*{SM-sec:supp-nogo-proof}\par
\hspace*{1.5em}\hyperref[SM-sec:supp-nogo-numerics]{\ref*{SM-sec:supp-nogo-numerics}\quad Matched-prefix benchmarks}\hfill\pageref*{SM-sec:supp-nogo-numerics}\par
\hyperref[SM-sec:supp-quantum-channel]{\ref*{SM-sec:supp-quantum-channel}\quad Boundary quantum-channel discrimination}\hfill\pageref*{SM-sec:supp-quantum-channel}\par
\hyperref[SM-sec:supp-a1-identity]{\ref*{SM-sec:supp-a1-identity}\quad Site-1 identity for the end Majorana}\hfill\pageref*{SM-sec:supp-a1-identity}\par
\hyperref[SM-sec:supp-finite-shot-inversion]{\ref*{SM-sec:supp-finite-shot-inversion}\quad Validated inference and three-site benchmarks}\hfill\pageref*{SM-sec:supp-finite-shot-inversion}\par
\hspace*{1.5em}\hyperref[SM-sec:supp-shot-model]{\ref*{SM-sec:supp-shot-model}\quad Validated response enclosures}\hfill\pageref*{SM-sec:supp-shot-model}\par
\hspace*{1.5em}\hyperref[SM-sec:supp-few-site-benchmarks]{\ref*{SM-sec:supp-few-site-benchmarks}\quad Three-site benchmarks}\hfill\pageref*{SM-sec:supp-few-site-benchmarks}\par
\hspace*{1.5em}\hyperref[SM-sec:supp-readout-noise]{\ref*{SM-sec:supp-readout-noise}\quad Three-site inference with calibrated readout noise}\hfill\pageref*{SM-sec:supp-readout-noise}\par
\hyperref[SM-sec:supp-beyond]{\ref*{SM-sec:supp-beyond}\quad Persistence beyond the observation window and the cost of depth}\hfill\pageref*{SM-sec:supp-beyond}\par
\hspace*{1.5em}\hyperref[SM-sec:supp-sweet-spot]{\ref*{SM-sec:supp-sweet-spot}\quad Three-site benchmark near the boundary sweet spot}\hfill\pageref*{SM-sec:supp-sweet-spot}\par
\hspace*{1.5em}\hyperref[SM-sec:supp-tail-independent]{\ref*{SM-sec:supp-tail-independent}\quad Site-1 identity benchmark}\hfill\pageref*{SM-sec:supp-tail-independent}\par
\hspace*{1.5em}\hyperref[SM-sec:supp-robustness]{\ref*{SM-sec:supp-robustness}\quad Robustness of the long-time site-1 certificate}\hfill\pageref*{SM-sec:supp-robustness}\par
\hspace*{1.5em}\hyperref[SM-sec:supp-ramsey-comparison]{\ref*{SM-sec:supp-ramsey-comparison}\quad Comparison with isolated onsite Ramsey calibration}\hfill\pageref*{SM-sec:supp-ramsey-comparison}\par
\hspace*{1.5em}\hyperref[SM-sec:supp-truncation]{\ref*{SM-sec:supp-truncation}\quad Truncation bound and three-site inference with unknown tails}\hfill\pageref*{SM-sec:supp-truncation}\par
\hspace*{1.5em}\hyperref[SM-sec:supp-depth-cost]{\ref*{SM-sec:supp-depth-cost}\quad Depth cost of certification}\hfill\pageref*{SM-sec:supp-depth-cost}\par
\hspace*{1.5em}\hyperref[SM-sec:supp-interacting]{\ref*{SM-sec:supp-interacting}\quad Interacting continuations}\hfill\pageref*{SM-sec:supp-interacting}\par
\hspace*{1.5em}\hyperref[SM-sec:supp-contractor]{\ref*{SM-sec:supp-contractor}\quad Affine contractor and inference at depth four and five}\hfill\pageref*{SM-sec:supp-contractor}\par
\hyperref[SM-sec:supp-gate]{\ref*{SM-sec:supp-gate}\quad IBM processor records}\hfill\pageref*{SM-sec:supp-gate}\par
\hspace*{1.5em}\hyperref[SM-sec:supp-ibm-three-site]{\ref*{SM-sec:supp-ibm-three-site}\quad Three-site joint inference on IBM Marrakesh}\hfill\pageref*{SM-sec:supp-ibm-three-site}\par
\hspace*{1.5em}\hyperref[SM-sec:supp-ibm-six-site]{\ref*{SM-sec:supp-ibm-six-site}\quad Six-site inference by one-coordinate reference scans}\hfill\pageref*{SM-sec:supp-ibm-six-site}\par
\hyperref[SM-sec:supp-experiments]{\ref*{SM-sec:supp-experiments}\quad Relation to experiments}\hfill\pageref*{SM-sec:supp-experiments}\par
\hspace*{1.5em}\hyperref[SM-sec:supp-pfaffian]{\ref*{SM-sec:supp-pfaffian}\quad Splittings for reported couplings}\hfill\pageref*{SM-sec:supp-pfaffian}\par
\hspace*{1.5em}\hyperref[SM-sec:supp-experiment-guarantees]{\ref*{SM-sec:supp-experiment-guarantees}\quad Static-offset persistence bounds from dot-level precision}\hfill\pageref*{SM-sec:supp-experiment-guarantees}\par
\hspace*{1.5em}\hyperref[SM-sec:supp-off-line-dot-bound]{\ref*{SM-sec:supp-off-line-dot-bound}\quad Off-line leakage and simultaneous coefficient uncertainty}\hfill\pageref*{SM-sec:supp-off-line-dot-bound}\par
\hspace*{1.5em}\hyperref[SM-sec:supp-thermal-spectroscopy]{\ref*{SM-sec:supp-thermal-spectroscopy}\quad Thermally broadened end spectroscopy}\hfill\pageref*{SM-sec:supp-thermal-spectroscopy}\par
\hyperref[SM-sec:supp-related-learning]{\ref*{SM-sec:supp-related-learning}\quad Relation to conservation-law and Krylov learning}\hfill\pageref*{SM-sec:supp-related-learning}\par
}
\medskip
\section{Finite-resolution topology obstruction}
\label{SM-sec:supp-nogo-section}

We prove main-text Eqs.~\eqref{eq:moment_matching},
\eqref{eq:first_unmatched_moment}, \eqref{eq:locality_nogo}, and
\eqref{eq:bessel_window} for the matched pair of main-text
\secref{sec:nogo}.

\subsection{Moment and response bounds}
\label{SM-sec:supp-nogo-proof}

\noindent\textit{Moments.}
In the ordering $(a_1,b_1,a_2,b_2,\ldots)$ the equal-line generator graph
is a path, and $PK^nP^T$ sums length-$n$ boundary walks. The shortest
return walk through the first differing edge $(a_{m+1},b_{m+1})$ is
unique, with $(2m-1)+2+(2m-1)=4m$ steps, and up to sign its weights differ
by the squared common path weight times
$(\mu_{m+1}^{\rm top})^2-(\mu_{m+1}^{\rm triv})^2\ne0$. This proves
main-text Eqs.~\eqref{eq:moment_matching} and
\eqref{eq:first_unmatched_moment}.

\noindent\textit{Response bounds.}
With $R_m$ the projector onto the tail Majoranas, the common outgoing bond
gives $V=K_{\rm top}-K_{\rm triv}=R_mVR_m$, and Duhamel's identity gives
\begin{equation}
P(e^{K_{\rm top}t}-e^{K_{\rm triv}t})P^T
=\int_0^tPe^{K_{\rm top}(t-s)}R_mVR_m e^{K_{\rm triv}s}P^T\,ds.
\label{SM-eq:compact-nogo-duhamel}
\end{equation}
Both boundary-to-tail graph distances are $2m-1$. The finite-range bound
$\|P_Xe^{Kt}P_Z\|_2\le\exp[-\zeta d+e^{\zeta}\|K\|_2t]$ for projections at
graph distance $d$~\cite{Iserles2000}, applied to both factors, bounds the
integral by $t\|V\|_2e^{-2\zeta(2m-1)+e^{\zeta}gt}$ with
$g=\max_\sigma\|K_\sigma\|_2$. Each row has one onsite and one $2J$ bond
entry, and antisymmetry equates row and column sums, so $g\le G_\star$,
while $\|V\|_2=\Lambda_\star$. This proves main-text
Eq.~\eqref{eq:locality_nogo}. Bounding both factors instead by the
Bessel envelope of main-text \secref{sec:truncated-inference}, with
$G_\sigma=2J[1+\max(r_{\rm pre},r_\sigma)]$, gives main-text
Eq.~\eqref{eq:bessel_window}. The convolution of nonnegative
nondecreasing factors is nondecreasing in $T$, so this bound covers the
whole window.

\noindent\textit{Numerical evaluation.}
All Bessel-majorant integrals of this work are outward-rounded upper
Riemann sums that use the monotonicity of every term of the envelope on
$0\le x\le d$~\cite{DLMF}, 80-digit directed arithmetic for each
$\operatorname{BesJ}_n$, and the tail bound
$\sum_{n>N}|\operatorname{BesJ}_n(x)|\le(x/2)^{N+1}/[(N+1)!(1-q)]$ for
$q=x/[2(N+2)]<1$.

\subsection{Matched-prefix benchmarks}
\label{SM-sec:supp-nogo-numerics}
\label{SM-sec:supp-stress-test}

The pair of main-text Fig.~\ref{fig:matched_prefix_summary} has $L=64$,
$J=\Delta=1$, and
$(r_{\rm pre},r_{\rm top},r_{\rm triv})=(0.70,0.60,1.40)$, so
$G_\star T=38.4$ at $TJ=8$. For $m=8$, $10$, and $14$ the largest sampled
response distances, confirmed by 90-digit Taylor evaluation, are
$1.72\times10^{-2}$, $2.71\times10^{-5}$, and $1.63\times10^{-13}$, below
the upward-rounded Bessel bounds $1.80$, $3.55\times10^{-2}$, and
$8.80\times10^{-9}$. The optimized exponential bound first falls below one
at $m=28$.

For the decision bound of main-text
Eq.~\eqref{eq:learner-success-bound}, the pairs with $m=4,6,\ldots,14$
are sampled in all four channels at $tJ=0.5,1.0,\ldots,8.0$ ($D=64$,
$y_{\max}=0.78364$), with $N_{\rm sgn}=10^3$ and $10^6$; $L=80$ changes no
coordinate by more than $5.4\times10^{-15}$. Over $2\times10^4$ balanced
test records, the Bayes-optimal rule, the sign of the exact binomial
log-likelihood ratio, has accuracies $0.6198\pm0.0067$ at $m=8$,
$N_{\rm sgn}=10^3$, and $0.5097\pm0.0069$ at $m=10$, $N_{\rm sgn}=10^6$
($95\%$ intervals), against pair-specific chi-squared Pinsker bounds of
$0.7048$ and $0.50972$, and $0.50031$ at $m=10$, $N_{\rm sgn}=10^3$. The
small-signal success $\Phi(\chi/2)$, with $\Phi$ the standard normal
distribution function and
$\chi^2=N_{\rm sgn}\sum_\iota(y_{{\rm top},\iota}-y_{{\rm triv},\iota})^2/(1-y_{{\rm triv},\iota}^2)$,
gives $0.614$, $0.5055$, and $0.50017$, so the Bayes advantage is about
$56\%$ of the Pinsker advantage.

\section{Boundary quantum-channel discrimination}
\label{SM-sec:supp-quantum-channel}

Each interrogation of equal-length truncations with $L\ge m+1$ starts in
$\rho_{BR}\otimes I_E/d_E$, with $B$ site 1, $E$ the other qubits,
$d_E=2^{L-1}$, and $R$ a noninteracting reference. With
$H_K=(i/4)\bm\gamma^TK\bm\gamma$ and $U_K(t)=e^{-iH_Kt}$,
\begin{equation}
\Phi_K^t(\rho_B)=\Tr_E\!\left[
U_K(t)(\rho_B\otimes I_E/d_E)U_K(t)^\dagger\right].
\label{SM-eq:supp-boundary-channel}
\end{equation}
For $\rho_B=(I+xX_1+yY_1+zZ_1)/2$ the channel is unital with Bloch matrix
\begin{equation}
\mathsf B_K(t)=Y(t)\oplus\det Y(t),\qquad
(x',y',z')^T=\mathsf B_K(t)(x,y,z)^T.
\label{SM-eq:supp-channel-bloch}
\end{equation}
The images of $X_1=a_1$ and $Y_1=b_1$ are the rows of $Y$, since hidden
Majoranas have zero partial trace, and $Z_1=-ia_1b_1$ maps to $\det Y$ by
row orthogonality.

Input and output rotations reduce the Bloch-matrix difference
$\Delta\mathsf B=\mathsf B_0-\mathsf B_1$ to $\diag(\lambda_i)$, with
$|\lambda_i|=s_i$ its singular values $s_1\ge s_2\ge s_3\ge0$. The
difference map $\Delta\Phi^t=\Phi_0^t-\Phi_1^t$ is then Pauli diagonal
with coefficients $q_0=\sum_i\lambda_i/4$ and
$q_k=(2\lambda_k-\sum_i\lambda_i)/4$, and its diamond norm, without a
factor one half, is $\sum_j|q_j|$, attained by a maximally entangled
input~\cite{Sacchi2005,Watrous2018}. This gives
\begin{equation}
d_\diamond(t)=\|\Delta\Phi^t\|_\diamond
=\max_{\rho_{BR}}\|(\Delta\Phi^t\otimes\operatorname{id}_R)(\rho_{BR})\|_1
=\max\{s_1,(s_1+s_2+s_3)/2\},\qquad \dim R=2.
\label{SM-eq:supp-diamond-exact}
\end{equation}
For the boundary channels, with singular values $a\ge b\ge0$ of $Y_0-Y_1$
and $c=\det Y_0-\det Y_1$,
\begin{equation}
d_\diamond(t)=\max\{a,|c|,(a+b+|c|)/2\}
\le\min\{2,2\|Y_0(t)-Y_1(t)\|_2\}.
\label{SM-eq:supp-diamond-response}
\end{equation}
The determinant difference of two contractions obeys
$|c|\le2\|Y_0-Y_1\|_2$, which gives main-text
Eq.~\eqref{eq:boundary-channel-distance}; its factor 2 is approached by
$Y_0=I$ and $Y_1=(1-\epsilon)I$. By the product rule, the channel
derivatives first differ at order $4m$.

For the pair of Sec.~\ref{SM-sec:supp-nogo-numerics} at $tJ=8$, the optimal
single-use successes $1/2+d_\diamond/4$ are $0.5043086$ and $0.5000068$
for $m=8$ and $10$. Full-Hilbert-space calculations on two to six sites
agree with the Bloch reduction to $10^{-14}$, and exact integer arithmetic
confirms the first channel-moment mismatch at $4m$ for $m=1,\ldots,8$.

\section{Site-1 identity for the end Majorana}
\label{SM-sec:supp-a1-identity}

Write $H=H_1+H_{\ge2}$, with $H_1$ the site-1 onsite and bond-1 terms of
main-text Eq.~\eqref{eq:H} and $H_{\ge2}$ any self-adjoint operator on
sites $j\ge2$, parity even for fermions. For qubits, $H_{\ge2}$ may
include parity-breaking $X_j$ fields, but no $X_1$ or $Y_1$ field may act
on the probed qubit. Since $H_{\ge2}$ commutes with $a_1=X_1$ and
$b_1=Y_1$,
\begin{equation}
\dot a_1=-\mu_1b_1+(\Delta_1-J_1)b_2,\qquad
\dot b_1=\mu_1a_1+(J_1+\Delta_1)a_2,
\label{SM-eq:supp-a1-heisenberg}
\end{equation}
with continuation-dependent $b_2(t)$ and $a_2(t)$. A real combination of
two anticommuting Majoranas has the norm of its coefficient vector, so
$\|[H,a_1]\|=[\mu_1^2+(J_1-\Delta_1)^2]^{1/2}$ and
$\|[H,b_1]\|=[\mu_1^2+(J_1+\Delta_1)^2]^{1/2}\equiv\Omega_1$ for every
such continuation.

On an equal-line first bond, the first relation, linearity of the signed
half differences, and $Y_{a_1b_1}(0)=0$ give main-text
Eq.~\eqref{eq:a1_identity}, $Y_{a_1b_1}(t)=-\mu_1I(t)$ with
$I(t)=\int_0^tY_{b_1b_1}(s)\,ds$. Main-text
Eq.~\eqref{eq:operator_overlap} applied to $b_1$ gives
$\frac12\{b_1(s),b_1\}\succeq(1-\frac12\Omega_1^2s^2)I$, whose expectation
in $\frac12I\otimes\rho_E$ is $Y_{b_1b_1}(s)$, so
$\max\{-1,1-\frac12\Omega_1^2s^2\}\le Y_{b_1b_1}(s)\le1$ and
\begin{equation}
I_-(t)\le I(t)\le t,\qquad
I_-(t)=\begin{cases}t-\tfrac16\Omega_1^2t^3,&\Omega_1t\le2,\\[3pt]
\tfrac{8}{3\Omega_1}-t,&\Omega_1t>2.\end{cases}
\label{SM-eq:supp-a1-lower}
\end{equation}
The lower bound is positive for $t<8/(3\Omega_1)$ and sharp to third
order, since $\kappa_1=0$ gives
$I(t)=\sin(\mu_1t)/\mu_1=t-\mu_1^2t^3/6+O(t^5)$. Off the equal line,
quadratic continuations add $(\Delta_1-J_1)\int_0^t[e^{Ks}]_{b_2b_1}ds$,
bounded by $|\Delta_1-J_1|\Omega_1t^2/2$ because
$[e^{Ks}]_{b_2b_1}=[e^{-Ks}]_{b_1b_2}$ is a coefficient of $b_1(-s)-b_1$.

For an observation interval $[\widehat y-\xi,\widehat y+\xi]$ of
$Y_{a_1b_1}(t)$ with $I_-(t)>0$ and $|\widehat y|\le\xi$, the identity
gives
\begin{equation}
-\frac{\xi+\widehat y}{I_-(t)}\le\mu_1\le\frac{\xi-\widehat y}{I_-(t)} .
\label{SM-eq:supp-a1-interval}
\end{equation}
If $\xi-\widehat y<0$ or $\xi+\widehat y<0$, the corresponding endpoint
uses $t$ in place of $I_-(t)$. Intersecting over sample times and the
declared domain gives a confidence interval for $\mu_1$. We evaluate
$\Omega_1$ with the declared bounds $|\mu_1|\le2J_{\rm ref}$ and
$\kappa_1\le2.4J_{\rm ref}$, for which $8/(3\Omega_1)=0.85/J_{\rm ref}$
admits all sample times $tJ_{\rm ref}\le0.75$, and then iterate with the
current interval for $\mu_1$. On the coverage event,
$\tau_\eta(a_1)\ge\sqrt{2(1-\eta)}/\sup|\mu_1|$ for every continuation
(Sec.~\ref{SM-sec:supp-tail-independent}).

\paragraph{Bounded additional boundary terms.}
If $H=H_0+V_{\partial}$, with $H_0=\mu_1Z_1/2-\kappa_1X_1X_2/2+H_{\ge2}$
and a calibrated bound $\|V_{\partial}\|\le w_{\partial}$, then
$\|[V_{\partial},O]\|\le2w_{\partial}\|O\|$ and the unit-norm signed
preparation functional give
\begin{align}
Y_{a_1b_1}(t)&=-\mu_1I(t)+R_{\partial}(t),
&|R_{\partial}(t)|&\le2w_{\partial}|t|,\nonumber\\
\|[H,a_1]\|&\le|\mu_1|+2w_{\partial},
&\|[H,b_1]\|&\le\Omega_1+2w_{\partial}.
\label{SM-eq:supp-boundary-perturbation}
\end{align}
Here $I(t)$ uses the actual responses. Replacing $\Omega_1$ by
$\Omega_1+2w_{\partial}$ in Eq.~\eqref{SM-eq:supp-a1-lower} and enlarging
each tolerance by $2w_{\partial}|t|$ preserves the inference, and on the
joint coverage event
\begin{equation}
\tau_{0.9}(a_1)\ge
\frac{\sqrt{0.2}}{\sup|\mu_1|+2w_{\partial}}.
\label{SM-eq:supp-boundary-perturbation-time}
\end{equation}

\paragraph{Shot-to-shot coefficient variation.}
If $\mu_1$ is redrawn independently before each shot and held fixed during
it, with $\kappa_1$, $H_{\ge2}$, and the reset fixed, conjugation by $X_1$
maps $\mu_1\to-\mu_1$ and swaps the two $Y_1$ preparations. Hence
$Y_{a_1b_1}$ is odd in $\mu_1$ and averages to zero for a symmetric law,
even when every shot has a nonzero residual. The diagonal response bounds
the second moment instead. For a fixed offset,
$a_1(t)-a_1=-\mu_1\int_0^tb_1(s)\,ds$ and
$\frac12\{b_1(s),b_1(u)\}\succeq[1-\frac12\Omega_1^2(s-u)^2]I$ give
$t^2I\succeq[\int_0^tb_1(s)\,ds]^2\succeq(t^2-\Omega_1^2t^4/12)I$, so
\begin{equation}
\left(1-\frac{\mu_1^2t^2}{2}\right)I
\preceq\frac12\{a_1(t),a_1\}
\preceq\left[1-\frac{\mu_1^2t^2}{2}
\left(1-\frac{\Omega_1^2t^2}{12}\right)\right]I
\label{SM-eq:supp-mixture-finite-time}
\end{equation}
for every initial state and time-independent hidden Hamiltonian consistent
with the cut, with $1/12$ sharp at fourth order for $\kappa_1=0$. The
diagonal response is the expectation of the middle term. With declared
$|\mu_1|\le\mu_{\max}$ and $|\kappa_1|\le\kappa_{\max}$,
$\Omega_{\max}^2=\mu_{\max}^2+\kappa_{\max}^2$, and
$q_{\rm mix}(t)=1-\Omega_{\max}^2t^2/12$, its shot average therefore obeys
\begin{equation}
1-\frac{t^2}{2}\mathbb E\mu_1^2
\le\overline Y_{a_1a_1}(t)
\le1-\frac{t^2q_{\rm mix}(t)}{2}\mathbb E\mu_1^2,
\label{SM-eq:supp-mixture-response-envelope}
\end{equation}
informative for $0<t<\sqrt{12}/\Omega_{\max}$, which includes
$tJ_{\rm ref}=0.25,0.5,0.75$ for $\mu_{\max}=2J_{\rm ref}$ and
$\kappa_{\max}=2.4J_{\rm ref}$. Simultaneous Hoeffding intervals
$[\widehat y^{xx}_\ell-\xi_\ell,\widehat y^{xx}_\ell+\xi_\ell]$ for
independent static-offset shots from one law then imply
\begin{equation}
\mathbb E\mu_1^2\le U_{\mu^2}:=
\min\left\{\mu_{\max}^2,
\min_{\ell:\,q_{\rm mix}(t_\ell)>0}
\frac{2[1-\widehat y^{xx}_\ell+\xi_\ell]}
{t_\ell^2q_{\rm mix}(t_\ell)}\right\},
\label{SM-eq:supp-mixture-moment-bound}
\end{equation}
and hence $\overline C_{a_1}(t)\ge0.9$ for $|t|\le\sqrt{0.2/U_{\mu^2}}$; a
negative right-hand side returns inconclusive. This extension needs
additional $\pm x$ preparations with $X_1$ readout.
{\color{black} It is evaluated in the offline benchmark of
Sec.~\ref{SM-sec:supp-robustness}. The processor analyses retain their
fixed-coefficient transfer model.}

{\color{black}
\paragraph{Joint mismatch and quasistatic onsite variation.}
The fluctuation bound extends to a bounded first-bond mismatch without
identifying one Hamiltonian for the whole record. For each shot, let
\[
H_d=\frac{\mu_1}{2}Z_1-\frac{s_1}{2}X_1X_2
       +\frac{d_1}{2}Y_1Y_2+H_{\ge2},
\qquad s_1=J_1+\Delta_1,\quad d_1=\Delta_1-J_1.
\]
The coefficients are constant within a shot, and their joint law is the
same for the two preparation signs and for all sampled times. Assume
$|\mu_1|\le\mu_{\max}$, $|s_1|\le s_{\max}$ and
$|d_1|\le d_{\max}$. The same hidden reset is used for both signs.
Let $a_d(t)$ and $a_0(t)$ denote the Heisenberg evolutions of $a_1$
under $H_d$ and the Hamiltonian with $d_1$ set to zero, respectively,
with all other coefficients held at their values in that shot.
Duhamel's identity and $\|d_1Y_1Y_2/2\|=|d_1|/2$ give
\begin{equation}
\|a_d(t)-a_0(t)\|\le |d_1|\,|t|.
\label{SM-eq:supp-mismatch-displacement}
\end{equation}
For $\rho_0=I_1/2\otimes\rho_E$, the positive quadratic form
$\|A\|_{\rho_0}^2=\Tr(\rho_0A^\dagger A)$ obeys the triangle inequality.
Equation~\eqref{SM-eq:supp-mixture-finite-time} implies
$\|a_0(t)-a_1\|_{\rho_0}\ge
|\mu_1t|\sqrt{q(t)}$, where
$q(t)=1-(\mu_{\max}^2+s_{\max}^2)t^2/12>0$.
Taking the root mean square over the quasistatic law and applying the
triangle inequality once more therefore yields
\begin{equation}
\sqrt{q(t)\,\mathbb E\mu_1^2}
\le \frac{\sqrt{2[1-\overline Y_{a_1a_1}(t)]}}{|t|}
       +d_{\max}.
\label{SM-eq:supp-mixture-mismatch-moment}
\end{equation}
The identity
$\|a_d(t)-a_1\|_{\rho_0}^2
=2[1-Y_{a_1a_1}(t)]$ connects this displacement to the measured
diagonal response. This argument also allows a time-independent
interacting $H_{\ge2}$.

Suppose simultaneous response intervals, enlarged by prescribed
systematic and numerical allowances, give lower endpoints
$\underline y_\ell$ for the shot-averaged diagonal responses.
Endpoints below $-1$ are replaced by $-1$, while an empty physical
response interval returns inconclusive. Equation~\eqref{SM-eq:supp-mixture-mismatch-moment}
then gives
\begin{equation}
U_{\mu^2}^{(d)}=
\min\left\{\mu_{\max}^2,
\min_{\ell:\,q(t_\ell)>0}
\frac{\left[\sqrt{2(1-\underline y_\ell)}/t_\ell+d_{\max}\right]^2}
     {q(t_\ell)}\right\}.
\label{SM-eq:supp-mixture-mismatch-inference}
\end{equation}
The exact commutator norm of $a_1$ is
$\sqrt{\mu_1^2+d_1^2}$. Consequently, for every fixed initial state,
averaging the operator persistence inequality over the same law gives
\begin{equation}
\overline C_{a_1}(t)\ge
1-\frac{t^2}{2}\left[U_{\mu^2}^{(d)}+d_{\max}^2\right],
\qquad
\overline\tau_{0.9}\ge
\sqrt{\frac{0.2}{U_{\mu^2}^{(d)}+d_{\max}^2}}.
\label{SM-eq:supp-mixture-mismatch-time}
\end{equation}
The zero-denominator case gives an unlimited interval in this static
model. This is a guarantee for the autocorrelation averaged over a
stationary distribution of Hamiltonians. A guarantee for every
realization instead requires a bound on the support of that
distribution. Time-dependent fluctuations during an evolution require
a different dynamical model.
}


\section{Validated inference and three-site benchmarks}
\label{SM-sec:supp-finite-shot-inversion}

\subsection{Validated response enclosures}
\label{SM-sec:supp-shot-model}
\label{SM-sec:supp-pilot}

The admissible domain of main-text \secref{sec:finite_data} is a
finite union of compact boxes. Site signs are fixed by making the first
nonzero member of $(J_j+\Delta_j,J_j-\Delta_j)$ positive, which is $J_j>0$
on the equal line, and sets that cross a gauge boundary are split or
reported inconclusive. A validated oracle $[f_\iota](B)$ deterministically
encloses the response range on a box $B$~\cite{Moore1966,Rump2010}. With
midpoint $\bm c_B$, half-diagonal $s_B$, and verified generator Lipschitz
constant $c_K$, orthogonality and Duhamel's identity give
$|f_\iota(\bm\theta)-f_\iota(\bm c_B)|\le|t_\ell|c_Ks_B$ on $B$, and the
center series through order $q_{\rm ser}$ is evaluated with outward
rounding and the tail bound
\begin{equation}
S_{q_{\rm ser}}(t)=\sum_{n=0}^{q_{\rm ser}}\frac{(K(\bm c_B)t)^n}{n!},
\qquad
\varrho_{q_{\rm ser}}(x)=e^x\frac{x^{q_{\rm ser}+1}}{(q_{\rm ser}+1)!},
\qquad x\ge |t|\|K(\bm c_B)\|_2.
\label{SM-eq:pilot-series-tail}
\end{equation}
Adding the tail and $[-|t_\ell|c_Ks_B,|t_\ell|c_Ks_B]$ to the enclosed
entry of $PS_{q_{\rm ser}}P^T$ gives $[f_\iota](B)$. All benchmarks and
processor analyses round every floating-point operation outward, and in
the label benchmarks a search that reaches a computational limit returns
an inconclusive label.

\subsection{Three-site benchmarks}
\label{SM-sec:supp-few-site-benchmarks}

\subsubsection{End-to-end three-site qubit benchmark}
\label{SM-sec:supp-three-site}

The benchmark of main-text \secref{sec:three-site} assumes that the
chain ends at site 3; Sec.~\ref{SM-sec:supp-truncation} relaxes this
assumption. Calibration isolates bonds 1--2 and 2--3 in turn, with
$\kappa_j/J_{\rm ref}\in[1.6,2.4]$, $H_j=-\kappa_jX_jX_{j+1}/2$, and $Z_j$
readout at $t_{\rm cal}J_{\rm ref}=1$, giving means
$\cos(\kappa_jt_{\rm cal})$. Simultaneous radii
$[2\log(4/\delta_{\rm cal})/N_{\rm cal}]^{1/2}$ are inverted on the
monotone cosine branch by validated bisection, assuming ideal local
preparation, isolation, and readout. The boundary stage samples all four
responses at $t_\ell J_{\rm ref}=\ell/4$, $\ell=1,\ldots,12$ ($D=48$),
with a maximally mixed hidden reset. The settings $N_\pm=20000$,
$N_{\rm cal}=200000$, and $\delta_{\rm pil}=\delta_{\rm cal}=0.01$ are
fixed before the exact-binomial records are generated, for
$2DN_\pm+2N_{\rm cal}=2320000$ outcomes. A checked $10^{-9}$ mean-bias
allowance covers the conversion of simulated probabilities, and count
rounding is enclosed separately.

\noindent\textit{Validated forward computation.}
With $K_0$ the exact binary center generator, $Q$ numerical real Schur
vectors, and $D_0=-D_0^T$ the retained rotation blocks, variation of
constants gives, without assuming exact factorization or orthogonality,
\begin{equation}
\|e^{K_0t}-Qe^{D_0t}Q^T\|_2
\le \|I-QQ^T\|_2+
|t|\,\|K_0Q-QD_0\|_2\,\|Q\|_2.
\label{SM-eq:supp-schur-validation}
\end{equation}
For largest onsite and bond half-widths $r_\mu,r_\kappa$, each
perturbation consists of disjoint skew blocks, so
\begin{equation}
\|K-K_0\|_2\le r_\mu+r_\kappa,
\qquad
|f_\iota(K)-f_\iota(K_0)|\le |t_\ell|(r_\mu+r_\kappa).
\label{SM-eq:supp-three-site-box-oracle}
\end{equation}
The search bisects the longest side of the onsite cube down to
$3J_{\rm ref}/128$, with limits of $10^5$ visited boxes and 300 seconds.

\begin{figure}[tp]
\centering
\includegraphics[width=0.75\linewidth,height=0.72\textheight,keepaspectratio]{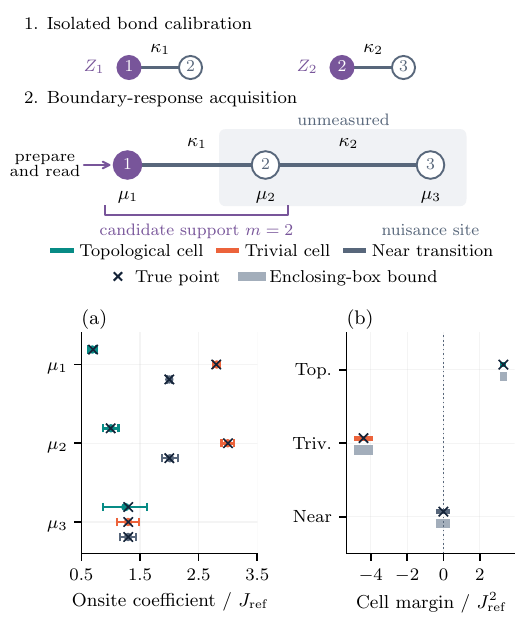}
\caption{Synthetic three-site inference. Top, isolated-bond calibration
uses pair preparation and $Z_1$ or $Z_2$ readout, and boundary acquisition
prepares and measures only qubit $1$. The bracket marks the candidate
support $m=2$, with outgoing bond $\kappa_2$ and fitted nuisance site $3$.
(a) Onsite hulls for the three devices of
Table~\ref{SM-tab:compact-three-site}, with true values as crosses. (b)
Periodic-cell margin intervals over the retained sets (colored) and their
enclosing boxes (grey).}
\label{SM-fig:supp-three-site}
\end{figure}

\noindent\textit{Frozen operator.}
The two-site operator $\Gamma=(a_1-\alpha a_2)/\sqrt{1+\alpha^2}$ of
main-text \secref{sec:three-site} uses
$\alpha=\widetilde\mu_1/\widetilde\kappa_1$ from the hull and calibration
centers, clipped to $[-0.49,0.49]$ so that $W_1>0.8$. Its outgoing term
vanishes on the equal line, and
\begin{equation}
\varepsilon(\bm\mu,\bm\kappa)^2
=\frac{(\mu_1-\alpha\kappa_1)^2+\alpha^2\mu_2^2}
{1+\alpha^2}.
\label{SM-eq:supp-three-site-residual}
\end{equation}
Interval evaluation over the retained boxes gives $\varepsilon^{\rm cert}$
and $\tau_{0.9}\ge\sqrt{0.2}/\varepsilon^{\rm cert}$ for all compatible
parity-even continuations without extra cross-cut terms. The three
examples share $\bm\kappa/J_{\rm ref}=(1.92,2.08)$
(Table~\ref{SM-tab:compact-three-site} and Fig.~\ref{SM-fig:supp-three-site}).
For the first, $W_1=1/(1+\alpha^2)\simeq0.88614$,
$\varepsilon^{\rm cert}\le0.3885J_{\rm ref}$, and
$\tau_{0.9}J_{\rm ref}\ge1.151$, despite onsite-hull widths
$(0.140625,0.2578125,0.75)J_{\rm ref}$. The repeated-cell label of
main-text \secref{sec:certificate} has per-record failure probability
at most $0.02$, and unions and hulls give the same labels.

\begin{table}[t]
\caption{Calibrated three-site records. Margin intervals are
outward-rounded enclosures of the cell margin
$(\kappa_1\kappa_2-\mu_1\mu_2)/J_{\rm ref}^2$. All records use the same
calibration and boundary-shot budgets.}
\label{SM-tab:compact-three-site}
\centering
\begin{ruledtabular}
\begin{tabular}{c c c c}
$\bm\mu/J_{\rm ref}$ & Margin interval & Output & Retained boxes \\
\hline
$(0.7,1.0,1.3)$ & $[3.100,3.467]$ & Topological & 841 \\
$(2.8,3.0,1.3)$ & $[-4.923,-3.910]$ & Trivial & 293 \\
$(2,2,1.3)$ & $[-0.427,0.346]$ & Inconclusive & 304
\end{tabular}
\end{ruledtabular}
\end{table}

Twenty independent repetitions per example give 20 correct strict labels
for each of the first two and 20 inconclusive outputs for the third,
with no coverage miss or computation stop. Independent matrix-series, full-qubit-likelihood, perturbation, and
forced-stop checks test the implementation.

\subsubsection{Persistence baselines for the saved record}
\label{SM-sec:supp-persistence-baselines}

For the first record we freeze $\alpha=0.35846$ in
$\Gamma_2=(a_1-\alpha a_2)/\sqrt{1+\alpha^2}$ and evaluate
Eq.~\eqref{SM-eq:supp-three-site-residual} on the declared onsite cube with
the original bond domain, the same cube with the calibrated bonds, the
841-box retained union, and the true coefficients. The two squared terms
depend on disjoint coefficients, so each box supremum is exact and is
evaluated in rational arithmetic. The resulting times are those of
main-text Fig.~\ref{fig:persistence_window}(a); the first two are
retrospective evaluations of the data-selected operator, not separate
protocols.

\subsubsection{Independently uncertain hopping and pairing}
\label{SM-sec:supp-independent-bonds}

This benchmark keeps the three-site model and onsite domain but declares
independent bonds $J_j/J_{\rm ref}\in[0.8,1.2]$ and
$\Delta_j/J_{\rm ref}\in[0.6,1.0]$. The simulated profile
$\bm\mu/J_{\rm ref}=(0.7,1.0,1.3)$, $\bm J/J_{\rm ref}=(0.96,1.04)$, and
$\bm\Delta/J_{\rm ref}=(0.86,0.89)$ enters only data generation and
diagnostics. An isolated bond,
$H_j=-(J_j+\Delta_j)X_jX_{j+1}/2+(\Delta_j-J_j)Y_jY_{j+1}/2$, couples
$|01\rangle$ to $|10\rangle$ with amplitude $-J_j$ and $|00\rangle$ to
$|11\rangle$ with $-\Delta_j$, so $Z_j$ readout at
$t_{\rm cal}J_{\rm ref}=1$ after preparing $|01\rangle$ or $|00\rangle$
gives means $\cos(2J_jt_{\rm cal})$ and $\cos(2\Delta_jt_{\rm cal})$,
which do not fix the declared positive signs. Four calibration records of
$2\times10^5$ outcomes, with simultaneous radius
$[2\log(8/\delta_{\rm cal})/N_{\rm cal}]^{1/2}$ and
$\delta_{\rm cal}=0.01$, give the $J_j$ and $\Delta_j$ intervals. With the
equal-line boundary settings, a record has
$4N_{\rm cal}+2DN_\pm=2.72\times10^6$ outcomes and joint failure
probability at most $0.02$, and the forward model is the full affine
generator including $\mu_3$.

With onsite half-widths $r_{\mu,j}$ and bond half-widths
$r_{J,j},r_{\Delta,j}$, the hopping and pairing perturbations each form
two skew paths, so
\begin{equation}
\|K-K_0\|_2\le
\max_j r_{\mu,j}
+\sqrt{r_{J,1}^2+r_{J,2}^2}
+\sqrt{r_{\Delta,1}^2+r_{\Delta,2}^2}.
\label{SM-eq:supp-independent-box}
\end{equation}
Equation~\eqref{SM-eq:supp-schur-validation} also encloses the rounding of
$J_j\pm\Delta_j$. From the retained hull and calibration centers we freeze
$\alpha=\widetilde\mu_1/(\widetilde J_1+\widetilde\Delta_1)$, clipped to
$[-0.49,0.49]$, for which $\Gamma_2$ has the exact squared residual
\begin{align}
\varepsilon^2(\bm\mu,\bm J,\bm\Delta)
={}&\frac{[\mu_1-\alpha(J_1+\Delta_1)]^2
+[\alpha\mu_2+\Delta_1-J_1]^2}{1+\alpha^2}
\nonumber\\
&+\frac{\alpha^2(\Delta_2-J_2)^2}{1+\alpha^2}.
\label{SM-eq:supp-independent-residual}
\end{align}
All terms, including the outgoing last term that is nonzero off the equal
line, are enclosed over each retained box and both calibrated bond
intervals.

The search visits 6527 boxes and retains 1174 without reaching a limit,
giving $\alpha=0.38418$, $W_1\ge0.8713$,
$\varepsilon_2\le0.3348J_{\rm ref}$, and $\tau_{0.9}J_{\rm ref}\ge1.3358$,
with outgoing norm in $[0.04992,0.05628]J_{\rm ref}$. At the exact
coefficients the same candidate gives $\tau_{0.9}J_{\rm ref}=1.652$, and
the simulated autocorrelation first falls below $0.9$ at $tJ_{\rm ref}\simeq4.39$.

\subsection{Three-site inference with calibrated readout noise}
\label{SM-sec:supp-readout-noise}

We add calibrated symmetric readout noise to the benchmark of
Sec.~\ref{SM-sec:supp-three-site}, keeping its unitary dynamics.

\subsubsection{Observation model and simultaneous coverage}

Each outcome is independently flipped with an unknown stationary
probability $p_{\rm ro}\in[0,0.49]$, common to all bases and stages of a
record, so the contrast is $c_{\rm ro}=1-2p_{\rm ro}\in[0.02,1]$;
preparation, reset, isolation, and evolution stay ideal. Measurement
effects $(I\pm c_{\rm ro}\sigma_\alpha)/2$ give
\begin{align}
\Pr(O=+1\mid s,\iota)&=\frac{1+s c_{\rm ro}Y_\iota}{2},\qquad s=\pm1,
\nonumber\\
\Pr(O=+1\mid j,\mathrm{cal})&=
\frac{1+c_{\rm ro}\cos(\kappa_jt_{\rm cal})}{2}.
\label{SM-eq:supp-noisy-likelihood}
\end{align}
An independent reference batch measures ideally prepared $\pm Z$ states
through the same site-1 readout and estimates $c_{\rm ro}$. With
$N_{\rm ro}$ shots per reference
sign, $N_{\rm cal}$ per bond, and $N_\pm$ per boundary sign and
coordinate, the simultaneous radii are
\begin{align}
u_{\rm ro}&=\sqrt{\frac{\log(2/\delta_{\rm ro})}{N_{\rm ro}}},
&u_{\rm cal}&=\sqrt{\frac{2\log(4/\delta_{\rm cal})}{N_{\rm cal}}},
\nonumber\\
u_Y&=\sqrt{\frac{\log(2D/\delta_Y)}{N_{\pm}}},
&D&=48.
\label{SM-eq:supp-noisy-radii}
\end{align}
The intervals include the $10^{-9}$ simulation bias and count-to-mean
rounding. With enclosures $[c_{\rm ro}]\subset[0.02,1]$, $[m_j]$, and
$[d_\iota]$ of the contrast, the noisy bond means, and the boundary
half-differences, joint coverage gives
\begin{equation}
\cos(\kappa_jt_{\rm cal})\in[m_j]/[c_{\rm ro}],
\qquad Y_\iota(\bm\theta_\star)\in[d_\iota]/[c_{\rm ro}].
\label{SM-eq:supp-noisy-projection}
\end{equation}
Outward-rounded division and cosine inversion give the bond intervals, and
the projected response intervals enter the unchanged onsite search; taking
the interval product of these correlated constraints only enlarges the
compatible set. With
$(\delta_{\rm ro},\delta_{\rm cal},\delta_Y)=(0.005,0.005,0.010)$, a union
bound gives coefficient coverage at least $0.98$, which carries over to
the frozen-operator residual and the repeated-cell label of each record.
This is neither joint coverage over the scan nor error conditioned on
conclusive records, and empty calibration intersections are inconclusive.

\subsubsection{Executed records}

Each record uses $1.92\times10^6$ boundary, $2\times10^5$
readout-reference, and $4\times10^5$ isolated-bond outcomes. Twenty
records of the topological example at $p_{\rm ro}=0.15$ give
$\tau_{0.9}J_{\rm ref}\in[1.043,1.081]$, with no coverage miss or search
stop and 20 strict labels. Single records of the three examples at
$p_{\rm ro}=0$, $0.05$, $0.15$, $0.30$, and $0.45$ all cover the true
coefficients. They give correct strict labels for the topological and
trivial cells through $p_{\rm ro}=0.30$, while the near-transition cell
stays inconclusive. At $p_{\rm ro}=0.45$ every search stops at the
$10^5$-box limit and returns inconclusive, a computational limit rather
than a noise threshold. The contrast is inferred even at zero true noise,
and full-qubit calculations reproduce the noisy likelihood.

\section{Persistence beyond the observation window and the cost of depth}
\label{SM-sec:supp-beyond}

This section documents the analyses of main-text
\secref{sec:beyond-window}, \secref{sec:a1-identity}, and
\secref{sec:truncated-inference}.

\subsection{Three-site benchmark near the boundary sweet spot}
\label{SM-sec:supp-sweet-spot}
The original device has single-particle energies
$E_k/J_{\rm ref}\simeq0.175,1.88,2.76$, and the autocorrelation of any
normalized linear Majorana operator is a convex combination of
$\cos(E_kt)$, all below $0.9$ just after
$t_*=\arccos(0.9)/E_1\simeq2.6/J_{\rm ref}$. Certificates far beyond the
window therefore need a different device.

We rerun the pipeline for $\bm\mu/J_{\rm ref}=(0.005,1.0,1.3)$ and
$\bm\kappa/J_{\rm ref}=(1.92,2.08)$ on the domain
$[-1,2]\times[0.5,3.5]^2$, with Eq.~\eqref{SM-eq:supp-schur-validation}
extended to $1\times1$ blocks for exact zero modes and a never-reached
limit of $4\times10^5$ boxes. Calibration uses $N_{\rm cal}=10N_\pm$ shots
per bond, and the terminal box width is the largest $3/2^k$ not exceeding
$(3/128)(2\times10^4/N_\pm)^{1/2}$. We certify $a_1$, whose equal-line
residual is $|\mu_1|$, and the two-site operator of
Sec.~\ref{SM-sec:supp-three-site}, here with $\alpha\in[-0.002,0.010]$. All
25 records of Table~\ref{SM-tab:supp-sweet-spot} cover the true coefficients
and give the topological repeated-cell label. Log-log fits of the median
times against the total outcomes give exponents $0.44$ for $a_1$ and
$0.49$ for the two-site operator, and the $a_1$ values are quantized
because $\sup|\mu_1|$ is attained at box endpoints.

\begin{table*}[t]
\caption{Sweet-spot benchmark, five records per budget. Certified times
are medians with ranges in brackets, in units of $J_{\rm ref}^{-1}$; the
observation window is $TJ_{\rm ref}=3$.}
\label{SM-tab:supp-sweet-spot}
\centering
\begin{ruledtabular}
\begin{tabular}{c c c c c c c}
$N_\pm$ & Total outcomes & Box width & Median $\mu_1$ hull width & $\tau_{0.9}(a_1)$ & $\tau_{0.9}$ (two-site) & Visited boxes\\
\hline
$2.00\times10^{4}$ & $2.32\times10^{6}$ & $3/128$ & 0.1172 & 5.7 [5.7, 8.2] & 7.5 [6.3, 9.5] & 3921--4267 \\
$6.32\times10^{4}$ & $7.34\times10^{6}$ & $3/256$ & 0.0703 & 10.4 [10.4, 14.3] & 12.7 [10.9, 15.3] & 4953--5653 \\
$2.00\times10^{5}$ & $2.32\times10^{7}$ & $3/512$ & 0.0293 & 22.9 [17.6, 22.9] & 30.0 [21.6, 30.0] & 5557--6213 \\
$6.32\times10^{5}$ & $7.34\times10^{7}$ & $3/1024$ & 0.0205 & 32.7 [26.9, 32.7] & 42.9 [37.3, 48.8] & 6257--8777 \\
$2.00\times10^{6}$ & $2.32\times10^{8}$ & $3/2048$ & 0.0117 & 41.6 [41.6, 41.6] & 69.9 [69.9, 81.0] & 7403--9103
\end{tabular}
\end{ruledtabular}
\end{table*}

\subsection{Site-1 identity benchmark}
\label{SM-sec:supp-tail-independent}
Records of $Y_{a_1b_1}$ at $tJ_{\rm ref}=0.25$, $0.5$, and $0.75$ ($D=3$)
use $N_\pm$ shots per sign, failure allocation $0.01$, and the $10^{-9}$
simulation bias. A bare $L=3$ chain and three 40-site chains share the
first three sites of Sec.~\ref{SM-sec:supp-sweet-spot}. The tails have
$\kappa_j=2J_{\rm ref}$ and $\mu_j=1.0J_{\rm ref}$ (topological) or
$3.0J_{\rm ref}$ (trivial), or $\mu_j$ and $\kappa_j$ uniform in
$[0.5,3.5]J_{\rm ref}$ and $[1.6,2.4]J_{\rm ref}$ (disordered). Validated
response enclosures check the bias allowance, and the inference uses none
of these coefficients. Table~\ref{SM-tab:supp-identity} summarizes five
records per budget and continuation, and all 100 records cover $\mu_1$.
Scanning the bare-device window over $TJ_{\rm ref}=0.25$--$1.1$, with
times $T/3$, $2T/3$, and $T$, the certificate grows up to
$TJ_{\rm ref}\approx0.5$--$0.75$ and then declines, because the relative
loss $\Omega_1^2t^2/6$ in Eq.~\eqref{SM-eq:supp-a1-lower} grows with time.
With the declared bounds, $\sqrt6/\Omega_1=0.78/J_{\rm ref}$, and times
beyond $8/(3\Omega_1)=0.85/J_{\rm ref}$ are discarded. The medians at
$T=0.75/J_{\rm ref}$ scale as $N_{\rm tot}^{0.40}$.

\begin{table}[t]
\caption{Site-1 identity certificate $\tau_{0.9}(a_1)J_{\rm ref}$: medians of
five records for the bare device and the topological, trivial, and
disordered 40-site continuations, and the pooled range. The window is
$TJ_{\rm ref}=0.75$.}
\label{SM-tab:supp-identity}
\centering
\begin{ruledtabular}
\begin{tabular}{c c c c c c c}
$N_\pm$ & Outcomes & Bare & Top. & Triv. & Dis. & Range\\
\hline
$2.0\times10^{4}$ & $1.2\times10^{5}$ & 8.3 & 8.8 & 7.3 & 8.3 & [6.3, 11.1] \\
$6.3\times10^{4}$ & $3.8\times10^{5}$ & 13.9 & 14.8 & 14.0 & 15.1 & [9.8, 24.5] \\
$2.0\times10^{5}$ & $1.2\times10^{6}$ & 22.1 & 24.4 & 23.5 & 21.8 & [17.4, 32.5] \\
$6.3\times10^{5}$ & $3.8\times10^{6}$ & 31.2 & 36.2 & 31.1 & 34.8 & [28.9, 41.3] \\
$2.0\times10^{6}$ & $1.2\times10^{7}$ & 43.4 & 45.7 & 45.8 & 46.3 & [40.1, 52.9]
\end{tabular}
\end{ruledtabular}
\end{table}

{\color{black}
\subsection{Robustness of the long-time site-1 certificate}
\label{SM-sec:supp-robustness}
\label{SM-sec:supp-robustness-benchmark}
The new synthetic study fixes the observable $a_1$ and the sample times
$t_\ell J_{\rm ref}=0.25,0.5,0.75$. Its six-site data-generating chain
has onsite values
$\bm\mu/J_{\rm ref}=(0.005,1,1.3,0.7,1.5,0.9)$, sum couplings
$s_j=J_j+\Delta_j=2J_{\rm ref}$ and a maximally mixed hidden reset.
The other bonds stay on the equal line while the first-bond mismatch
$d_1=\Delta_1-J_1$ is varied. The learner receives only the counts and
the declared bounds $|\mu_1|\le2J_{\rm ref}$,
$|s_1|\le2.4J_{\rm ref}$ and $|d_1|\le d_{\max}$.
The generating continuation is excluded from the inference.

Each of the three coordinates and two preparation signs receives
$N=N_{\rm tot}/6$ shots. With per-record failure allocation $0.01$,
the half-width is $\sqrt{\log(6/0.01)/N}$, enlarged by a prescribed
contrast-bias bound $b_{\max}$ and numerical allowance $10^{-12}$.
For the error scenarios the generating contrast is shifted by
$-b_{\max}$, with clipping at physical binary endpoints where needed.
The learner allows every bias in $[-b_{\max},b_{\max}]$ independently
at each coordinate.

The fixed-Hamiltonian records use $Y_{a_1b_1}$.
The quadratic mismatch remainder of Sec.~\ref{SM-sec:supp-a1-identity},
bounded by $d_{\max}\Omega_1t^2/2$, enlarges the response interval
before the onsite interval is inferred. The current upper bound
$U_\mu$ for $|\mu_1|$ updates $\Omega_1$ in the integral bound.
The final residual and time are enclosed as
\begin{equation}
\varepsilon_{a_1}\le\sqrt{U_\mu^2+d_{\max}^2},\qquad
\tau_{0.9}\ge\frac{\sqrt{0.2}}{\sqrt{U_\mu^2+d_{\max}^2}}.
\label{SM-eq:supp-robustness-fixed-time}
\end{equation}
Thus mismatch enters both inference and persistence. General boundary
perturbations instead use Eq.~\eqref{SM-eq:supp-boundary-perturbation}.

Figure~\ref{SM-fig:supp-robustness}(a) uses four budgets from
$1.2\times10^5$ to $1.2\times10^8$ outcomes, with 25 independent
records per condition. The nonzero widths are
$d_{\max}/J_{\rm ref}=0.005$ and $b_{\max}=0.001$.
At $1.2\times10^7$ outcomes the median fixed-Hamiltonian times,
in units of $J_{\rm ref}^{-1}$, are $45.46$, $31.02$, $28.59$ and
$21.30$ for the ideal case, mismatch, bias and both, respectively.
At $1.2\times10^8$ outcomes they are $63.69$, $42.29$, $37.17$ and
$25.86$. Fixed uncertainty allowances reduce the improvement obtained
from increasing the number of shots.

For Fig.~\ref{SM-fig:supp-robustness}(b), the onsite value is redrawn
independently as $\mu_1/J_{\rm ref}=+0.05$ or $-0.05$ with equal
probability and held constant during each evolution. The remaining
coefficients and reset are fixed. On the equal line the averaged cross
response vanishes, while the diagonal response retains information
about $\mathbb E\mu_1^2$. The study therefore uses $\pm x$
preparation and $X_1$ readout, with the same budgets, error widths and
confidence allocation. Equations~\eqref{SM-eq:supp-mixture-mismatch-inference}
and \eqref{SM-eq:supp-mixture-mismatch-time} handle the combined mismatch
and fluctuation case. At $1.2\times10^7$ outcomes the median averaged
times are $3.581$, $3.395$, $2.642$ and $2.537$ in the same order of
conditions. These numbers concern the averaged autocorrelation under
the same quasistatic law.

The mismatch and bias envelopes are supplied assumptions in these
simulations. If obtained by experimental calibration, their failure
allocations must enter the joint coverage budget. All 1250 generated
records, including the fixed-budget width sweeps, enclose the generating
coefficient or second moment. This is an implementation diagnostic.
The statistical coverage follows from the simultaneous response
intervals and the derived bounds, conditional on the model assumptions.

The executable analysis replays the archived integer counts without
using the generating coefficients. Scalar operations use 50-digit
interval arithmetic with outward export. Independent order-65 interval
Taylor sums and an analytic exponential remainder bound the response
error below $2.34\times10^{-16}$ for all 36 generating parameter-and-time
settings, within the $10^{-12}$ allowance. Eighty full-qubit comparisons
check the Majorana convention, and 40 cases check the mismatch integral
bound. An additional independent full-Hilbert-space calculation checks
the combined moment and persistence inequalities in 720 cases with
interacting hidden Hamiltonians. The proof is given above rather than
inferred from these checks.

\begin{figure*}[t]
\centering
\includegraphics[width=\textwidth]{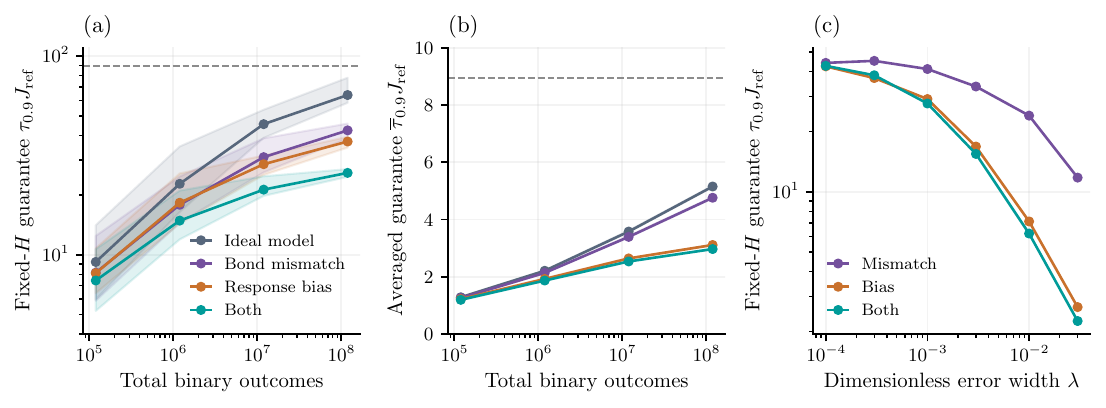}
\caption{\color{black}Synthetic robustness of site-1 persistence inference.
(a) Fixed-Hamiltonian guarantees from the cross response.
(b) Averaged guarantees from the diagonal response for independent
quasistatic onsite values $\mu_1/J_{\rm ref}=\pm0.05$.
The mismatch and bias scenarios use
$d_{\max}/J_{\rm ref}=0.005$ and $b_{\max}=0.001$.
Points are medians of 25 records and shaded bands show their full range.
Dashed lines give the commutator bounds using the generating
$|\mu_1|/J_{\rm ref}=0.005$ in (a) and second moment $0.05^2$ in (b),
with zero mismatch.
(c) Sensitivity at $1.2\times10^7$ outcomes to width $\lambda$.
Mismatch alone uses $d_{\max}/J_{\rm ref}=\lambda$, bias alone uses
$b_{\max}=\lambda$, and both uses both equalities.
All records have window $TJ_{\rm ref}=0.75$ and simultaneous per-record
failure allocation $0.01$.}
\label{SM-fig:supp-robustness}
\end{figure*}
}


{\color{black}
\subsection{Comparison with isolated onsite Ramsey calibration}
\label{SM-sec:supp-ramsey-comparison}
The additional controls used for isolated-bond calibration also motivate
a direct onsite comparator~\cite{Ramsey1950}. When the neighboring coupling can be removed
without changing the onsite coefficient, opposite $Y_1$ preparations and
$X_1$ readout give the isolated contrast
$y_{\rm R}(t)=-\sin(\mu_{\rm iso}t)$. On the declared domain
$|\mu_{\rm iso}|\le2J_{\rm ref}$ and at
$tJ_{\rm ref}=0.25,0.5,0.75$, this response is monotone in the coefficient.
Simultaneous binary-shot intervals can therefore be inverted directly.
An independently justified transfer allowance
$|\mu_1-\mu_{\rm iso}|\le r$ enlarges the inferred isolated interval by
$[-r,r]$ before the persistence bound is evaluated for the coupled device.

The matched simulations use the same observable $a_1$, true coefficient
$\mu_1=0.005J_{\rm ref}$, domain, time grid, binary-outcome allocation,
and simultaneous failure probability $0.01$ for both protocols. The
declared first-bond sum bound is $2.4J_{\rm ref}$. Each budget is divided
equally between three times and two preparation signs. The coupled data
are generated from a three-site device with onsite coefficients
$(0.005,1,1.3)J_{\rm ref}$ and bond sums $(2,2)J_{\rm ref}$, while the
site-1 inversion receives only the boundary data and the declared bounds.
The same mismatch allowance enters both final residuals as
$\sqrt{U_\mu^2+d_{\max}^2}$, where $U_\mu$ bounds $|\mu_1|$.

For an ideal equal-line target, the median guaranteed times from 25 records
at $1.2\times10^7$ additional binary outcomes are
$43.42/J_{\rm ref}$ for the coupled site-1 protocol and
$60.66/J_{\rm ref}$ for isolated Ramsey with $r=0$.
Allowing $r=0.005J_{\rm ref}$ reduces the Ramsey median to
$36.15/J_{\rm ref}$. With a first-bond mismatch bounded by
$0.005J_{\rm ref}$ and response bias bounded by $0.001$, the corresponding
medians are $21.66/J_{\rm ref}$, $40.64/J_{\rm ref}$, and
$28.63/J_{\rm ref}$. The actual response distortion is $-0.001$ in this
stress test, and the actual transfer shift is zero.
The nonzero values of $r$ describe uncertainty about transfer rather than
a measured hardware effect.

Both methods hold the available calibration bounds fixed, and the plotted
budgets count additional learning outcomes. Establishing isolation and
coefficient transfer in a laboratory can require further calibration
measurements, whose cost is not evaluated here. Ramsey gives tighter
statistical bounds when isolation and transfer are sufficiently accurate.
On an equal-line first bond, the site-1 identity operates with the chain
connected and uses no model or measurement of the continuation. The
mismatch comparison uses the quadratic-chain correction to the
cross-response identity. Figure~\ref{SM-fig:supp-ramsey-comparison} displays both comparisons.
Across the 200 simulated
records, the true coefficient lies in all coupled intervals and 199
Ramsey intervals at $r=0$; the one statistical miss remains in the saved
data and plot. This finite-record audit is separate from the analytical
coverage guarantee.
}

{\color{black}
\begin{figure*}[t]
\centering
\includegraphics[width=\textwidth]{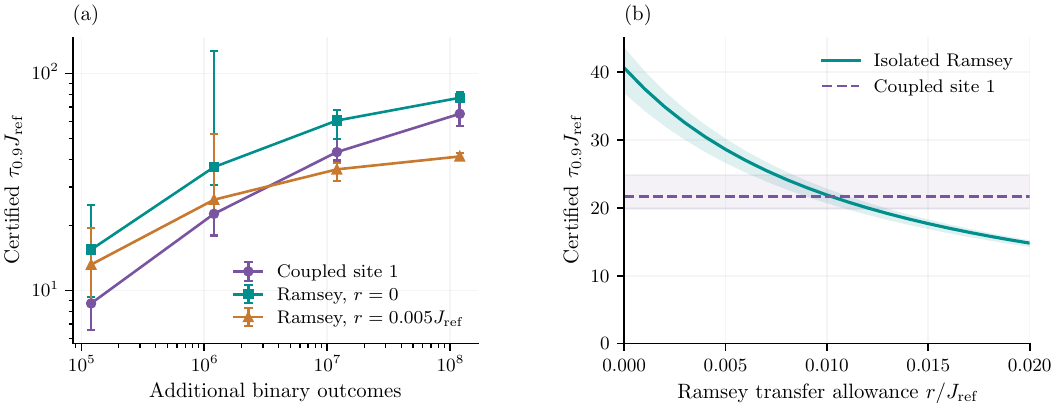}
\caption{\color{black}Matched synthetic comparison for the same end Majorana $a_1$.
(a) Median certified times versus additional binary outcomes for the
coupled site-1 identity and isolated onsite Ramsey calibration, with zero
and nonzero coefficient-transfer allowances. Bars span 25 independent
records at each budget. The target is on the equal line and has no added
systematic response distortion. (b) At $1.2\times10^7$ outcomes, the
Ramsey time varies with its declared transfer allowance $r$. The
first-bond mismatch is bounded by $0.005J_{\rm ref}$ and the additive
response bias by $0.001$ for both methods. The dashed line is the coupled
site-1 median, and shaded bands span the records. All inversions use
$tJ_{\rm ref}=0.25,0.5,0.75$ and simultaneous failure probability $0.01$.
The Ramsey comparator assumes coupling-isolation controls. Its transfer
allowances are declared sensitivity parameters rather than measured IBM
calibrations. Common pre-existing calibration information is held fixed.
The displayed ranges retain one statistical coverage miss among the 200
ideal and perturbed Ramsey records.}
\label{SM-fig:supp-ramsey-comparison}
\end{figure*}
}


\subsection{Truncation bound and three-site inference with unknown tails}
\label{SM-sec:supp-truncation}
Table~\ref{SM-tab:supp-truncation} evaluates main-text
Eq.~\eqref{eq:truncation_bound} with the grid majorants of
Sec.~\ref{SM-sec:supp-nogo-proof} on 2001 points for
$\mu_{\max}=3.5J_{\rm ref}$ and $\kappa_{\max}=2.4J_{\rm ref}$, so
$G=5.9J_{\rm ref}$.

\begin{table}[t]
\caption{Truncation bound of main-text Eq.~\eqref{eq:truncation_bound}
and the smallest number of modeled sites $n$ that keeps it below a
tolerance, for $G=5.9J_{\rm ref}$.}
\label{SM-tab:supp-truncation}
\centering
\begin{ruledtabular}
\begin{tabular}{c c c c c}
$TJ_{\rm ref}$ & $GT$ & $E_3(T)$ & $n$ ($2\times10^{-3}$) & $n$ ($10^{-4}$)\\
\hline
0.25 & 1.48 & $4.7\times10^{-8}$ & 2 & 3 \\
0.50 & 2.95 & $4.8\times10^{-5}$ & 3 & 3 \\
0.75 & 4.43 & $2.5\times10^{-3}$ & 4 & 4 \\
1.00 & 5.90 & $3.5\times10^{-2}$ & 4 & 5 \\
2.00 & 11.80 & $1.9$ & 6 & 6 \\
3.00 & 17.70 & $4.3$ & 8 & 8
\end{tabular}
\end{ruledtabular}
\end{table}

For inference with unknown tails we sample at $tJ_{\rm ref}=k/24$,
$k=1,\ldots,12$ ($D=48$), and add $E_3(0.5)\le4.71\times10^{-5}$,
evaluated on 4001 points, to every radius. All other settings follow
Sec.~\ref{SM-sec:supp-sweet-spot} at $N_\pm=2\times10^4$, for
$2.32\times10^6$ outcomes per record, with data from the 40-site devices
of Sec.~\ref{SM-sec:supp-tail-independent}. For the topological, trivial, and
disordered tails, the $\mu_1$ hulls are $[-0.0625,0.0547]$,
$[-0.0625,0.0547]$, and $[-0.0625,0.0781]J_{\rm ref}$, certifying
$\tau_{0.9}(a_1)J_{\rm ref}\ge7.155$, $7.155$, and $5.724$ and, for the
two-site operator, $7.608$, $7.608$, and $6.304$. Each search visits about
$1.3\times10^5$ boxes, retains $4.6$--$5.6\times10^4$, reaches no limit,
and covers the coefficients of the first three sites. The short window
leaves $\mu_3$ unconstrained and $\mu_2$ weakly constrained, which these
certificates do not need.

\subsection{Depth cost of certification}
\label{SM-sec:supp-depth-cost}
We estimate the shots needed to certify a depth-$m$ recurrence candidate
in a local Gaussian approximation, for an equal-line chain of $L=m+1$
sites with $\mu_j=2rJe^{0.05u_j}$, $\kappa_j=2Je^{0.04u'_j}$, and
$u_j,u'_j$ uniform on $[-1,1]$, sampled in all four entries at $tJ=0.2k$
up to $TJ=2L$. Bounding each binary variance by one reproduces the
Hoeffding scaling, and $2N_\pm$ shots per coordinate give Fisher
information $N_\pm F_1$ with $F_1=2\mathcal J^T\mathcal J$ for the
response Jacobian $\mathcal J$. The candidate frozen at the nominal
coefficients has affine residual $K_m\bm g=\bm R_0+A\,\delta\bm\theta$,
with $\partial(K_m\bm g)_{b_j}/\partial\mu_j=g_{a_j}$ and
$\partial(K_m\bm g)_{b_j}/\partial\kappa_j=g_{a_{j+1}}$ for $j<m$. On the
asymptotic region $\delta\bm\theta^TF\delta\bm\theta\le q^2$, with $q^2$
the $0.99$ quantile of $\chi^2_p$, this residual is at most
$\|\bm R_0\|_2+q\,s_m/\sqrt{N_\pm}$ with $s_m=\|AF_1^{-1/2}\|_2$.
Boundary-only access learns all $p=2L-1$ coefficients, and calibrated
access only the $L$ onsite ones. A target $\tau_{0.9}$ thus needs
$N_{\rm tot}=2Dq^2s_m^2/(\sqrt{0.2}/\tau_{0.9}-\varepsilon_m)^2$ outcomes
for $D$ coordinates, with $s_m$ in Table~\ref{SM-tab:supp-depth-cost}.

Per added site, the smallest singular value of $\mathcal J$ drops by
factors of $7.5$, $2.6$, and $2.0$ for $r=0.3$, $0.5$, and $0.7$ when all
coefficients are learned, and by $3.0$, $1.3$, and $1.1$ with fixed bonds.
A weakly disordered profile with $\Delta_j\simeq0.6J_j$ gives about $5.1$
and $1.4$, which motivates the bond calibration of the benchmarks. These
local asymptotic estimates omit the box-enclosure inefficiency,
roughly an order of magnitude at depth 2.

\begin{table}[t]
\caption{Exact residual $\varepsilon_m$, exact-coefficient horizon, and
residual sensitivity $s_m$ with calibrated bonds (cal.) and boundary-only
access (bd.), for nominal ratios $r=0.5$ and $0.3$.}
\label{SM-tab:supp-depth-cost}
\centering
\begin{ruledtabular}
\begin{tabular}{c c c c c c c c c}
& \multicolumn{4}{c}{$r=0.5$} & \multicolumn{4}{c}{$r=0.3$}\\
$m$ & $\varepsilon_m/J$ & $\tau_{0.9}J$ & cal. & bd. & $\varepsilon_m/J$ & $\tau_{0.9}J$ & cal. & bd.\\
\hline
1 & 1.04 & 0.431 & 0.200 & 0.209 & 0.623 & 0.718 & 0.235 & 0.244 \\
2 & 0.445 & 1.00 & 0.164 & 0.244 & 0.173 & 2.59 & 0.195 & 0.272 \\
3 & 0.208 & 2.15 & 0.159 & 0.245 & 0.0492 & 9.08 & 0.191 & 0.286 \\
4 & 0.105 & 4.25 & 0.149 & 0.229 & 0.0151 & 29.5 & 0.167 & 0.310 \\
5 & 0.0533 & 8.39 & 0.143 & 0.238 & 0.00459 & 97.5 & 0.166 & 0.531 \\
6 & 0.0257 & 17.4 & 0.134 & 0.243 & 0.00133 & 336 & 0.160 & 1.24 \\
7 & 0.0129 & 34.7 & 0.132 & 0.446 & 0.000399 & 1120 & 0.158 & 4.41 \\
8 & 0.00607 & 73.7 & 0.130 & 0.932 & 0.000113 & 3970 & 0.158 & 11.6
\end{tabular}
\end{ruledtabular}
\end{table}

\subsection{Interacting continuations}
\label{SM-sec:supp-interacting}
We test the identity certificate on ten-qubit chains
$H=\mu_1Z_1/2-(\kappa_1/2)X_1X_2+H_{\rm tail}$ with
$\mu_1=0.005J_{\rm ref}$, $\kappa_1=1.92J_{\rm ref}$, $\mu_2=1.0$,
$\mu_3=1.3$, and $\kappa_2=2.08$, as in the sweet-spot device. The
continuations on $j\ge2$ are (i) $\mu_j=1.0$, $\kappa_j=2.0$, and
interactions $U_jZ_jZ_{j+1}$ with $U_j=0.5$; (ii) the same with
$\mu_j=3.0$; (iii) continuation (i) with parity-breaking fields $0.3X_j$,
admissible only for qubits, and parity-even terms $0.3Y_jY_{j+1}$
($Y_jY_{j+1}=ia_jb_{j+1}$) that leave the equal line; and (iv) random
$\mu_j\in[0.5,3.5]$, $\kappa_j\in[1.6,2.4]$, $U_j\in[-0.5,0.5]$, and
$h_j\in[-0.3,0.3]$, in units of $J_{\rm ref}$. The labels of (i) and (ii)
refer to their quadratic parts. With a maximally mixed hidden register,
exact diagonalization gives the signed responses
$G(t)=\Tr[Y_1X_1(t)]/2^{10}$ and $F(t)=\Tr[Y_1Y_1(t)]/2^{10}$. They obey
$G=-\mu_1\int_0^tF$ to $4\times10^{-7}$ (trapezoidal rule, grid
$0.025/J_{\rm ref}$) and $F(t)\ge1-\Omega_1^2t^2/2$ on
$[0,2/J_{\rm ref}]$, and twelve qubits change them by less than
$10^{-16}$. These floating-point correlators are diagnostics, not
validated enclosures.

\begin{figure}[t]
\centering
\includegraphics[width=0.75\columnwidth]{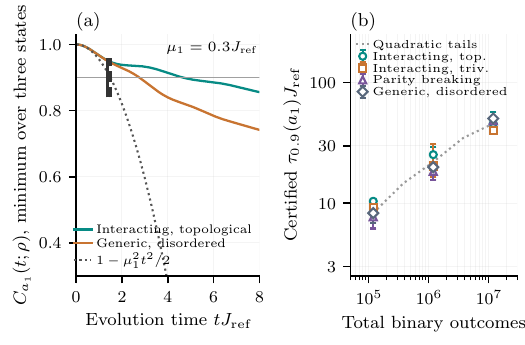}
\caption{Site-1 identity certificate with interacting continuations. (a)
Autocorrelation of $a_1$, minimized over the infinite-temperature, ground,
and a random product state, for ten-qubit chains with
$\mu_1=0.3J_{\rm ref}$ and an interacting topological or a generic
disordered continuation. The dotted curve is the envelope
$1-\mu_1^2t^2/2$, and ticks mark the learned (solid) and exact-coefficient
(dashed) certificates. (b) Certified $\tau_{0.9}(a_1)$ at
$\mu_1=0.005J_{\rm ref}$ for four interacting continuations (medians and
ranges over three records, failure allocation $0.01$ each) and the medians
for the quadratic tails of main-text
Fig.~\ref{fig:persistence_window}(c). Topological and trivial refer to
the quadratic parts.}
\label{SM-fig:supp-interacting-main}
\end{figure}

Records at $tJ_{\rm ref}=0.25,0.5,0.75$ give median certificates from
$40.2/J_{\rm ref}$ to $50.4/J_{\rm ref}$ at $1.2\times10^7$ outcomes
[Fig.~\ref{SM-fig:supp-interacting-main}(b)], and all 36 records cover
$\mu_1$. The $a_1$ autocorrelation, minimized over the
infinite-temperature, ground, and a random product state, stays above
$0.9885$ for $tJ_{\rm ref}\le60$ in all four chains and above
$1-\mu_1^2t^2/2$ throughout. At $\mu_1=0.3J_{\rm ref}$, three records with
$N_\pm=2\times10^5$ give $\sup|\mu_1|\in[0.325,0.337]$ and
$\tau_{0.9}J_{\rm ref}\in[1.33,1.37]$ for (i) and (iv), against
$\sqrt{0.2}/0.3=1.49$ at the exact coefficients, while the minimal
autocorrelation first crosses $0.9$ at $4.7/J_{\rm ref}$ and
$2.8/J_{\rm ref}$ [Fig.~\ref{SM-fig:supp-interacting-main}(a)]. The
admissible device with $\kappa_1=0$, for which $C_{a_1}(t)=\cos(\mu_1t)$,
crosses $0.9$ at $\arccos(0.9)/\mu_1\simeq1.50/J_{\rm ref}$, just above
the exact-coefficient guarantee. A guarantee uniform over continuations
can thus be much shorter than the decay of a particular device.

\subsection{Affine contractor and inference at depth four and five}
\label{SM-sec:supp-contractor}
On a box $B$ with center $\bm c$ and half-widths $\bm r$, the second-order
Dyson expansion gives
$f_\iota(\bm\theta)=f_\iota(\bm c)+\bm J_\iota(\bm c)(\bm\theta-\bm c)+R_\iota$
with
$|R_\iota|\le\int_0^{t_\ell}\!\!\int_0^s\|Pe^{K(\bm c)(t_\ell-s)}\delta K\|\,\|\delta Ke^{K(\bm\theta)u}P^T\|\,du\,ds$
and $\delta K=K(\bm\theta)-K(\bm c)$. Each factor is at most
$\rho=\max_jr_{\mu_j}+\max_jr_{\kappa_j}$, giving the isotropic remainder
$\frac12t_\ell^2\rho^2$. Each factor is also at most
$\sum_kr_k\overline{\mathfrak B}_{d_k}(G\tau)$, because each unit
coefficient perturbation $E_k$ obeys
$\|Pe^{K\tau}E_k\|_2\le\overline{\mathfrak B}_{d_k}(G\tau)$, with $d_k$
the graph distance from $\{a_1,b_1\}$ to coefficient $k$ ($2j-3$ for
$\mu_j$ with $j\ge2$, $2j-2$ for $\kappa_j$) and $G\ge\|K\|_2$ on the
domain. The light-cone remainder integrates the smaller bound on a
$0.0125/J_{\rm ref}$ grid with the majorants of
Sec.~\ref{SM-sec:supp-nogo-proof}, and it is small when the wide coefficients
lie outside the light cone of the sampling time. Forward differences of
the validated propagator with step $2^{-18}$ enclose the Jacobian with
remainder $\frac12\cdot2^{-18}t_\ell^2$. Every compatible $\bm\theta\in B$
then obeys
$|f_\iota(\bm c)+\bm J_\iota(\bm\theta-\bm c)-\widehat y_\iota|\le\xi+S_\iota$,
with $S_\iota$ the rounded-up sum of the remainder, the enclosure widths,
and the Jacobian error times the half-widths.

Linear programming gives the coordinate hull of this polytope in $B$, with
upper bounds from the numerical dual evaluated in rational
arithmetic~\cite{NeumaierShcherbina2004}, so the floating-point solution
is never trusted; the dual of $\min s$ subject to $A\bm x-s\le\bm b$, with
$A\bm x\le\bm b$ the affine tests, certifies emptiness in the same way.
Boxes are contracted until the widest side shrinks by less than $2\%$ and
then bisected along the widest onsite or learned-bond side down to a
predeclared width of $0.01$--$0.025J_{\rm ref}$. No box is dropped without
an emptiness certificate, and a box whose linear program fails is kept
uncontracted.

The depth-four and depth-five records use the profile of
Sec.~\ref{SM-sec:supp-depth-cost} at $r=0.5$ with $L=5$ and $6$, onsite
domain $[0.5,1.5]J_{\rm ref}$, bonds calibrated as in
Sec.~\ref{SM-sec:supp-three-site} with $N_{\rm cal}=10N_\pm$, and all four
responses at $tJ_{\rm ref}=0.25k$ up to $TJ_{\rm ref}=2L$ with
$N_\pm=2\times10^5$. A four-site record instead learns the bonds within
$[1.8,2.2]J_{\rm ref}$ from the boundary data. Each recurrence candidate
is frozen at the hull center and its squared residual maximized over every
retained box in rational arithmetic, and every retained union covers the
true coefficients. The certificates of Table~\ref{SM-tab:supp-deep} lie
within their acquisition windows, at a substantial measurement cost.

\begin{table}[t]
\caption{Top-depth certificates from the affine contractor: certified and
exact-coefficient residuals and $\tau_{0.9}$ for the recurrence candidate of
depth $m$, and the first crossing of $0.9$ by the simulated
autocorrelation. Certified residuals are rounded upward and certified times
downward. Outcomes include bond calibration where used, and $TJ_{\rm ref}$
is the observation window.}
\label{SM-tab:supp-deep}
\centering
\begin{ruledtabular}
\begin{tabular}{l c c c c c c c c}
Record & \makecell{Outcomes\\($10^6$)} & $TJ_{\rm ref}$ & $m$ & \makecell{$\varepsilon^{\rm cert}$\\($J_{\rm ref}$)} & \makecell{$\varepsilon^{\rm exact}$\\($J_{\rm ref}$)} & \makecell{$\tau_{0.9}^{\rm cert}$\\($J_{\rm ref}^{-1}$)} & \makecell{$\tau_{0.9}^{\rm exact}$\\($J_{\rm ref}^{-1}$)} & \makecell{Crossing\\($J_{\rm ref}^{-1}$)}\\
\hline
$L=5$, calibrated bonds & 80.0 & 10 & 4 & 0.1080 & 0.1051 & 4.14 & 4.26 & 9.60\\
$L=6$, calibrated bonds & 96.8 & 12 & 5 & 0.0577 & 0.0531 & 7.76 & 8.41 & 20.0\\
$L=4$, bonds learned & 51.2 & 8 & 3 & 0.2138 & 0.2080 & 2.09 & 2.15 & 4.95
\end{tabular}
\end{ruledtabular}
\end{table}

\section{IBM processor records}
\label{SM-sec:supp-gate}

\subsection{Three-site joint inference on IBM}
\label{SM-sec:supp-ibm-three-site}

The three-site record of main-text Fig.~\ref{fig:ibm-learning} was
acquired on \texttt{ibm\_marrakesh}, on physical qubits
$(95,94,93)$ in logical chain order. Both studies program
\begin{equation}
H_0=\frac{J_{\rm ref}}2\left[
\sum_{j=1}^{L}x_jZ_j-\sum_{j=1}^{L-1}k_jX_jX_{j+1}\right],
\qquad x_j=\frac{\mu_j}{J_{\rm ref}},\quad
k_j=\frac{\kappa_j}{J_{\rm ref}},
\label{SM-eq:supp-marrakesh-model}
\end{equation}
with $\kappa_j=2J_j=2\Delta_j$, onsite domain $[0.5,1.5]^L$, and bond
domain $[1.6,2.4]^{L-1}$. Here $L=3$, and all five operating coefficients
are withheld from the learner.

\paragraph*{Transfer hypotheses.}
The inference assumes a fixed coherent target Hamiltonian, independent
shots, a common hidden reset for both preparation signs, known onsite
programming, fixed native templates with audited affine-angle exposures,
stationary coefficient-independent CPTP noise, and coefficients shared by
target and reference circuits. Bond calibration also assumes local
preparation and readout and one coefficient-dependent native rotation.
Controls and coverage do not test these hypotheses, and the guarantees
concern the programmed continuous coherent Hamiltonian, not the noisy
circuits that implement it.

\paragraph*{Acquisition.}
The target uses opposite $X_1$ or $Y_1$ preparations, $X_1$ or $Y_1$
readout, and the intended hidden reset $|00\rangle$. Each symmetric step
applies $R_Z(\mu_j\delta t/2)$ on every site, $R_{XX}(-\kappa_j\delta t)$
on every bond, and $R_Z(\mu_j\delta t/2)$ again, with
$R_Z(\varphi)=e^{-i\varphi Z/2}$,
$R_{XX}(\varphi)=e^{-i\varphi X\otimes X/2}$, and
$\delta tJ_{\rm ref}=0.25$, so $n=6,12,16$ steps reach
$u_\ell=t_\ell J_{\rm ref}=1.5,3,4$. Each of the $D=12$ time-and-channel
coordinates has 4096 shots per sign. Known onsite controls sample
$\{0.5,0.75,1,1.25,1.5\}^3$ with 256 shots per sign and coordinate at the
unchanged unknown bonds, which isolated-pair measurements calibrate.

\paragraph*{Derivative control.}
For the fixed unknown bonds $\bm k_\star$, let $h_\iota(\bm x)$ be the
mean signed contrast at coordinate $\iota$, sampled directly by the known
onsite settings. The coefficient-dependent native gates are Pauli
rotations with affine angles. If the audited sum of absolute angle
derivatives with respect to $x_j$ is at most $u_\ell^+$, then
\begin{equation}
|\partial_{x_j}^3h_\iota(\bm x)|\le(u_\ell^+)^3.
\label{SM-eq:supp-marrakesh-derivative}
\end{equation}
The derivative of a rotation channel costs at most
$|\partial_{x_j}\varphi|$ in induced trace norm and CPTP maps contract
this norm, so the product rule bounds all third-derivative terms by the
cubed total exposure, independently of the bonds.

\paragraph*{Interpolation and coverage.}
Each predeclared cell $[v_q,v_{q+1}]$, $q=0,1,2,3$, of the node grid
$v_0,\ldots,v_4$ uses the nodes $v_s$, $v_{s+1}$, and $v_{s+2}$, with
$s=\min\{2,\max\{0,q-1\}\}$. With the quadratic Lagrange basis
$L_{j,r}(x_j)$ of these nodes in coordinate $j$, the tensor weights
$\lambda_{\bm r}(\bm x)=\prod_{j=1}^3L_{j,r_j}(x_j)$, $r_j\in\{0,1,2\}$,
applied to the 27 stencil contrasts give
$\widehat q_\iota(\bm x)=\sum_{\bm r}\lambda_{\bm r}(\bm x)\widehat d_{\iota,\bm r}$.
With $\pi_j=\prod_r(x_j-v_{j,r})$ and $\mathcal L_j=\sum_r|L_{j,r}(x_j)|$,
the univariate Lagrange remainder~\cite{DLMF} and telescoping over the
three coordinates give
\begin{equation}
|h_\iota(\bm x)-q_\iota(\bm x)|\le R_\iota(\bm x),\qquad
R_\iota=\frac{(u_\ell^+)^3}{6}
\min_{(i,j,k)}\bigl[|\pi_i|+\mathcal L_i|\pi_j|
+\mathcal L_i\mathcal L_j|\pi_k|\bigr],
\label{SM-eq:supp-marrakesh-remainder}
\end{equation}
where the minimum runs over the six coordinate orders and $q_\iota$
interpolates the exact control means. At a fixed true vector the weights
are deterministic, and the target-minus-reference half difference has
squared shot range
$2(1/N_{\rm t}+\sum_{\bm r}\lambda_{\bm r}^2/N_{\iota,\bm r})$ for target
and reference budgets $N_{\rm t}$ and $N_{\iota,\bm r}$, so Hoeffding's
inequality~\cite{Hoeffding1963} gives the radius
\begin{equation}
r_\iota(\bm x)=\sqrt{\log\!\left(\frac{2D}{\delta_{\rm bd}}\right)
\left[\frac1{N_{\rm t}}+
\sum_{\bm r}\frac{\lambda_{\bm r}(\bm x)^2}{N_{\iota,\bm r}}\right]},
\qquad \delta_{\rm bd}=0.01.
\label{SM-eq:supp-marrakesh-radius}
\end{equation}
A union bound over the $D$ coordinates encloses the true onsite vector by
the tests
\begin{equation}
|\widehat d_{\iota,\rm t}-\widehat q_\iota(\bm x)|
\le r_\iota(\bm x)+R_\iota(\bm x)+2\times10^{-10}
\quad\text{for every }\iota.
\label{SM-eq:supp-marrakesh-compatible}
\end{equation}
The last term covers numerical evaluation and angle rounding. Coverage is
pointwise in the fixed true parameter and uniform over the domain, so no
simultaneous event over hypothetical vectors is needed.

\paragraph*{Bond calibration and retained set.}
An isolated pair starts in $|00\rangle$, evolves under
$R_{XX}(-\kappa_jt_{\rm cal})$ with $t_{\rm cal}J_{\rm ref}=1$, and is
read out in $Z$. With exactly one coefficient-dependent rotation in the
audited template, the noisy mean is exactly
\begin{equation}
g_j(k)=\mathsf a_j+\mathsf b_j\cos k+\mathsf c_j\sin k.
\label{SM-eq:supp-marrakesh-calibration}
\end{equation}
Controls at $k=1.6,2,2.4$ enclose the three coefficients by interval
inversion, and their correlated uncertainty propagates to midpoint checks
at $k=1.8,2.2$ and to the operating bond. Each setting has 4096 shots, and
the ten control means and the two operating-bond means each share failure
allocation $0.005$. All midpoint checks pass, giving
$k_1\in[1.75,2.0421875]$ and $k_2\in[1.6,2.4]$. Outward branch and bound
over Eq.~\eqref{SM-eq:supp-marrakesh-compatible}, with Bernstein bounds for
the tensor polynomial and interval bounds for the radius and remainder,
subdivides the 64 grid cells to onsite width $0.05$. It evaluates 23318
boxes and retains 9709, none unresolved, with hull
\begin{equation}
x_1\in[0.5,1.09375],\quad x_2\in[0.5,1.40625],\quad
x_3\in[0.5,1.5],
\label{SM-eq:supp-marrakesh-hull}
\end{equation}
together with the calibrated bonds, so $x_3$ and $k_2$ remain unresolved.
The boundary and calibration allocations give conditional coefficient
coverage at least $0.98$.

\paragraph*{The same frozen operator.}
The retained hull selects
\begin{equation}
\Gamma_2=\frac{a_1-\alpha a_2}{\sqrt{1+\alpha^2}},\qquad
\alpha=0.34427471685503347,
\label{SM-eq:supp-marrakesh-candidate}
\end{equation}
from the smallest eigenvector of the nominal two-site residual form, with
the ratio restricted to $[0,0.5]$. It stays fixed in every comparison and
has $W_1\ge0.8940$, and its exact coherent equal-line residual is
\begin{equation}
\frac{\|[H_0,\Gamma_2]\|}{J_{\rm ref}}
=\sqrt{\frac{(x_1-\alpha k_1)^2+\alpha^2x_2^2}{1+\alpha^2}}.
\label{SM-eq:supp-marrakesh-residual}
\end{equation}
The outgoing equal-line bond does not contribute for this $a$-chirality
candidate, although $x_3$ and $k_2$ remain in the circuit and its
inference model. Maximizing the residual over the initial, calibrated, and
learned sets gives $\tau_{0.9}^{\rm coh}J_{\rm ref}\ge0.43771$, $0.45676$,
and $0.77543$ from main-text Eq.~\eqref{eq:certified_persistence_time}.

\subsection{Six-site inference by one-coordinate reference scans}
\label{SM-sec:supp-ibm-six-site}

\begin{figure*}[t]
\centering
\includegraphics[width=\textwidth]{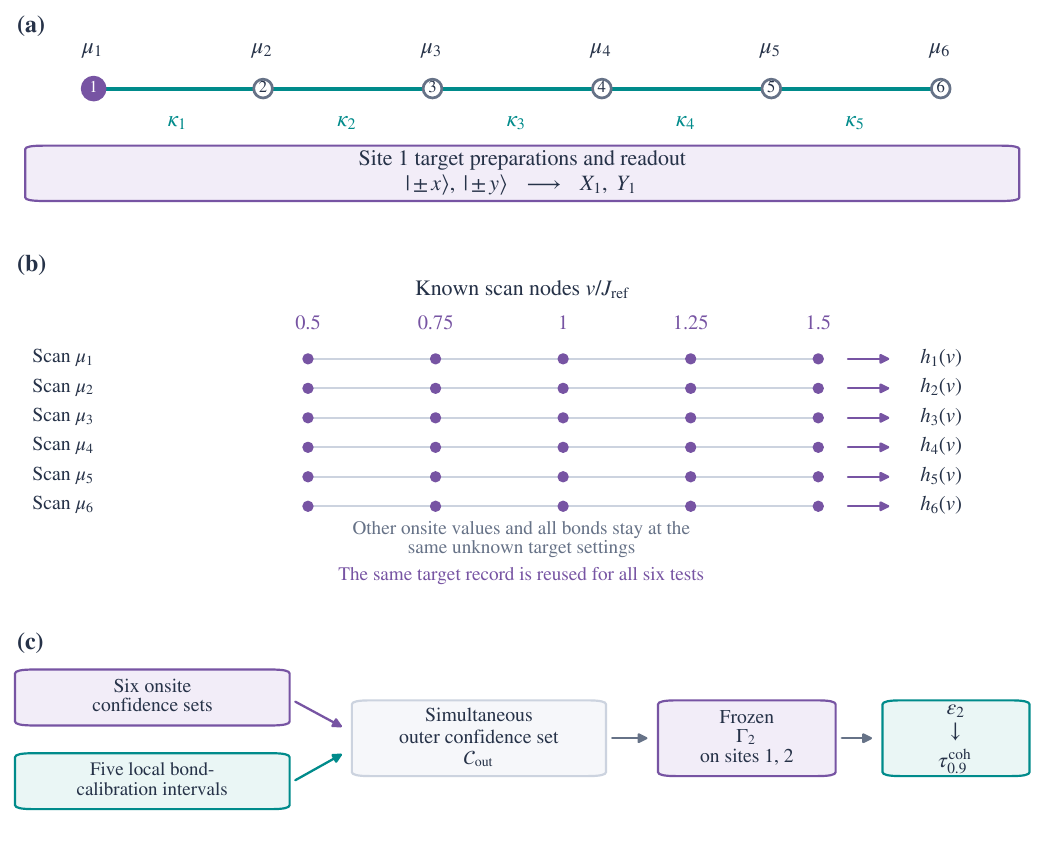}
\caption{Six-site coordinate-scan protocol. (a) The target has six unknown
onsite values and five uncertain equal-line bonds, and opposite $X_1/Y_1$
preparations with $X_1/Y_1$ readout give its boundary record. (b) Each
scan replaces one onsite value by five known nodes and leaves the other
settings unchanged and unknown; $h_j(v)$ denotes its response coordinates,
and the same target record serves all six tests. The references need known
programming controls throughout the register, and bond calibration needs
local access. (c) Coordinate confidence sets and bond intervals form a
simultaneous outer set, over which the residual of the frozen $\Gamma_2$
gives its coherent-model time bound. Some coordinates can remain
unresolved.}
\label{SM-fig:supp-coordinate-scans}
\end{figure*}

The second acquisition treats all six onsite values and five bonds of
Eq.~\eqref{SM-eq:supp-marrakesh-model} as unknown, with the same domains,
times, step, and channels, and intended reset $|00000\rangle$. Its
references are six one-coordinate scans rather than a six-dimensional grid
(Fig.~\ref{SM-fig:supp-coordinate-scans}). Scan $j$ overrides only $x_j$ by a
known node $v$, leaving every other onsite value and bond at its unknown
target setting.

\paragraph*{Acquisition and frozen design.}
Each of the $6\times5$ reference settings has $D=12$ coordinates with 256
shots per sign, the target has 2048 shots per sign and coordinate, and
each bond has the six calibration settings above with 2048 shots each. The
720 reference, 24 target, and 30 calibration circuits give 184320, 49152,
and 61440 outcomes, for 774 circuits and 294912 outcomes. Job
\texttt{db8baumil99s73fpm1r0} ran on \texttt{ibm\_marrakesh} on 15 October
2026, on physical qubits $(95,94,93,92,91,98)$ in logical chain order. Its
29 symbolic native templates, compiled before coefficient binding, contain
at most 160 CZ gates, and the templates, angle-exposure audits, analysis
settings, and $\alpha$ of Eq.~\eqref{SM-eq:supp-marrakesh-candidate} were
frozen before submission.

\paragraph*{Univariate response.}
For target settings $(\bm x_\star,\bm k_\star)$, the scan mean is
\begin{equation}
h_{\iota j}(v)=h_\iota(x_{1,\star},\ldots,x_{j-1,\star},v,
x_{j+1,\star},\ldots,x_{6,\star};\bm k_\star).
\label{SM-eq:supp-scan-context}
\end{equation}
Since $h_{\iota j}(x_{j,\star})$ equals the target mean for every $j$, the
learner compares the target with the measured curve without knowing its
context. With the nodes and stencil rule of
Sec.~\ref{SM-sec:supp-ibm-three-site}, the quadratic Lagrange basis $L_r(v)$
of the selected nodes $v_0,v_1,v_2$ gives
$\widehat q_{\iota j}(v)=\sum_{r=0}^{2}L_r(v)\widehat d_{\iota jr}$, and
the univariate Lagrange remainder with
Eq.~\eqref{SM-eq:supp-marrakesh-derivative} gives
\begin{equation}
|h_{\iota j}(v)-q_{\iota j}(v)|
\le R_{\iota j}(v)
=\frac{(u_\ell^+)^3}{6}\prod_{r=0}^{2}|v-v_r|.
\label{SM-eq:supp-scan-remainder}
\end{equation}

\paragraph*{Simultaneous coverage.}
The six scans share the target batch, and at the true $x_{j,\star}$ the
comparison with independent reference shots obeys the weighted Hoeffding
bound of Sec.~\ref{SM-sec:supp-ibm-three-site}. A union bound over $LD=72$
tests gives
\begin{equation}
r_{\iota j}(v)=
\sqrt{\log\!\left(\frac{2LD}{\delta_{\rm bd}}\right)
\left[\frac1{N_{\rm t}}+
\sum_{r=0}^{2}\frac{L_r(v)^2}{N_{\iota j r}}\right]},
\quad L=6,\quad N_{\rm t}=2048,\quad
N_{\iota j r}=256,\quad \delta_{\rm bd}=0.01.
\label{SM-eq:supp-scan-radius}
\end{equation}
The retained set for coordinate $j$ is
\begin{equation}
\mathcal C_j=\left\{v\in[0.5,1.5]\ \middle|\
|\widehat d_{\iota,\rm t}-\widehat q_{\iota j}(v)|
\le r_{\iota j}(v)+R_{\iota j}(v)+2\times10^{-10}
\ \text{for every }\iota\right\}.
\label{SM-eq:supp-scan-set}
\end{equation}
Hence $\Pr[x_{j,\star}\in\mathcal C_j\text{ for every }j]\ge0.99$ under
the transfer model, and the shared target data correlate these events
without affecting the union bound. The 25 known-control means and five
operating-bond means have allocations $0.005$ each and simultaneous
half-widths at most $0.094839196$ and $0.086155420$ including rounding,
and the one-rotation transfer of Eq.~\eqref{SM-eq:supp-marrakesh-calibration}
and all ten midpoint checks are retained. With these bond sets
$\mathcal K_j$, the simultaneous outer set is
\begin{equation}
\mathcal C_{\rm out}=
\left(\prod_{j=1}^{6}\mathcal C_j\right)
\times\left(\prod_{j=1}^{5}\mathcal K_j\right),\qquad
\Pr[(\bm x_\star,\bm k_\star)\in\mathcal C_{\rm out}]\ge0.98.
\label{SM-eq:supp-scan-coverage}
\end{equation}
This product is a conservative enclosure, and every operator bound is
evaluated throughout it.

\paragraph*{Validated inversion and measured result.}
The search encloses the quadratic means, squared weights, and cubic
remainders by directed rounding and bisects to width $2^{-10}$, leaving no
budget-unresolved intervals. The onsite hulls are
\begin{equation}
\begin{split}
x_1&\in[0.5,0.9267578125],\qquad
x_2\in[0.6171875,1.3564453125],\\
x_j&\in[0.5,1.5],\qquad j=3,4,5,6.
\end{split}
\label{SM-eq:supp-scan-measured-onsites}
\end{equation}
Only the first two onsite ranges narrow. The calibrated bond hulls,
displayed outward, are
\begin{equation}
\begin{split}
k_1&\in[1.682812,2.131251],\quad k_2\in[1.6,2.4],\\
k_3&\in[1.759374,2.242188],\quad
k_4\in[1.801562,2.300001],\quad
k_5\in[1.751562,2.234375].
\end{split}
\label{SM-eq:supp-scan-measured-bonds}
\end{equation}
For the frozen operator of Eq.~\eqref{SM-eq:supp-marrakesh-candidate},
Eq.~\eqref{SM-eq:supp-marrakesh-residual} gives residual bounds
$1.021697J_{\rm ref}$, $0.998100J_{\rm ref}$, and $0.550340J_{\rm ref}$
for the initial, calibration-only, and learned sets, and the unrounded
enclosures give
\begin{equation}
\tau_{0.9}^{\rm coh}J_{\rm ref}\ge
\begin{cases}
0.43771,&\text{initial domain},\\
0.44806,&\text{bond calibration},\\
0.81261,&\text{calibration and boundary learning}.
\end{cases}
\label{SM-eq:supp-scan-measured-times}
\end{equation}
The boundary record strengthens the calibration-only guarantee by about
$81\%$ without changing the candidate or the programmed device. The two
studies use different grids, budgets, and enclosures and are not a
scaling comparison. Both acquisitions keep the programmed target values
out of the learning input, so any comparison with them is a post-analysis
audit.

\section{Relation to experiments}
\label{SM-sec:supp-experiments}

This section evaluates the quantum-dot and spectroscopy statements of
main-text \secref{sec:experiments}.

\subsection{Splittings for reported couplings}
\label{SM-sec:supp-pfaffian}

Random equal-line chains with $2\le L\le8$ obey main-text
Eq.~\eqref{eq:pfaffian_identity} to relative precision
$4\times10^{-12}$. At $\mu=0$ each pair $(b_j,a_{j+1})$ forms a mode of
energy $\kappa_j$, so by that identity a small common detuning $\delta$ of
three sites gives
$E_1=\delta^3/(E_2E_3)\simeq\delta^3/(\kappa_1\kappa_2)$, asymptotically
attaining the bound of main-text Eq.~\eqref{eq:splitting_hierarchy}. The
couplings reported at the sweet spot, $\mu_j=0$ and $J_j=\Delta_j$, are
$J_1=\Delta_1\simeq20\,\mu$eV and $J_2=\Delta_2\simeq25\,\mu$eV in an
InSbAs two-dimensional electron gas~\cite{tenHaaf2025} and
$|2J_1|\simeq30\,\mu$eV and $|2J_2|\simeq20\,\mu$eV in an InSb
nanowire~\cite{Bordin2026}, so that $(\kappa_1,\kappa_2)=(40,50)$ and
$(30,20)\,\mu$eV. Of the seven patterns with $\mu_j\in\{0,\delta\}$ and
$0\le\delta\le10\,\mu$eV, those with one or two detuned dots give
$E_1<5\times10^{-14}\,\mu$eV, zero to rounding. Detuning all three dots by
$5\,\mu$eV gives $E_1=0.0617$ and $0.199\,\mu$eV, or $98.7\%$ and $95.5\%$
of $\delta^3/(\kappa_1\kappa_2)$, and the two-dot chain on the first bond
gives $0.616$ and $0.811\,\mu$eV, close to $\delta^2/\kappa_1$.

\subsection{Static-offset persistence bounds from dot-level precision}
\label{SM-sec:supp-experiment-guarantees}

Table~\ref{SM-tab:supp-experiment-guarantees} lists the static-offset bounds
of main-text \secref{sec:three-site-experiments},
$\tau_{0.9}\ge0.29436\,\mathrm{ns}/(\varepsilon/\mu\mathrm{eV})$ with
$\varepsilon\le\delta^m/\prod_{j<m}\kappa_j$; for $m=1$ the candidate is
$a_1$, which needs only the first bond on the equal line. The exact
residual $|\mu_m\alpha_m|/\mathcal N_m$ increases with every $|\mu_j|$, so
its worst case, at $|\mu_j|=\delta$, lies within $2\%$ of the product
bound in every entry.

\begin{table}[t]
\caption{Static-offset bounds $\tau_{0.9}$ for profile-dependent depth-$m$
operators on the exact equal line with every dot level within
$|\mu_j|\le\delta$, for the couplings of Refs.~\cite{tenHaaf2025} and
\cite{Bordin2026}. For $m\ge2$, $W_1\ge1-(\delta/\kappa_{\min})^2$. The
times bound the model autocorrelation horizon from below, and all entries
are rounded conservatively.}
\label{SM-tab:supp-experiment-guarantees}
\centering
\begin{ruledtabular}
\begin{tabular}{c c c c c c c c}
& & \multicolumn{3}{c}{$\kappa=(40,50)\,\mu$eV \cite{tenHaaf2025}}
& \multicolumn{3}{c}{$\kappa=(30,20)\,\mu$eV \cite{Bordin2026}}\\
$\delta$ ($\mu$eV) & $m$ & $\varepsilon$ bound ($\mu$eV) & $\tau_{0.9}$ bound & $W_1\ge$
& $\varepsilon$ bound ($\mu$eV) & $\tau_{0.9}$ bound & $W_1\ge$\\
\hline
1 & 1 & 1 & 0.2943 ns & 1 & 1 & 0.2943 ns & 1\\
1 & 2 & 0.025 & 11.77 ns & 0.9993 & 0.0334 & 8.830 ns & 0.9975\\
1 & 3 & $5.0\times10^{-4}$ & 588.7 ns & 0.9993 & $1.67\times10^{-3}$ & 176.6 ns & 0.9975\\
3 & 1 & 3 & 98.12 ps & 1 & 3 & 98.12 ps & 1\\
3 & 2 & 0.225 & 1.308 ns & 0.9943 & 0.300 & 0.9812 ns & 0.9775\\
3 & 3 & 0.0135 & 21.80 ns & 0.9943 & 0.0450 & 6.541 ns & 0.9775\\
5 & 1 & 5 & 58.87 ps & 1 & 5 & 58.87 ps & 1\\
5 & 2 & 0.625 & 0.4709 ns & 0.9843 & 0.834 & 0.3532 ns & 0.9375\\
5 & 3 & 0.0625 & 4.709 ns & 0.9843 & 0.209 & 1.412 ns & 0.9375
\end{tabular}
\end{ruledtabular}
\end{table}

\subsection{Off-line leakage and simultaneous coefficient uncertainty}
\label{SM-sec:supp-off-line-dot-bound}

For real coefficients, let $s_j=J_j+\Delta_j>0$ and $d_j=\Delta_j-J_j$,
and let $K^{(0)}$ share the onsite terms with equal-line couplings
$J_j^{(0)}=\Delta_j^{(0)}=s_j/2$. Its depth-$m$ candidate
$\overline{\bm g}=(g_{a_1},0,\ldots,g_{a_m},0,0,\ldots)^T$, with
$g_{a_j}=\alpha_j/\mathcal N_m$, $\alpha_1=1$, and
$\alpha_{j+1}=-\mu_j\alpha_j/s_j$, has reference residual
$|\mu_mg_{a_m}|$. Since $D=K-K^{(0)}$ has only the entries
$D_{a_jb_{j+1}}=-D_{b_{j+1}a_j}=d_j$, whose output coordinates are
distinct,
\begin{equation}
D\overline{\bm g}=-\sum_{j=1}^{q}d_jg_{a_j}\bm e_{b_{j+1}},
\qquad
\|D\overline{\bm g}\|_2^2=\sum_{j=1}^{q}d_j^2g_{a_j}^2,
\qquad q=\min(m,L-1).
\label{SM-eq:supp-dot-leakage-norm}
\end{equation}
The triangle inequality then gives the nonperturbative estimate
\begin{equation}
E_1\le\varepsilon_m\le|\mu_mg_{a_m}|+
\left(\sum_{j=1}^{q}d_j^2g_{a_j}^2\right)^{1/2}.
\label{SM-eq:supp-off-line-reference}
\end{equation}
For $m<L$, $d_mg_{a_m}$ is leakage onto $b_{m+1}$ through the outgoing
bond. With mismatches, a zero onsite coefficient no longer forces $E_1=0$.

For a simultaneous uncertainty set at the operating point with
$|\mu_j|\le\delta$, $s_j\ge s_{j,-}>0$, and $|d_j|\le d_{\max}$, and for
$m\ge2$ with $s_{\min}=\min_{j<m}s_{j,-}$ and $r_+=\delta/s_{\min}<1$, the
bounds $\mathcal N_m\ge1$, $\sum_jg_{a_j}^2\le1$, and
$|\alpha_{j+1}/\alpha_j|\le r_+$ give, for each compatible profile,
\begin{equation}
E_1\le\varepsilon_m\le
\frac{\delta^m}{\prod_{j<m}s_{j,-}}+d_{\max},
\qquad W_\ell\ge1-r_+^{2\ell},\quad 1\le\ell<m.
\label{SM-eq:supp-off-line-dot-box}
\end{equation}
For $m=1$, $W_1=1$ and the residual is at most $\delta+d_{\max}$, or
exactly $|\mu_1|$ for $L=1$. The time $\hbar\sqrt{0.2}/\varepsilon_m$
concerns this profile-dependent operator. The full chiral recurrence
sharpens this bound for a terminal three-site
chain.

\paragraph{Full recurrence for a terminal three-site chain.}
For $L=m=3$ and $\bm g=(\alpha_1,0,\alpha_2,0,\alpha_3,0)^T/\mathcal N_3$,
the $b_j$ component of $K\bm g$ is
$(\mu_j\alpha_j+s_j\alpha_{j+1}-d_{j-1}\alpha_{j-1})/\mathcal N_3$, with
missing endpoint terms absent. Cancelling the $b_1$ and $b_2$ components
gives
\begin{equation}
\alpha_1=1,\qquad
\alpha_2=-\frac{\mu_1}{s_1},\qquad
\alpha_3=\frac{d_1+\mu_1\mu_2/s_1}{s_2},\qquad
\mathcal N_3=\sqrt{1+\alpha_2^2+\alpha_3^2}.
\label{SM-eq:supp-three-site-full-recurrence}
\end{equation}
Only the terminal $b_3$ component remains, so
\begin{equation}
\varepsilon_3=
\frac{|\mu_3\alpha_3-d_2\alpha_2|}{\mathcal N_3},
\qquad W_1=\frac{1}{1+\alpha_2^2+\alpha_3^2}.
\label{SM-eq:supp-three-site-full-residual}
\end{equation}
The numerator equals
$|\mu_1\mu_2\mu_3/(s_1s_2)+\mu_3d_1/s_2+\mu_1d_2/s_1|$, so
$\mathcal N_3\ge1$ and the simultaneous bounds give main-text
Eq.~\eqref{eq:off-line-dot-bound}. With $1/(1+x)\ge1-x$, the same box
bounds the weight,
\begin{equation}
W_1\ge1-\left(\frac{\delta}{s_{1,-}}\right)^2
-\left(\frac{d_{\max}+\delta^2/s_{1,-}}{s_{2,-}}\right)^2.
\label{SM-eq:supp-three-site-full-localization}
\end{equation}
When this exceeds $w_{\min}$, the profile-dependent construction also
bounds the localized existence residual.

For $\delta=d_{\max}=1\,\mu$eV and $(s_{1,-},s_{2,-})=(40,50)\,\mu$eV, the
box residual is at most $0.0455\,\mu$eV, so $\tau_{0.9}\ge6.46\,$ns with
$W_1\ge0.9989$, against $1.0005\,\mu$eV and $0.294\,$ns from
Eq.~\eqref{SM-eq:supp-off-line-dot-box}. Direct diagonalization at
$\mu_j=d_j=1\,\mu$eV agrees with the full-recurrence residual to about
$0.05\%$. The terminal three-site mismatch contribution scales as
$\delta d_{\max}$, whereas the two-site residual
$|\mu_1\mu_2/s_1+d_1|/\sqrt{1+(\mu_1/s_1)^2}$ can remain of order $|d_1|$
at zero offsets. A three-site prefix of a longer chain adds the outgoing
component $-d_3\alpha_3/\mathcal N_3$.

\begin{samepage}
A candidate $\overline{\bm g}$ frozen at nominal equal-line coefficients
$\bar\mu_j,\bar s_j$ must also absorb their operating-point uncertainty.
For joint bounds $|\mu_j-\bar\mu_j|\le\eta_{\mu,j}$,
$|s_j-\bar s_j|\le\eta_{s,j}$, and $|d_j|\le\eta_{d,j}$, the triangle
inequality over the onsite, sum-coupling, and mismatch perturbations
gives, without assuming independent errors,
\begin{align}
\sup_{K\in\mathcal C_m}\|K\overline{\bm g}\|_2
\le{}&|\bar\mu_mg_{a_m}|
+\left(\sum_{j=1}^{m}\eta_{\mu,j}^2g_{a_j}^2\right)^{1/2}\nonumber\\
&+\left(\sum_{j=1}^{m-1}\eta_{s,j}^2g_{a_{j+1}}^2\right)^{1/2}
+\left(\sum_{j=1}^{q}\eta_{d,j}^2g_{a_j}^2\right)^{1/2}.
\label{SM-eq:supp-frozen-dot-box}
\end{align}
\end{samepage}

\subsection{Thermally broadened end spectroscopy}
\label{SM-sec:supp-thermal-spectroscopy}

For two chains, main-text Eq.~\eqref{eq:thermal_spectrum} with
$|v^\dagger\Delta Y(t)v|\le\|\Delta Y(t)\|_2\le\min\{2,\mathcal D_m(t)\}$
and the conjugate symmetry $v^\dagger Y(-t)v=\overline{v^\dagger Y(t)v}$
gives
$|\Delta\mathcal A_{T_e}(E)|\le(\pi\hbar)^{-1}\int_0^\infty\min\{2,\mathcal D_m(t)\}\phi_{T_e}(t)\,dt$.
Since the site-1 spectral function is nonnegative with unit weight,
$\mathcal A_{T_e}\le1/(4k_BT_e)$, the normalization of main-text
Eq.~\eqref{eq:spectral_obstruction}. With
$\int_0^\infty\phi_T\,dt=\pi\hbar/(4k_BT)$ and
$\int_X^\infty x/\sinh x\,dx\le2(1+X)e^{-X}/(1-e^{-2X})$, splitting the
integral at any $T>0$ gives
\begin{equation}
4k_BT_e\sup_E|\Delta\mathcal A_{T_e}(E)|
\le\mathcal D_m(T)+\frac{16}{\pi^2}\,\frac{(1+X)e^{-X}}{1-e^{-2X}},
\qquad X=\frac{\pi k_BT_eT}{\hbar}.
\label{SM-eq:supp-two-term}
\end{equation}
Lorentzian lifetime broadening of half-width $\gamma$ multiplies the time
weight by $e^{-\gamma|t|/\hbar}$ and preserves the normalization, so both
bounds still hold.

We evaluate $\mathcal D_m(t)$ for the pair of main-text
Fig.~\ref{fig:matched_prefix_summary}, with $G_{\rm top}=3.4J$,
$G_{\rm triv}=4.8J$, and $\Lambda_\star=1.6J$, using the majorants of
Sec.~\ref{SM-sec:supp-nogo-proof} with step $0.005/J$ up to $tJ=40$ and
$\min\{2,\mathcal D_m\}=2$ with the closed-form tail beyond. At $tJ=8$
this reproduces the bounds of Sec.~\ref{SM-sec:supp-nogo-numerics}. The
actual spectra use exact diagonalization of $L=600$ truncations on an
energy grid of spacing $k_BT_e/25$ over $|E|\le5.2J$, which contains the
band edge $4.8J$. Main-text Eq.~\eqref{eq:pfaffian_identity} excludes a
finite-$L$ zero eigenvalue, since all onsite coefficients are nonzero, and
$L=800$ changes every spectrum by less than $10^{-14}$.
Table~\ref{SM-tab:supp-spectroscopy} compares the actual and bounded
differences. The bound falls below $10^{-2}$ ($10^{-3}$) for $m\ge24$
(33), 13 (17), and 7 (9) at $k_BT_e/J=0.1$, $0.2$, and $0.4$, and stays
above $10^{-2}$ through $m=40$ at $0.05$ and $0.025$.

\begin{table}[t]
\caption{Largest relative difference $4k_BT_e\sup_E|\Delta\mathcal A_{T_e}|$
of the thermally broadened end spectra of the matched-prefix pair: actual
value for $L=600$ truncations / bound of main-text
Eq.~\eqref{eq:spectral_obstruction}. Bound columns are rounded upward.}
\label{SM-tab:supp-spectroscopy}
\centering
\begin{ruledtabular}
\begin{tabular}{c c c c c c}
$m$ & $k_BT_e=0.025J$ & $0.05J$ & $0.1J$ & $0.2J$ & $0.4J$\\
\hline
2 & 0.31 / 1 & 0.31 / 1 & 0.30 / 1 & 0.24 / 1 & 0.10 / 0.64 \\
4 & 0.26 / 1 & 0.24 / 1 & 0.15 / 1 & $5.2\times10^{-2}$ / 0.58 & $6.9\times10^{-3}$ / 0.12 \\
6 & 0.24 / 1 & 0.14 / 1 & $4.6\times10^{-2}$ / 0.80 & $1.2\times10^{-2}$ / 0.24 & $3.9\times10^{-4}$ / $1.6\times10^{-2}$ \\
8 & 0.14 / 1 & $4.8\times10^{-2}$ / 1 & $2.4\times10^{-2}$ / 0.53 & $2.8\times10^{-3}$ / $9.0\times10^{-2}$ & $2.0\times10^{-5}$ / $1.9\times10^{-3}$ \\
10 & $4.8\times10^{-2}$ / 1 & $3.1\times10^{-2}$ / 0.94 & $1.2\times10^{-2}$ / 0.34 & $6.3\times10^{-4}$ / $3.3\times10^{-2}$ & $9.9\times10^{-7}$ / $2.1\times10^{-4}$ \\
12 & $2.8\times10^{-2}$ / 1 & $2.3\times10^{-2}$ / 0.77 & $5.9\times10^{-3}$ / 0.22 & $1.3\times10^{-4}$ / $1.2\times10^{-2}$ & $4.7\times10^{-8}$ / $2.2\times10^{-5}$ \\
14 & $2.5\times10^{-2}$ / 1 & $1.7\times10^{-2}$ / 0.63 & $2.8\times10^{-3}$ / 0.14 & $2.7\times10^{-5}$ / $3.9\times10^{-3}$ & $2.2\times10^{-9}$ / $2.2\times10^{-6}$ \\
16 & $2.1\times10^{-2}$ / 1 & $1.2\times10^{-2}$ / 0.51 & $1.3\times10^{-3}$ / $7.9\times10^{-2}$ & $5.4\times10^{-6}$ / $1.3\times10^{-3}$ & $9.6\times10^{-11}$ / $2.2\times10^{-7}$
\end{tabular}
\end{ruledtabular}
\end{table}

{\color{black}
\section{Relation to conservation-law and Krylov learning}
\label{SM-sec:supp-related-learning}

\paragraph*{Conservation-law learning.}
Zhan et al. learn sums of geometrically local Pauli observables
using randomized single-qubit measurements across the measured region
\cite{ZhanConservation2024}. Their framework admits closed and
open dynamics. Its finite-shot tests concern expectation-value
conservation for a selected initial-state ensemble, with derivative
bounds supporting continuous-time interpolation. Our residual controls
operator-norm persistence for every initial state throughout a coefficient
confidence set. This uses the declared quadratic-chain structure or the
separately stated site-1 cut. The validated set inversion has no general
polynomial resource guarantee.

\paragraph*{Probe spectral learning.}
Quantum Krylov Learning obtains a probe spectral measure and its
operator-Krylov Jacobi matrix from opposite probe preparations, free
evolution, and probe readout \cite{MatsuuraQuantumKrylov2026}.
The unknown Hamiltonian is closed and time independent, without a
prescribed microscopic parameterization. The exact construction uses
a maximally mixed hidden system, and finite-shot resolution depends
on spectral spacing and weights. Krylov coordinates label operator
space. Our candidates have physical-site support and explicit
localization and outgoing-coupling bounds. The quadratic response
protocol permits a common unknown
hidden state initially uncorrelated with the prepared boundary qubit.
Both spectral and coefficient learning can constrain autocorrelation.
Our benchmarks also use the stated calibration and reference controls.

\paragraph*{Majorana locality and protection.}
Samuelson et al. use fermionic partial traces to quantify Majorana
locality and bound environmental couplings and spectral splittings
\cite{Samuelson2026}. The main interacting construction uses a gapped
opposite-parity low-energy pair and weak coupling. Appendix G also
treats a noninteracting subsystem coupled
to an interacting environment in the full Hilbert space, including
arbitrarily strong coupling. Our contribution is the inference of a
covered residual from finite uncertain measurements and its
propagation into a persistence time. Their many-body locality
measures and our linear-Majorana coefficient weights serve different
operator classes. Persistence alone does not establish protected
qubit storage or braiding performance.
}


\end{document}